\documentclass[a4paper,fleqn,usenatbib]{mnras}

\usepackage[T1]{fontenc}
\usepackage{ae,aecompl}

\usepackage{graphicx}   
\usepackage{amsmath}    
\usepackage{amssymb}    
\usepackage{txfonts}
\usepackage{longtable}
\usepackage{morefloats}

\title[Chemical abundances of symbiotic giants -- III]{Chemical abundance
analysis of symbiotic giants -- III.  Metallicity and CNO abundance patterns
in 24 southern systems}
\author[C. Ga{\l}an et al.]{Cezary Ga{\l}an,$^{1}$\thanks{E-mail: cgalan@camk.edu.pl} Joanna Miko{\l}ajewska,$^{1}$\thanks{E-mail: mikolaj@camk.edu.pl} Kenneth H.
Hinkle$^{2}$\thanks{E-mail: hinkle@noao.edu} and Richard R. Joyce$^{2}$
\\
$^{1}$N. Copernicus Astronomical Center, Bartycka 18, PL-00-716 Warsaw, Poland\\
$^{2}$National Optical Astronomy Observatory, PO Box 26732, Tucson, AZ 85726, USA}

\begin{document}

\date{Accepted 2015 October 9. Received 2015 October 8; in original form 2014 December 23}

\pagerange{\pageref{firstpage}--\pageref{lastpage}} \pubyear{2015}

\maketitle

\label{firstpage}

\begin{abstract}
The elemental abundances of symbiotic giants are essential to address the
role of chemical composition in the evolution of symbiotic binaries, to map
their parent population, and to trace their mass transfer history.  However,
the number of symbiotic giants with fairly well determined photospheric
composition is still insufficient for statistical analyses.
This is the third in a series of papers on the chemical composition of
symbiotic giants determined from high resolution ($R$\,$\sim$\,50000),
near-infrared spectra.  Here we present results for 24 S-type systems. 
Spectrum synthesis methods employing standard local thermal equilibrium
analysis and atmosphere models were used to obtain photospheric abundances
of CNO and elements around the iron peak (Fe, Ti, Ni, and Sc).
Our analysis reveals metallicities distributed in a wide range from slightly
supersolar ([Fe$/$H]$\sim+0.35$\,dex) to significantly subsolar
([Fe$/$H]$\sim-0.8$\,dex) but principally with near-solar and slightly
subsolar metallicity ([Fe$/$H]$\sim -0.4$ to $-0.3$\,dex).  The enrichment in
$^{14}$N isotope, found in all these objects, indicates that the giants have
experienced the first dredge-up.  This was confirmed in a number of objects
by the low $^{12}$C/$^{13}$C ratio (5--23).  We found that the relative
abundance of [Ti/Fe] is generally large in red symbiotic systems.

\end{abstract}


\section{Introduction}

Symbiotic stars are long-period binary systems consisting of two stars
representing a late stage in stellar evolution: the cool primary and hot and
luminous secondary (typically white dwarf albeit a neutron star has been
found in a few cases) surrounded by an ionized nebula.  Based on their
near-infrared (IR) characteristics, symbiotic stars are divided into two
main classes: S-type with normal red giant ($\sim$80\,per\,cent), and D-type
with Mira variable embedded in an optically thick dust shell
($\sim$20\,per\,cent).  A strong interaction between components is driven by
mass loss from the cool donor that is partly accreted from the wind and$/$or
via Roche lobe overflow (\citealt{Pod2007,Mik2012}) on to the hot companion. 
In the past, when the present compact object underwent its red giant stage,
mass had to be transferred in the opposite direction from this star to the
star that is currently a red giant.  That mass transfer episode should have
left traces in the chemical composition of the red giant observed today. 
Indeed such chemical pollution has been detected in some red giant--white
dwarf binary systems \citep{SmLa1988}.

Knowledge of the atmospheric chemical composition of symbiotic giants is of
special significance as it can be used to track the mass exchange history as
well as their population origin.  However, at the moment reliable
measurements of photospheric compositions exist for only 10 symbiotic
systems with late-type (M) giants and about a dozen `yellow', i.e.  G or K
giant, symbiotic systems.  Prior to the current series of papers only four M
giants in S-type symbiotic systems had been analysed in the literature:
V2116\,Oph \citep{Hin2006}, T\,CrB, RS\,Oph \citep{Wal2008}, and CH\,Cyg
\citep{Sch2006}.  All of them had solar or nearly solar metallicities.  The
rarer symbiotic stars containing K-type giants are metal poor with s-process
elements overabundant (\citealt[1997]{Smi1996}; \citealt{Per1998,Per2009})
whereas those with G-type giants have solar metallicity and s-process
enhancement (\citealt{Smi2001b,Per2005}).

\begin{table*}
 \centering
  \caption{Journal of spectroscopic observations. Quadrature sums of the
projected rotational velocities and microturbulence$\,^a$ $(V_{\rmn{rot}}^2
\sin^2{i} + \xi^2_{\rmn{t}})^{0.5}$ shown have been obtained via
cross-correlation technique (CCF) and from measurement of full width at
half-maximum (FWHM) of $K$ band
\mbox{Ti\,{\sc i}}, \mbox{Fe\,{\sc i}}, and \mbox{Sc\,{\sc i}} absorption
lines.  Orbital phases have been calculated according to the referenced
literature ephemeris.  The full table with all 24 objects included is
shown in the online Appendix\,A (Table\,\ref{TA1}).}
\label{T1}
  \begin{tabular}{@{}lccccccc@{}}
  \hline
                      & \   Id. num.$^{b}$  & Sp.\,region               & Date         & HJD (mid) & \multicolumn{2}{c}{$(V_{\rmn{rot}}^2 \sin^2{i} + \xi^2_{\rmn{t}})^{0.5}$} & Orbital phase$^{c}$\\
                      &                     & band ($\lambda$[\micron]) & (dd.mm.yyyy) &           & CCF                            & FWHM                                     &                    \\
 \hline
                      &     & $H$ ($\sim$1.56)         & 16.02.2003 & 245 2686.7409 & ~6.08 & --                     & 0.30 \\
 BX\,Mon              &  23 & $K$ ($\sim$2.23)         & 20.04.2003 & 245 2749.5231 & ~7.58 & 8.67 $\pm$ 1.41        & 0.35 \\
                      &     & $K_{\rm r}$ ($\sim$2.36) & 03.04.2006 & 245 3828.5095 & ~8.44 & --                     & 0.20 \\
                      &     &                          &            &               &       & ~8.67 $\pm$ 1.41$^{d}$ &      \\
 \hline
                      &     & $H$ ($\sim$1.56)         & 16.02.2003 & 245 2686.7491 & ~4.19 & --                     & 0.39 \\
 V694\,Mon            &  24 & $K$ ($\sim$2.23)         & 20.04.2003 & 245 2749.5326 & ~6.34 & 8.42 $\pm$ 0.99        & 0.42 \\
                      &     & $K_{\rm r}$ ($\sim$2.36) & 03.04.2006 & 245 3828.5187 & ~7.21 & --                     & 0.98 \\
                      &     & $H_{\rm b}$ ($\sim$1.54) & 12.03.2010 & 245 5267.5052 & ~9.36 & --                     & 0.72 \\
                      &     &                          &            &               &       & ~8.42 $\pm$ 0.99$^{d}$ &      \\
...                   & ... & ...                      & ...        & ...           & ...   & ...                    & ...  \\
 \hline
\end{tabular}   
\begin{list}{}{}
\item[$Notes.$ $^{a}$Units $\rmn{km}\,\rmn{s}^{-1}$.]
\item[$^{b}$Identification number according to \citet{Bel2000}.]
\item[$^{c}$Orbital]\,phases are calculated from the following ephemerides:
BX\,Mon 2449796+1259$\times$E \citep{Fek2000}, V694\,Mon
2448080+1931$\times$E \citep{Gro2007a}, Hen\,3-461 2452063+635$\times$E
\citep{Gro2013}, SY\,Mus 2450176+625$\times$E \citep{Dum1999}, RW\,Hya
2445071.6+370.2$\times$E \citep{KeMi1995} or 2449512+370.4$\times$E
\citep{Sch1996}, Hen\,3-916 2452410+803$\times$E \citep{Gro2013},
Hen\,3-1213 2451806+514$\times$E \citep{Gro2013}, Hen\,2-173
2452625+911$\times$E \citep{Fek2007}.  KX\,TrA 2453053+1350$\times$E
\citep{Fer2003}, CL\,Sco 2452018+625$\times$E \citep{Fek2007}, V455\,Sco
2452641.5+1398$\times$E \citep{Fek2008}, Hen\,2-247 2452355+898$\times$E
\citep{Fek2008}, AE\,Ara 2453449+803.4$\times$E \citep{Fek2010}, AS\,270
2451633+671$\times$E \citep{Fek2007}, Y\,CrA 2454126+1619$\times$E
\citep{Fek2010}, Hen\,2-374 2453173+820$\times$E \citep{Fek2010}.
\item[$^{d}$Values]\,$(V_{\rmn{rot}}^2 \sin^2{i} + \xi^2_{\rmn{t}})^{0.5}$ obtained from all $K$-band spectra jointly -- used for synthetic spectra calculations.
\end{list} 
\end{table*}

The number of symbiotic giants with fairly well-determined photospheric
composition is too small to perform reliable statistical analysis.  To
improve this situation we have started a research program of chemical
composition measurements for southern S-type symbiotic systems.  The
motivation for this work and the first analysis of two classical S-type
symbiotic systems (RW\,Hya and SY\,Mus) were presented in
\citet[][hereafter Paper\,I]{Mik2014} and results for the next four systems
(AE\,Ara, BX\,Mon, KX\,Tra, and CL\,Sco) in \citet[][hereafter
Paper\,II]{Gal2015}.

This is the third in a series of papers on the chemical abundance analysis
of the symbiotic giants.  We present here the results obtained for 24 S-type
symbiotic systems observed from the Southern hemisphere.  The spectroscopic
observations and reductions are presented in Section 2.  The methods applied
to calculate abundances are discussed in Section 3 and the results in
Section 4.  In Section 5 we discuss briefly the obtained CNO, Fe, and Ti
abundances and compare them to the selected results from literature.  A
brief summary is given in Section 6.

\section[]{Observations and data reduction}

Table\,\ref{T1} lists information about the near-IR spectra that were
employed in this study.  All observations were acquired with Phoenix
cryogenic echelle spectrograph on the 8-m Gemini South telescope with high
resolving power ($R = \lambda/\Delta\lambda \sim 50000$) and, in most cases,
with a high signal-to-noise ratio (S/N $\ga$\,100).  The observed regions
were located in the $H$ and $K$ photometric bands centred at mean
wavelengths close to $\sim$1.54, 1.56, 2.23, and 2.36\,\micron (hereafter
$H$-, $H_{\rm b}$-, $K$-, and $K_{\rm r}$-band spectra, respectively).  All
the spectra cover narrow spectral ranges ($\sim$100\AA) that are typically
offset from each other, especially at $K$-band region, by several tens of
\AA\, from night to night mainly because of small differences in the grating
angle used, and to a lesser extent, due to the differences in the radial
velocities.  To extract and wavelength calibrate the spectra standard
reduction techniques were used \citep{Joy1992}.  In line with common
practice all the spectra were heliocentric corrected.  Telluric lines were
removed by reference to a hot standard star.  This was not necessary for
$H$-, $H_{\rm b}$-band regions that are free of telluric features.  The
Gaussian instrumental profile is about 6 km\,s$^{-1}$ full width at
half-maximum (FWHM), corresponding to an instrumental profile of
$\sim$0.31\AA\ in the case of the $H$- and $H_{\rm b}$-band spectra and
$\sim$0.44 and $\sim$0.47\AA\ in the case of the $K$- and $K_{\rm r}$-band
spectra, respectively.

We have from one to seven spectra for each target with $K$-band region
represented in all cases.  The $K$-band spectra were collected during four
observing runs in 2003 April, August, and December, and 2004 April.  This
spectral region contains moderately strong \mbox{Ti\,{\sc i}} lines as well
as a few other neutral atomic lines from \mbox{Fe\,{\sc i}} and
\mbox{Sc\,{\sc i}} all superimposed on the weak CN molecular lines from the
CN red system $\Delta\upsilon = -2$ transition.  The $H$-band spectra were
observed in 2003 February and during several observing runs in the years
2009--2010.  This region is dominated by first overtone OH lines and a
selection of neutral atomic lines \mbox{Fe\,{\sc i}}, \mbox{Ti\,{\sc i}},
\mbox{Ni\,{\sc i}} combined with weak red system $\Delta\upsilon = -1$ CN
lines and second-overtone CO vibration-rotation lines.  In 2010 March
several spectra in $H_{\rm b}$-band region were observed in poor weather
conditions but we found three of them to be suitable to include in our
analysis.  This region is dominated with OH and CN features with a small
admixture of Ti, Fe, Ni lines.  The selected absorption lines in $H$-,
$H_{\rm b}$-, and $K$-band spectra were useful to determine abundances of
carbon, nitrogen, and oxygen and elements around the iron peak: Sc, Ti, Fe,
Ni.  The $K_{\rm r}$-band spectra were acquired for 10 objects from our
sample in 2006 April.  This range is dominated by strong CO features that
are heavily blended.  Uncertainty in determining the continuum resulted in
our decision not to use these spectra to determine elemental abundances. 
However, we did use them to measure the $^{12}$C$/^{13}$C isotopic ratio.

\begin{figure}
  \includegraphics[width=84mm]{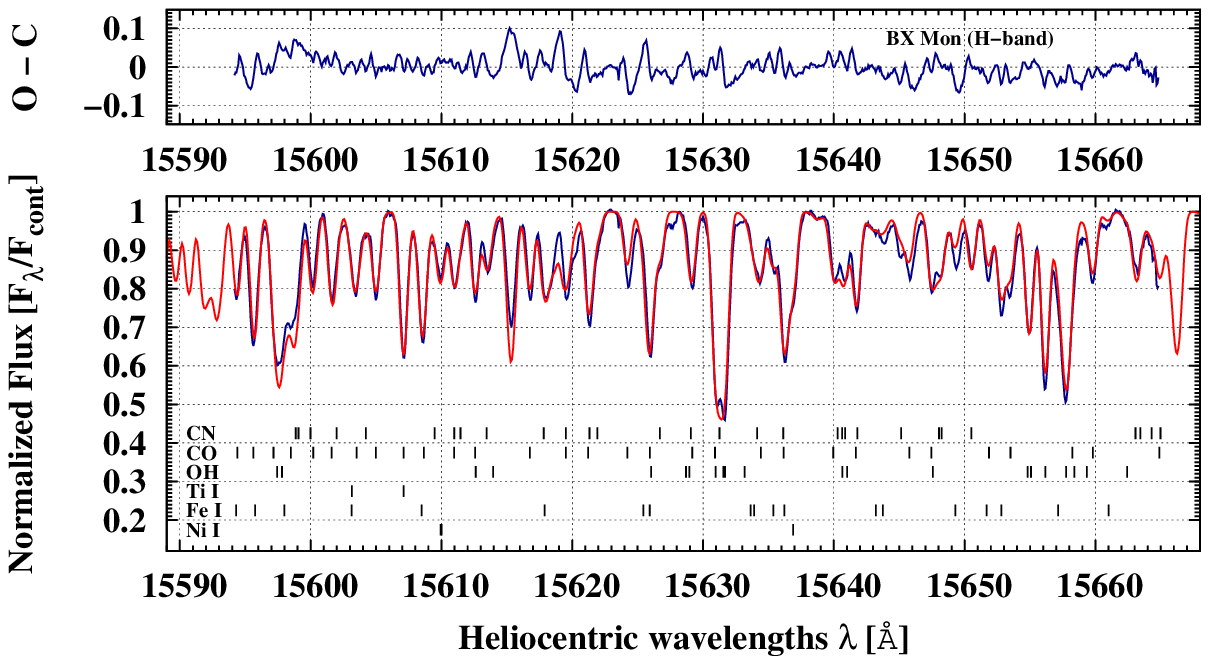}
  \caption{The $H$-band spectrum of BX\,Mon observed 2003 February  (blue
line) and a synthetic spectrum (red line) calculated using the final
abundances and $^{12}$C/$^{13}$C isotopic ratio (Table\,\ref{T4}).}
  \label{F1}
\end{figure}

\begin{figure}
  \includegraphics[width=84mm]{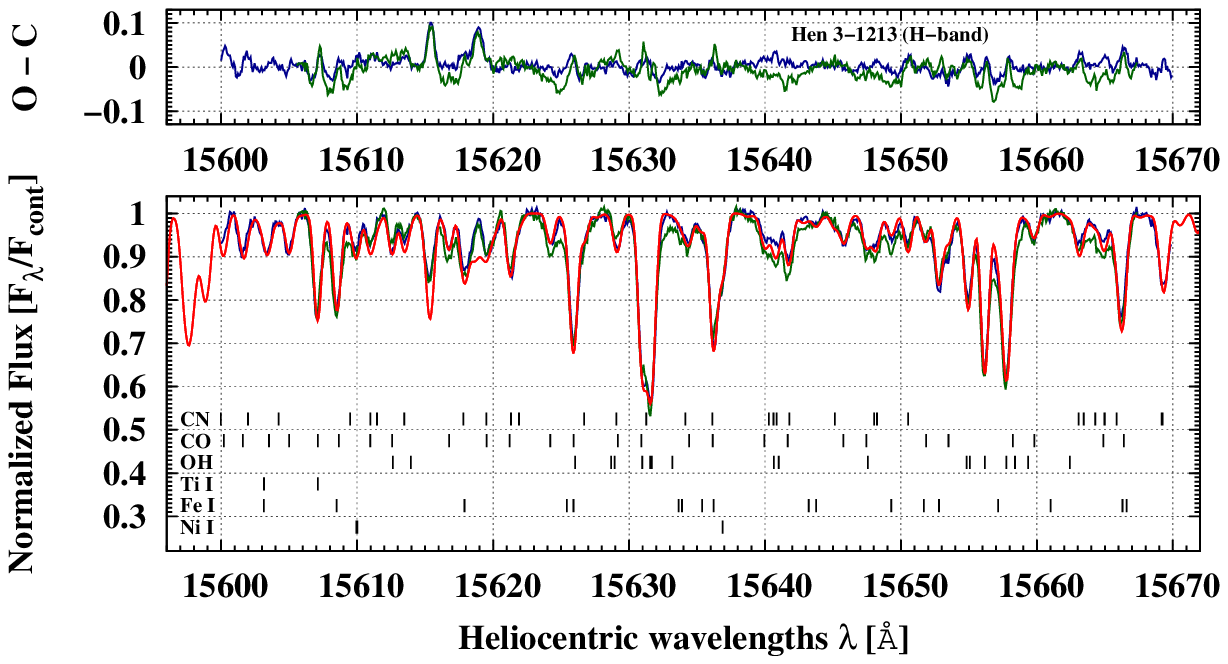}
  \caption{$H$-band spectra of Hen\,3-1213 observed 2003 February (blue
line), 2010 May (green line), and a synthetic spectrum (red line) calculated
using the final abundances (Table\,\ref{T4}).}
  \label{F2}
\end{figure}

\begin{figure}
  \includegraphics[width=84mm]{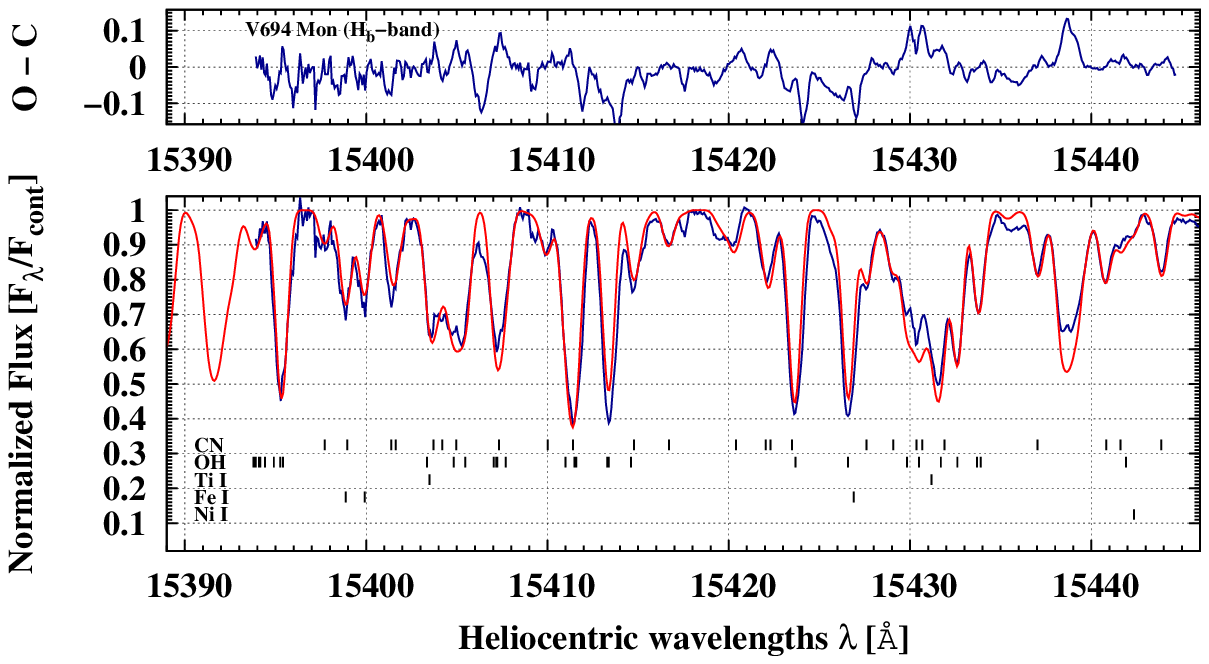}
  \caption{The $H_{\rm b}$-band spectrum of V694\,Mon observed 2010 March
(blue line) and a synthetic spectrum (red line) calculated using the final
abundances and $^{12}$C/$^{13}$C isotopic ratio (Table\,\ref{T4}).}
  \label{F3}
\end{figure}

\begin{figure}
  \includegraphics[width=84mm]{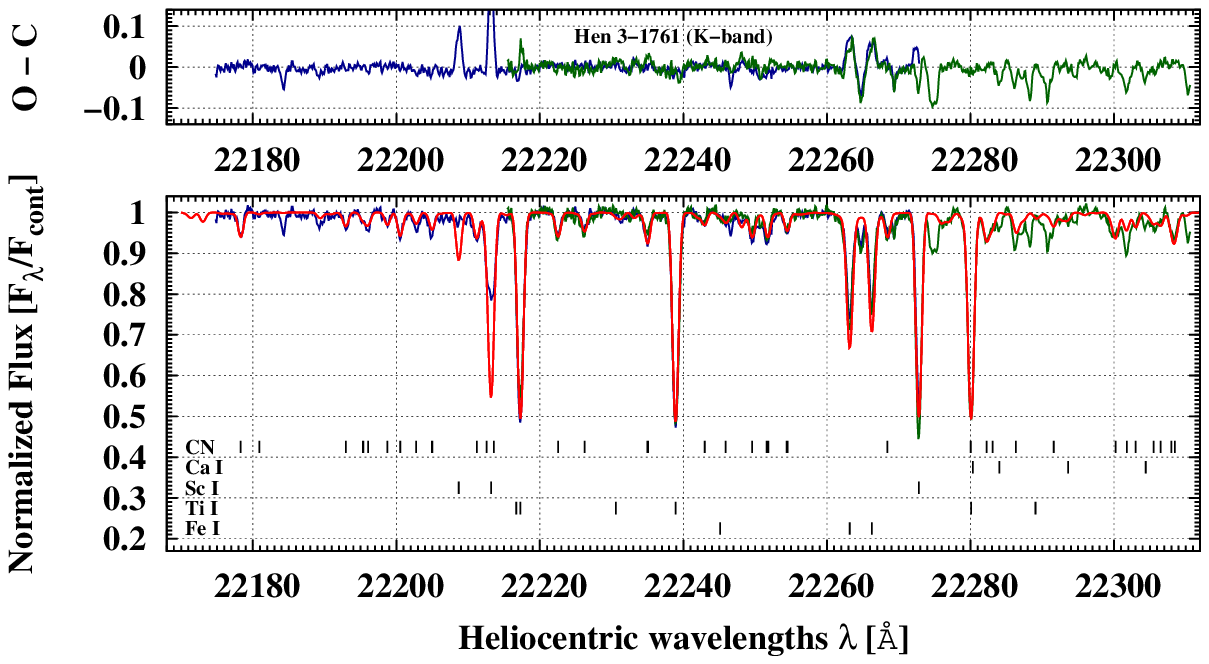} 
  \caption{$K$-band spectra of Hen\,3-1761 observed 2003 August (blue line),
2004 April (green line), and a synthetic spectrum (red line) calculated
using the final abundances (Table\,\ref{T4}).}
  \label{F4}
\end{figure}

\begin{figure}
  \includegraphics[width=84mm]{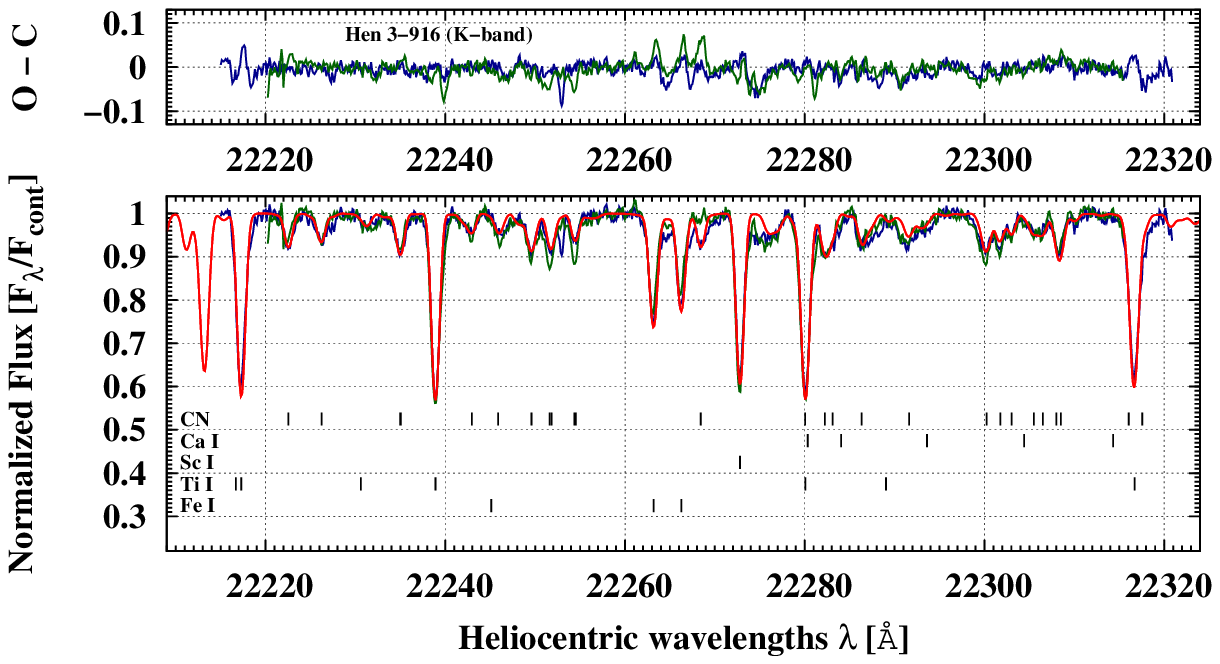} 
  \caption{$K$-band spectra of Hen\,3-916 observed 2003 April (blue line),
2004 April (green line), and a synthetic spectrum (red line) calculated
using the final abundances (Table\,\ref{T4}).}
  \label{F5}
\end{figure}

\begin{figure}
  \includegraphics[width=84mm]{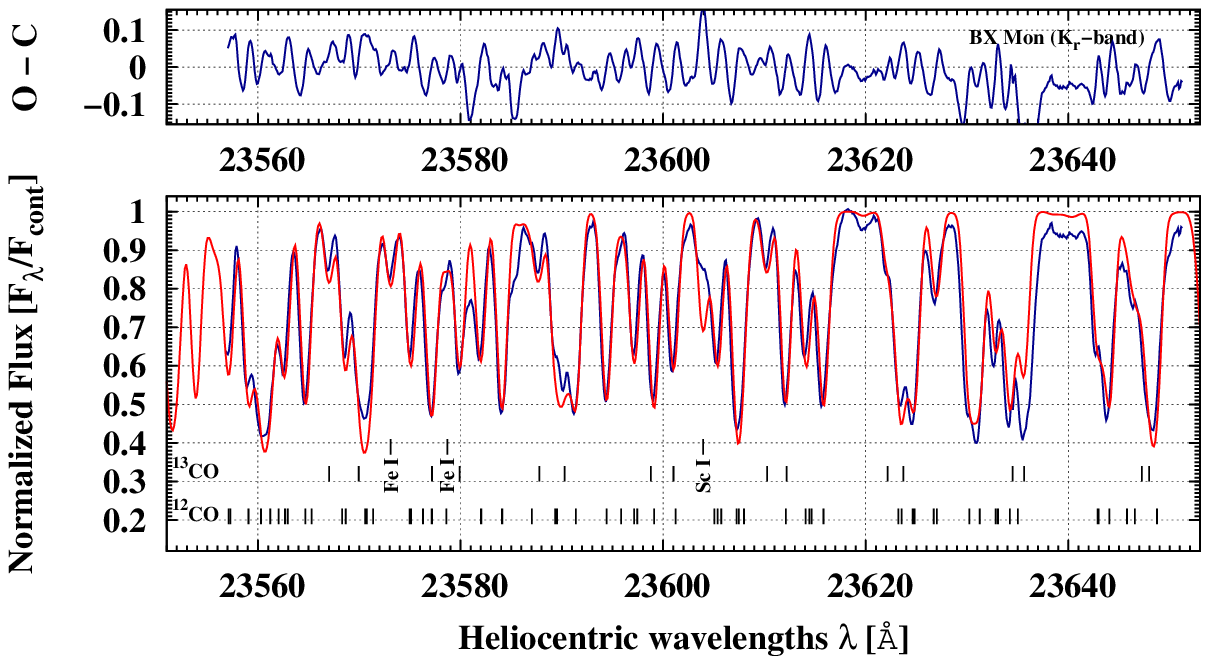} 
  \caption{The $K_{\rm r}$-band spectrum of BX\,Mon observed 2006 April
(blue line) and a synthetic spectrum (red line) calculated using the final
abundances and $^{12}$C/$^{13}$C isotopic ratio (Table\,\ref{T4}).}
  \label{F6}
\end{figure}

\begin{figure}
  \includegraphics[width=84mm]{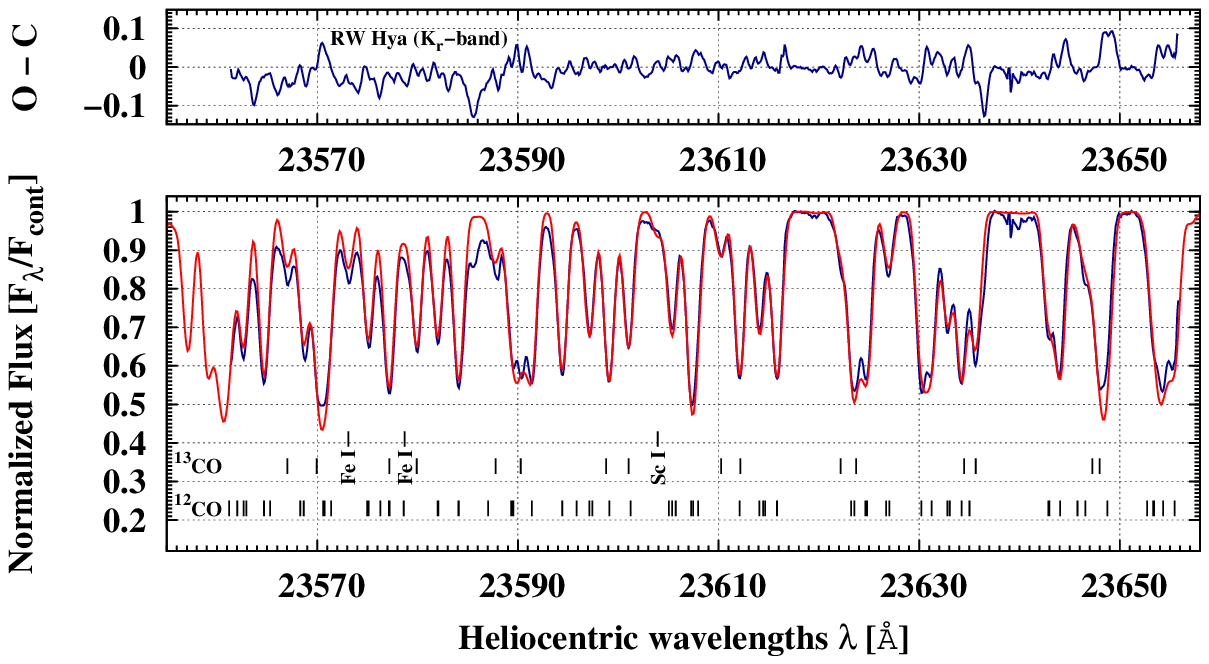} 
  \caption{The $K_{\rm r}$-band spectrum of RW\,Hya observed 2006 April
(blue line) and a synthetic spectrum (red line) calculated using the final
abundances and $^{12}$C/$^{13}$C isotopic ratio (Table\,\ref{T4}).}
  \label{F7}
\end{figure}

Representative spectra with synthetic fits are shown in
Figs\,\ref{F1}--\ref{F7}, which were selected to meet the following criteria:
(i) to span the whole wavelength ranges covered by the observations; (ii) to
show the spectra of possibly diverse sample of objects and with various
temperatures; and (iii) the spectra with lowest residuals were preferred
among those selected with criteria '(i)' and '(ii)'.

\section{Methods}\label{secmet}

The analysis technique we employ is the literature standard, i.e.  local
thermal equilibrium (LTE) analysis based on a 1D, hydrostatic model
atmosphere of the star.  Despite its shortcomings this remains the most
frequently used technique in chemical composition determination.  It is
known that this approach does not fully reflect reality.  The atmospheres of
cool giants/supergiants are complex, dynamic, and subject to stratification. 
A non-LTE (NLTE) approach combined with a 3D treatment of the atmosphere
would be a more appropriate model but remains computationally impractical.

Errors introduced by the LTE 1D approach can be qualitatively estimated. 
There are a very few studies of abundances in stars, especially giants and
supergiants, that use NLTE.  The technique has been restricted mainly to
objects with very low metallicity were NLTE effects are the most
significant.  Our targets have at most near-solar or barely subsolar
metallicities where NLTE effects are relatively weak.  It is possible to
estimate the magnitude of such corrections.  \citet{Mas2014} has calculated
NLTE--LTE differences for iron lines through a wide ranges of metallicities
and $\log{g}$.  While these results do not cover the parameters of our
program stars we can nonetheless extrapolate the corrections needed for the
target with largest $\log{g}$ and smallest metallicity in our sample.  This
correction would be of order hundredths of dex.  To our knowledge, the only
similar studies for near-solar metallicities and low surface gravities
characteristic for our giants are those for \mbox{Si\,{\sc i}}
\citep{Ber2013} and \mbox{Mg\,{\sc i}} \citep{Ber2015}, implying corrections
in the range from approximately $-0.1$ to $-0.4$ in the case of our cool
giants with lowest $\log{g}$.  Although significant progress in the analysis
of stellar spectra with 3D and NLTE models has been made since
\citet{Asp2005} described this field a decade ago, the corrections discussed
in detail by \citet{Ber2014} are available for single lines only, and they
do not correspond to the stellar parameters and the wavelength ranges used
in our work.\\

To determine chemical abundances we used LTE spectral synthesis techniques
particularly suited for strongly blended spectra.  Our spectra are heavily
blended.  For instance, only in the $K$-band region can atomic lines be
found that are significantly stronger than the background of weak molecular
CN lines.  With the exception of these lines the reliable measurement of
equivalent widths of individual lines is practically impossible.  The
synthetic spectra in our study were calculated with use of {\small{WIDMO}}
code \citep{Sch2006}.  1D hydrostatic {\small{MARCS}} model atmospheres by
\citet{Gus2008} were employed.  In selected cases our results were verified
with use of {\small{TURBOSPECTRUM}} spectral synthesis code
(\citealt{AlPl1998,Ple2012}).  {\small{TURBOSPECTRUM}} and {\small{WIDMO}}
produced almost identical synthetic spectra.  The method of fitting the
synthetic spectra to the observations is similar to that described earlier
in Papers\,I\,and\,II.  Some small changes were implemented and are
described below.

The line lists, with the excitation potentials and $gf$-values for
transitions, for the atomic and molecular lines are largely the same as in
our previous analyses (Papers\,I\,and\,II).  For $K$- and $K_{\rm r}$-band
regions the atomic data are from the Vienna Atomic Line Database (VALD)
\citep{Kup1999}.  For the $H$- and $H_{\rm b}$-band regions the list by
\citet{MeBa1999} was used.  For the molecular data we used the lists of
\citet{Goo1994} for CO and of \citet{Kur1999} for OH.  In the case of CN the
Kurucz compilation that we previously used was replaced with the recent line
list by \citet{Sne2014}.  The use of the \citet{Sne2014} list greatly
improved the fitting of the CN spectrum.


To perform the spectral synthesis the stellar parameters, effective
temperature $T_{\rm{eff}}$, surface gravity $\log{g}$, and the atmospheric
motion turbulence parameters, the micro ($\xi_{\rm t}$) and macro
($\zeta_{\rm t}$) turbulence velocities, must be specified and introduced as
inputs into the calculations.  To obtain $T_{\rm{eff}}$ and $\log{g}$ the
method traditionally employed uses neutral and ionized lines of the same
species, usually of iron.  Under the approximation of ionization equilibria
the abundances obtained from the two sets of lines should not depend on the
ionization stage or the excitation energy \citep[eg.][]{Ple2013}.  Similarly
the microturbulent velocities ($\xi_{\rm t}$) are commonly determined by
requiring that abundances resulting from individual lines of the same
species, but with differing line strengths, be independent of equivalent
width (e.g.  \citealt{Smi2002,Sch2006}).

However, our spectra do not have lines present from ionized elements. 
Similarly, we do not have a sample of unblended lines with different
intensities for the same elements.  The estimation of effective temperature
$T_{\rm{eff}}$ was based, instead, on spectral types (Table\,\ref{T2}).  The
spectral types we employ were derived by \citet{Mue1999} from TiO bands in
the near-IR.  The accuracy of the temperature classification is
approximately one spectral subclass.  The only object in our sample,
SS73\,96, for which \citet{Mue1999} have not performed spectral
classification was assigned spectral type M0 and M2 by \citet{MTSt1970} and
\citet{All1980}, respectively.  Our $K$-band spectrum casts doubt on these
classifications.  The CN lines resemble those in the spectra for the other
cool stars.  The hottest stars in our sample, RW\,Hya and Hen\,3-1213, have
much weaker CN lines.  Re-analysing the strength of TiO bands heads in the
$\sim$7000--9500\,\AA\ spectral region and the calcium triplet \ion{Ca}{II}
in the spectrum published by \citet{MTSt1970} we find that M5 is a more
suitable classification for this giant.  The calibrations of \citet{Ric1999}
and \citet{VBe1999} were used to translate the spectral types into effective
temperatures.  The effective temperatures are listed in Table\,\ref{T2}. 
The uncertainty in these temperatures are estimated as $\sim$100\,K.

\begin{table*}
 \centering
\caption{Estimates of the stellar effective temperature $T_{\rm{eff}}$ and
surface gravity $\log{g}$ using the techniques and calibrations indicated.}
  \label{T2}
  \begin{tabular}{@{}llllllllllll@{}}
  \hline
                    & Sp. type$^{[1]}$ & $T_{\rm{eff}}^{[2]}$   & $T_{\rm{eff}}^{[3]}$ & $J-K^{[4,5]}$ & $E$($B-V$)$^{[6]}$   & ($J-K$)$_0$      & $T_{\rm{eff}}^{[7]}$ & $\log{g}$$^{[7]}$ & $\log{g}$$^{[8]}$ & $T_{\rm{eff}}^{a}$ & $\log{g}^{a}$ \\
                    &                  & (K)                    & (K)                  & (mag)         & (mag)                & (mag)            & (K)                  &                   &                   & [K]                &               \\
  \hline
BX\,Mon             & M5               & 3355\,$\pm$\,75        & 3367                 & 1.37$\pm$0.06 & $<$0.14$\pm$0.02     & $>$1.30$\pm$0.07 & $<$3250$\pm$150      & $<$0.0$\pm$0.2    & 0.3--0.6          & 3400               & 0.0           \\
V694\,Mon           & M6               & 3240\,$\pm$\,75        & 3258                 & 1.42$\pm$0.07 & $<$0.22$\pm$0.01     & $>$1.32$\pm$0.08 & $<$3210$\pm$170      & $< -0.1\pm$0.3    & 0.1--0.4          & 3300               & 0.0           \\
Hen\,3-461          & M7               & 3100\,$\pm$\,80        & 3149                 & 1.41$\pm$0.32 & $<$0.42$\pm$0.02     & $>$1.21$\pm$0.34 & $\sim$3440           & $\sim$0.3         & 0.0--0.3          & 3200               & 0.0           \\
SY\,Mus             & M5               & 3355\,$\pm$\,75        & 3367                 & 1.37$\pm$0.07 & 0.4--0.5             & $\sim$1.20       & $<$3500              & $<$0.39           & 0.3--0.6          & 3400               & 0.5           \\
Hen\,2-87           & M5.5             & 3300\,$\pm$\,75        & 3312                 & 2.60$\pm$0.08 & $<\sim$6.1           & --               & --                   & --                & 0.2--0.5          & 3300               & 0.5           \\
Hen\,3-828          & M6               & 3240\,$\pm$\,75        & 3258                 & 1.48$\pm$0.07 & $<$0.37$\pm$0.02     & $>$1.30$\pm$0.09 & $<$3250$\pm$190      & $<$0.0$\pm$0.3    & 0.1--0.4          & 3300               & 0.0           \\
CD-36$^{\circ}$8436 & M5.5             & 3300\,$\pm$\,75        & 3312                 & 1.27$\pm$0.06 & $<$0.05$\pm$0.01     & $>$1.25$\pm$0.07 & $<$3360$\pm$70       & $<$0.1$\pm$0.1    & 0.2--0.5          & 3300               & 0.0           \\
RW\,Hya             & M2               & 3655\,$\pm$\,80        & 3695                 & 1.12$\pm$0.07 & $<$0.07$\pm$0.01     & $>$1.09$\pm$0.07 & $<$3690$\pm$150      & $<$0.7$\pm$0.2    & 0.8--1.1          & 3700               & 0.5           \\
Hen\,3-916          & M5               & 3355\,$\pm$\,75        & 3367                 & 1.63$\pm$0.08 & $<$1.25$\pm$0.05     & $>$1.02$\pm$0.13 & $<$3850$^\pm$280     & $<$1.0$\pm$0.5    & 0.3--0.6          & 3400               & 0.5           \\
Hen\,3-1092         & M5.5             & 3300\,$\pm$\,75        & 3312                 & 1.29$\pm$0.06 & $<$0.11$\pm$0.01     & $>$1.24$\pm$0.07 & $<$3380$\pm$140      & $<$0.2$\pm$0.2    & 0.2--0.5          & 3300               & 0.0           \\
WRAY\,16-202        & M6               & 3240\,$\pm$\,75        & 3258                 & 2.09$\pm$0.06 & $<\sim$3.1           & --               & --                   & --                & 0.1--0.4          & 3300               & 0.0           \\
Hen\,3-1213         & $\leq$K4         & $\leq$4080\,$\pm$\,120 & $\leq$4132           & 1.37$\pm$0.08 & $<$1.07$\pm$0.03     & $>$0.85$\pm$0.12 & $<$4240$\pm$285      & $<$1.8$\pm$0.6    & --                & 4100               & 1.5           \\
Hen\,2-173          & M4.5             & 3410\,$\pm$\,75        & 3421                 & 1.68$\pm$0.07 & $<$0.67$\pm$0.06     & $>$1.35$\pm$0.14 & $<$3150$\pm$290      & $< -0.2\pm$0.5    & 0.3--0.6          & 3400               & 0.5           \\
KX\,Tra             & M6               & 3240\,$\pm$\,75        & 3258                 & 1.39$\pm$0.07 & $<$0.17$\pm$0.01     & $>$1.30$\pm$0.06 & $<$3250$\pm$120      & $<$0.0$\pm$0.2    & 0.1--0.4          & 3300               & 0.0           \\
CL\,Sco             & M5               & 3355\,$\pm$\,75        & 3367                 & 1.29$\pm$0.06 & $<$0.28$\pm$0.01     & $>$1.15$\pm$0.07 & $<$3570$\pm$150      & $<$0.5$\pm$0.3    & 0.3--0.6          & 3400               & 0.5           \\
V455\,Sco           & M6.5             & 3170\,$\pm$\,80        & 3203                 & 1.62$\pm$0.07 & $<$0.67$\pm$0.03     & $>$1.29$\pm$0.10 & $<$3270$\pm$210      & $<$0.0$\pm$0.4    & 0.0--0.3          & 3200               & 0.0           \\
Hen\,2-247          & M6               & 3240\,$\pm$\,75        & 3258                 & 1.61$\pm$0.07 & $<$0.60$\pm$0.01     & $>$1.31$\pm$0.08 & $<$3230$\pm$160      & $< -0.1\pm$0.3    & 0.1--0.4          & 3300               & 0.0           \\
RT\,Ser             & M6               & 3240\,$\pm$\,75        & 3258                 & 1.56$\pm$0.07 & $<$0.48$\pm$0.01     & $>$1.32$\pm$0.08 & $<$3210$\pm$150      & $< -0.1\pm$0.3    & 0.1--0.4          & 3300               & 0.0           \\
AE\,Ara             & M5.5             & 3300\,$\pm$\,75        & 3312                 & 1.36$\pm$0.06 & $<$0.20$\pm$0.01     & $>$1.26$\pm$0.06 & $<$3330$\pm$150      & $<$0.1$\pm$0.2    & 0.2--0.5          & 3300               & 0.5           \\
SS73\,96            & M5$^b$           & 3355\,$\pm$\,75        & 3367                 & 1.81$\pm$0.07 & $<$1.23$\pm$0.03     & $>$1.21$\pm$0.10 & $<$3430$\pm$210      & $<$0.3$\pm$0.4    & 0.3--0.6          & 3400               & 0.5           \\
AS\,270             & M5.5             & 3300\,$\pm$\,75        & 3312                 & 1.75$\pm$0.07 & $<\sim$7.3           & --               & --                   & --                & 0.2--0.5          & 3300               & 0.5           \\
Y\,CrA              & M6               & 3240\,$\pm$\,75        & 3258                 & 1.33$\pm$0.06 & $<$0.12$\pm$0.01     & $>$1.27$\pm$0.07 & $<$3300$\pm$140      & $<$0.0$\pm$0.2    & 0.1--0.4          & 3300               & 0.0           \\
Hen\,2-374          & M5.5             & 3300\,$\pm$\,75        & 3312                 & 2.06$\pm$0.08 & $<$1.74$\pm$0.11     & $>$1.21$\pm$0.21 & $<$3440$\pm$450      & $<$0.3$\pm$0.8    & 0.2--0.5          & 3300               & 0.5           \\
Hen\,3-1761         & M5.5             & 3300\,$\pm$\,75        & 3312                 & 1.27$\pm$0.06 & $<$0.08$\pm$0.01     & $>$1.23$\pm$0.06 & $<$3390$\pm$130      & $<$0.2$\pm$0.2    & 0.2--0.5          & 3300               & 0.0           \\
  \hline
\end{tabular}
\begin{list}{}{}
\item[$Notes.$ References:] spectral types from $^{[1]}$\citet{Mue1999}, total
Galactic extinction adopted according to $^{[6]}$\citet{Sch2011} and
\cite{Sch1998}, infrared colours from 2MASS $^{[4]}$\citep{Phi2007}
transformed to $^{[5]}$\citet{BeBr1988} photometric system.
\item[Callibration by:] $^{[2]}$\citet{Ric1999}, $^{[3]}$\citet{VBe1999},
$^{[7]}$\citet{Kuc2005}, $^{[8]}$\citet{DuSc1998} for the red giant masses
in the range 1--2M$_{\sun}$.
\item[$^{a}$Adopted.]
\item[$^{b}$Spectral]\,type M5 adopted -- see the text for explanation.
\end{list}
\end{table*}

All our targets have Two Micron All Sky Survey (2MASS) $J$ and $K$
magnitudes \citep{Phi2007}.  Combining the 2MASS colours with colour
excesses (\citealt{Sch1998,Sch2011}) provides an estimate of the IR
intrinsic colours.  Upper limits to the effective temperature and surface
gravity were then derived according to the \citet{Kuc2005}
$T_{\rm{eff}}$--$\log{g}$--colour relation for late-type giants.  The
effective temperatures derived from spectral type calibrations fall below
these limits.  Independent estimates for $\log{g}$ have also been obtained
(Table\,\ref{T2}) by assuming that the masses of symbiotic giants are in the
range $\sim$1--2 M$_{\sun}$ \citep{Mik2003} and that their radii follow the
radius--spectral type relation from \citet[table 2]{DuSc1998}.  In some
cases we can also place additional constraints on $\log{g}$ from the orbital
solution of the red giant in the symbiotic binary (Table\,\ref{T3}).  A
limit can be set on the giant radius using the inclination.  The mass can be
estimated from the most probable orbital solutions.

The $T_{\rm{eff}}$ and $\log{g}$ adopted for the atmosphere models used in
our calculations are shown in the rightmost columns of Table\,\ref{T2}.  The
uncertainty in the $\log{g}$ is difficult to estimate.  However, the
limitations on this parameter obtained with the various methods discussed
above give consistent results.  The uncertainty in the adopted values should
not be larger than $\sim 0.5$ which is the resolution of the {\small{MARCS}}
model atmosphere grid used in our calculations.

In our previous studies (Papers\,I\,and\,II), as well as in the initial
phase of the current work, we tried holding the microturbulent velocity,
$\xi_{\rm{t}}$, as a free parameter.  We searched for its value by sampling
in the range 1.2--2.8\,km\,s$^{\rm{-1}}$.  We obtained microturbulences
close to 2 km\,s$^{\rm{-1}}$ with a dispersion $\sim \pm$0.35. 
$\xi_{\rm{t}} = 2.0$ km\,s$^{\rm{-1}}$ is typical for cool Galactic red
giants (\citealt[1986, 1990]{SmLa1985}) and is frequently used in studies of
chemical compositions \citep[e.g.][]{Ney2015}.  We subsequently decided to
use $\xi_{\rm{t}} = 2.0$ km\,s$^{\rm{-1}}$ as the input to our models. 
Similarly the macroturbulent velocity was set to the typical value for cool
red giants $\zeta_{\rm{t}}$ = 3\,km\,s$^{\rm{-1}}$ \citep[e.g.][]{Fek2003}.

\begin{table}
 \centering
  \caption{Estimates of surface gravities employing most probable mass and
radius from orbital solutions.} \label{T3}
  \begin{tabular}{@{}llllc@{}}
  \hline
  Object             & $M_{\rm{rg}}[$M$_{\sun}]$ & $R_{\rm{rg}}[$R$_{\sun}]$ & $\log{g}$            & Ref.$^a$ \\
  \hline
 BX\,Mon             & $1.5^b$                   & $160\pm50$                & $+0.5$--~$0.0$     & $[1]$    \\
 SY\,Mus             & $1.3\pm0.25$              & $135$                     & $+0.4$--$+0.2$     & $[2]$    \\ 
 CD-36$^{\circ}$8436 & $1.5$                     & $112$--$192$              & $+0.5$--~$0.0$     & $[7]$    \\
 RW\,Hya             & $3.4$                     & $145$                     & $+0.6$             & $[3]$    \\
 Hen\,3-1213         & $1.0$                     & $53$--$107$               & $+1.0$--$+0.4$     & $[7]$    \\
 Hen\,2-173          & $1.5$                     & $100$--$149$              & $+0.6$--$+0.3$     & $[4]$    \\
 CL\,Sco             & $1.5$                     & $75$--$114$               & $+0.9$--$+0.5$     & $[4]$    \\
 V455\,Sco           & $1.12$                    & $207\pm28$                &~~$0.0$--$-0.3$     & $[5]$    \\
 Hen\,2-247          & $1.2$                     & $137\pm15$                & $+0.3$--$+0.1$     & $[5]$    \\
 AE\,Ara             & $1.6$                     & $111$--$166$              & $+0.6$--$+0.2$     & $[6]$    \\
 SS73\,96            & $1.5$                     & $94\pm12$                 & $+0.8$--$+0.6$     & $[7]$    \\
 AS\,270             & $1.5$                     & $119$--$171$              & $+0.5$--$+0.1$     & $[4]$    \\
 Hen\,2-374          & $1.6$                     & $114$--$131$              & $+0.5$--$+0.4$     & $[6]$    \\
  \hline
\end{tabular}
\begin{list}{}{}
\item[$Notes.$ $^{a}$References]\,for masses and radii adopted for $\log{g}$
estimations: $[1]$ \citet{Dum1998}; $[2]$ \citet{Rut2007}; $[3]$
\citet{Otu2014}; $[4]$ \citet{Fek2007}; $[5]$ \citet{Fek2008}; $[6]$
\citet{Fek2010}; $[7]$ \citet{Fek2015}.
\item[$^{b}$The]\,mass adopted according to \citet{Bra2009} (see Paper\,II for the details).
\end{list}
\end{table}

The observed spectra are also affected by large-scale motions, e.g.  radial
and rotational velocities as well as velocity shifts introduced by the
instrument, e.g.  flexure.  These all have to be taken into account to
enable calculations of residuals ('observations minus model') in order to
perform the minimization.  The wavelength shifts originating from
combination of radial velocities and instrumental effects were determined
with the {\small{IRAF}} cross-correlation program {\small{FXCOR}}
\citep{Fit1993}.  The synthetic spectrum was used for the templates.  The
observed spectra were wavelength shifted to match the position of the
spectral lines in synthetic spectra.

In our sample the largest contribution to the broadening of the spectral
lines is the giant star rotational velocity, $V_{\rm{rot}}\sin{i}$.  To
measure $V_{\rm{rot}}\sin{i}$ we used two methods: (i) a cross-correlation
technique CCF similar to that adopted by \citet{Car2011} but using synthetic
spectra as the templates and (ii) direct measurement of the FWHM of the six
relatively strong unblended atomic lines (\mbox{Ti\,{\sc i}}, \mbox{Fe\,{\sc
i}}, \mbox{Sc\,{\sc i}}) present in the $K$-band region.  The measured
values are presented in Table\,\ref{T1}.  The rotational velocities obtained
with CCF method are generally somewhat smaller than those obtained with FWHM
method.  Spectral regions that are crowded with blended molecular lines, as
found in the $H$-band spectra, lead to an underestimate.  We compared
$V_{\rm{rot}}\sin{i}$ obtained with both these methods (CCF and FWHM) when
the rotational velocities were allowed to be free parameters in the solution
process (Paper\,I).  The differences were not significant with the changes
in $V_{\rm{rot}}\sin{i}$ having small impact in the resulting abundances. 
Our analysis uses $V_{\rm{rot}}\sin{i}$ obtained from atomic lines in
$K$-band spectra with the FWHM method and applied to all of spectra for a
given object (Table\,\ref{T1}).  This value of $V_{\rm{rot}}\sin{i}$ was
treated as a fixed parameter in our solution.

The method described in Papers\,I\,and\,II was adopted for the abundance
calculations with the difference, discussed above, that the microturbulence
was fixed at $\xi_{\rm{t}}$=2 km\,s$^{\rm {-1}}$.  The solutions were
performed in a semi-automatic way to improve the efficiency of the $\chi^2$
minimization and simultaneously too keep control of the parameter values
that were entered.  The simplex algorithm \citep{Bra1998} was applied to the
parameter space $\chi^2$ minimization.  The simplex algorithm is a
relatively slow least squares technique but it has notable advantages.  It
is known to be remarkably efficient in achieving convergence when more than
two or three parameters need to be adjusted.  In our case $n + 1 = 8$, where
'n' is the number of free parameters.  Calculations starting from this
number of places in parameter space are efficiently searched for the
minimum.

A brief outline of the analysis procedure follows.  Initial values for the
abundances of oxygen, carbon, and nitrogen were adjusted fitting by eye
alternately using OH, CO, and CN lines.  Next the abundances of elements
around the iron peak were adjusted using atomic lines.  This process was
repeated iteratively to find approximate parameters of the chemical
composition, around which the initial grid of the $n+1$, $n$ dimensional
sets of free parameters, the so-called simplex needed for the simplex
algorithm, was prepared.  Nine different randomly generated simplexes were
used to obtain best fits to $H$-, $H_{\rm b}$, and $K$-band spectra and the
standard deviations.  In the case of the 10 systems where the $K_{\rm
r}$-band spectrum had been observed, after we found the sets of parameters
that give the best fit to the $H$-, $H_{\rm b}$, and $K$-band spectra, the
abundances were then applied to the $K_{\rm r}$-band spectrum as fixed
values and a search for $^{12}$C$/^{13}$C isotopic ratio was performed. 
Reconciliation of carbon abundance and $^{12}$C$/^{13}$C required several
iterations.  Most frequently the $K_{\rm r}$-band region was not observed
and an isotopic ratio of $^{12}$C$/^{13}$C=10 was used in calculations. 
$^{12}$C$/^{13}$C=10 is a value close to the average, and median
simultaneously, among those stars for which we could obtain isotopic ratios. 
The above procedure was iteratively repeated, when needed, to choose the
model atmosphere with the best-matching metallicity.

\section{Results}\label{secres}

The abundances derived from CNO molecules and atomic lines (\mbox{Sc\,{\sc
i}}, \mbox{Ti\,{\sc i}}, \mbox{Fe\,{\sc i}}, \mbox{Ni\,{\sc i}}), on the
scale of $\log{\epsilon}(X) = \log{(N(X) N(H)^{-1})} + 12.0$, are summarized
in Table\,\ref{T4} together with the $^{12}$C/$^{13}$C isotopic ratio,
projected rotational velocities ($V_{\rm{rot}} \sin{i}$), and corresponding
uncertainties.  Synthetic fits to the observed spectra of BX\,Mon and
Hen\,3-1213 in $H$-band region, V694\,Mon in $H_{\rm b}$-band, Hen\,3-1761
and Hen\,3-916 in $K$-band, and BX\,Mon and RW\,Hya in $K_{\rm r}$-band
are shown in Figs\,\ref{F1}-\ref{F7}.  The molecular (OH, CO, CN) and
atomic (\mbox{Sc\,{\sc i}}, \mbox{Ti\,{\sc i}}, \mbox{Fe\,{\sc i}},
\mbox{Ni\,{\sc i}}) lines used in solving of the chemical composition are
identified.  The synthetic fits to all the observed spectra are shown in
Figs\,\ref{FB1}-\ref{FB59} in the online Appendix\,B. 
Systematic effects are possible due to the choice of model atmospheres.  We
made a comparison of abundances obtained with the {\small{PHOENIX}} model
atmospheres extracted from \citet{Hau1999} used for five selected cases of
BX\,Mon, Hen\,3-461, SY\,Mus, Hen\,2-173, and CL\,Sco.  The use of
{\small{PHOENIX}} models lead to somewhat higher abundances by on average
$\sim$0.1, $\sim$0.16, $\sim$0.22, $\sim$0.13, $\sim$0.1, $\sim$0.01, and
$\sim$0.02 dex for C, N, O, Sc, Ti, Fe, and Ni, respectively.

The atmospheric parameters have associated uncertainties of $\sim$100\,K in
effective temperature, up to 0.5 in $\log{g}$, and $\sim$0.25 km\,s$^{-1}$
in the case of microturbulence.  The latter is the largest uncertainty which
we previously obtained when microturbulence was considered as the free
parameter.  To investigate how these uncertainties manifest themselves as
abundance changes, we made additional fits with {\small{MARCS}} atmosphere
models varying the atmospheric parameters by the values of the uncertainties
($\Delta T_{\rmn{eff}} = \pm$100\,K, $\Delta \log{g} = \pm$0.5, $\Delta
\xi_{\rmn{t}} = \pm$0.25).  The changes in the abundance obtained for each
element as a function of each model parameter are listed in Table\,\ref{T5}
at the top for M giants and separately at the bottom for the yellow
symbiotic Hen\,3-1213.  In the case of Hen\,3-1213 the dependences on the
uncertainties in stellar parameters are significantly different than for the
M giants.  The final estimated uncertainty for each element is the
quadrature sum of each model uncertainty.  It is shown in the rightmost
column of Table\,\ref{T5}, marked with $\Delta$ symbol.  By comparing
Tables\,\ref{T4}\,and\,\ref{T5} we can see that the uncertainties of the
derived chemical composition come mainly from uncertainties in the
atmospheric parameters.  With a few exceptions, the uncertainties associated
with the fitting, originating mainly from line-to-line dispersion due to
line blending, continuum problems, uncertainties on oscillator strengths,
etc. are less important.  The uncertainty in $\log{g}$ can have large
impact on the carbon abundances and, in the case of cool M-type giants, the
uncertainty in the adopted microturbulence can have a large impact on the
abundances of scandium and titanium.

\begin{table*}
 \centering
  \caption{Calculated and relative abundances, rotational velocity, and
uncertainties$\,^a$ derived for 24 S-type symbiotic systems.}
  \label{T4}
  \begin{tabular}{@{}lrrrrrrrcr@{}}
  \hline
  \hline
                    & C                   & N         & O              & Sc$^b$         & Ti             & Fe             & Ni             & $^{12}$C/$^{13}$C & $V_{\rmn{rot}} \sin{i}$ \\
                    & $\log{\epsilon(X)}$ &           &                &                &                &                &                &          & $(\rmn{km}\,\rmn{s}^{-1})$       \\
                    & [$X$]$^c$           &           &                &                &                &                &                &          &                                  \\ 
  \hline
BX\,Mon             &  7.69$\pm$0.02 &  7.79$\pm$0.05 &  8.20$\pm$0.01 &  3.59$\pm$0.13 &  4.71$\pm$0.05 &  7.07$\pm$0.04 &  6.05$\pm$0.11 & ~8$\pm$1 & 8.4$\pm$1.4 \\
                    & -0.74$\pm$0.07 & -0.04$\pm$0.10 & -0.49$\pm$0.06 & +0.43$\pm$0.17 & -0.22$\pm$0.10 & -0.40$\pm$0.08 & -0.15$\pm$0.15 & &\\
V694\,Mon           &  8.08$\pm$0.01 &  7.90$\pm$0.02 &  8.40$\pm$0.01 &  4.01$\pm$0.07 &  4.54$\pm$0.05 &  7.12$\pm$0.04 &  5.86$\pm$0.06 & 23$\pm$2 & 8.2$\pm$1.0 \\
                    & -0.35$\pm$0.06 & +0.07$\pm$0.07 & -0.29$\pm$0.06 & +0.85$\pm$0.11 & -0.39$\pm$0.10 & -0.35$\pm$0.08 & -0.34$\pm$0.10 & &\\
Hen\,3-461          &  8.29$\pm$0.02 &  8.35$\pm$0.04 &  8.74$\pm$0.01 &  4.14$\pm$0.06 &  5.16$\pm$0.06 &  7.59$\pm$0.07 &  6.48$\pm$0.05 & 13$\pm$1 & 7.4$\pm$0.6 \\
                    & -0.14$\pm$0.07 & +0.52$\pm$0.09 & +0.05$\pm$0.06 & +0.98$\pm$0.10 & +0.23$\pm$0.11 & +0.12$\pm$0.11 & +0.28$\pm$0.09 & &\\
SY\,Mus             &  8.07$\pm$0.02 &  8.12$\pm$0.04 &  8.61$\pm$0.02 &  3.96$\pm$0.12 &  4.93$\pm$0.04 &  7.32$\pm$0.04 &  6.13$\pm$0.09 & ~8$\pm$2 & 6.6$\pm$0.6 \\
                    & -0.36$\pm$0.07 & +0.29$\pm$0.09 & -0.08$\pm$0.07 & +0.80$\pm$0.16 & ~0.00$\pm$0.09 & -0.15$\pm$0.08 & -0.07$\pm$0.13 & &\\
Hen\,2-87           &  8.61$\pm$0.02 &  8.30$\pm$0.05 &  8.99$\pm$0.02 &  3.93$\pm$0.10 &  4.74$\pm$0.04 &  7.64$\pm$0.05 &  6.41$\pm$0.07 & 18$\pm$2 & 9.6$\pm$0.6 \\
                    & +0.18$\pm$0.07 & +0.47$\pm$0.10 & +0.30$\pm$0.07 & +0.77$\pm$0.14 & -0.19$\pm$0.09 & +0.17$\pm$0.09 & +0.21$\pm$0.11 & &\\
Hen\,3-828          &  8.30$\pm$0.03 &  8.21$\pm$0.04 &  8.71$\pm$0.01 &  4.61$\pm$0.06 &  5.43$\pm$0.07 &  7.50$\pm$0.05 &  6.17$\pm$0.07 & 15$\pm$2 & 7.9$\pm$0.5 \\
                    & -0.13$\pm$0.08 & +0.38$\pm$0.09 & +0.02$\pm$0.06 & +1.45$\pm$0.10 & +0.50$\pm$0.12 & +0.03$\pm$0.09 & -0.03$\pm$0.11 & &\\
CD-36$^{\circ}$8436 &  7.74$\pm$0.01 &  7.99$\pm$0.03 &  8.36$\pm$0.01 &  3.66$\pm$0.07 &  4.71$\pm$0.06 &  7.17$\pm$0.03 &  6.00$\pm$0.09 & ~8$\pm$1 & 8.1$\pm$1.1 \\
                    & -0.69$\pm$0.06 & +0.16$\pm$0.08 & -0.33$\pm$0.06 & +0.50$\pm$0.11 & -0.22$\pm$0.11 & -0.30$\pm$0.07 & -0.20$\pm$0.13 & &\\
RW\,Hya             &  7.52$\pm$0.04 &  7.48$\pm$0.08 &  8.14$\pm$0.03 &  2.65$\pm$0.09 &  4.35$\pm$0.06 &  6.70$\pm$0.04 &  5.67$\pm$0.05 & 5.3$\pm$0.5 & 6.2$\pm$0.9 \\
                    & -0.91$\pm$0.09 & -0.35$\pm$0.13 & -0.55$\pm$0.08 & -0.51$\pm$0.13 & -0.58$\pm$0.11 & -0.77$\pm$0.08 & -0.53$\pm$0.09 & &\\
Hen\,3-916          &  7.99$\pm$0.02 &  7.90$\pm$0.05 &  8.30$\pm$0.01 &  3.39$\pm$0.09 &  4.67$\pm$0.07 &  7.01$\pm$0.04 &  5.77$\pm$0.07 & 6.6$\pm$0.6 & 8.5$\pm$0.9 \\
                    & -0.44$\pm$0.07 & +0.07$\pm$0.10 & -0.39$\pm$0.06 & +0.23$\pm$0.13 & -0.26$\pm$0.12 & -0.46$\pm$0.08 & -0.43$\pm$0.11 & &\\
Hen\,3-1092         &  7.41$\pm$0.02 &  7.47$\pm$0.03 &  8.04$\pm$0.02 &  3.33$\pm$0.18 &  4.21$\pm$0.08 &  6.68$\pm$0.07 &  5.57$\pm$0.13 & --       & 5.8$\pm$0.7 \\
                    & -1.02$\pm$0.07 & -0.36$\pm$0.08 & -0.65$\pm$0.07 & +0.17$\pm$0.22 & -0.72$\pm$0.13 & -0.79$\pm$0.11 & -0.63$\pm$0.17 & &\\
WRAY\,16-202        &  8.11$\pm$0.02 &  8.24$\pm$0.05 &  8.66$\pm$0.01 &  4.37$\pm$0.18 &  5.37$\pm$0.13 &  7.64$\pm$0.05 &  6.37$\pm$0.06 & 10$\pm$1 & 8.3$\pm$1.7 \\
                    & -0.32$\pm$0.07 & +0.41$\pm$0.10 & -0.03$\pm$0.06 & +1.21$\pm$0.22 & +0.44$\pm$0.18 & +0.17$\pm$0.09 & +0.17$\pm$0.10 & &\\
Hen\,3-1213         &  8.00$\pm$0.03 &  7.76$\pm$0.04 &  8.88$\pm$0.02 &  3.29$\pm$0.09 &  4.98$\pm$0.06 &  6.79$\pm$0.04 &  5.70$\pm$0.11 & --       & 7.4$\pm$0.3 \\
                    & -0.43$\pm$0.08 & -0.07$\pm$0.09 & +0.19$\pm$0.07 & +0.13$\pm$0.13 & +0.05$\pm$0.11 & -0.68$\pm$0.08 & -0.50$\pm$0.15 & &\\
Hen\,2-173          &  8.18$\pm$0.03 &  8.17$\pm$0.07 &  8.80$\pm$0.02 &  3.94$\pm$0.11 &  5.12$\pm$0.09 &  7.29$\pm$0.04 &  6.16$\pm$0.06 & --       & 8.4$\pm$0.6 \\
                    & -0.25$\pm$0.08 & +0.34$\pm$0.12 & +0.11$\pm$0.07 & +0.78$\pm$0.15 & +0.19$\pm$0.14 & -0.18$\pm$0.08 & -0.04$\pm$0.10 & &\\
KX\,TrA             &  8.02$\pm$0.03 &  7.90$\pm$0.09 &  8.62$\pm$0.03 &  3.87$\pm$0.12 &  4.99$\pm$0.15 &  7.13$\pm$0.07 &  6.09$\pm$0.11 & --       & 8.5$\pm$1.3 \\
                    & -0.41$\pm$0.08 & +0.07$\pm$0.14 & -0.07$\pm$0.08 & +0.71$\pm$0.16 & +0.06$\pm$0.20 & -0.34$\pm$0.11 & -0.11$\pm$0.15 & &\\
CL\,Sco             &  7.98$\pm$0.04 &  8.19$\pm$0.09 &  8.57$\pm$0.02 &  3.39$\pm$0.13 &  4.79$\pm$0.10 &  7.16$\pm$0.03 &  6.15$\pm$0.13 & --       & 7.8$\pm$0.8 \\
                    & -0.45$\pm$0.09 & +0.36$\pm$0.14 & -0.12$\pm$0.07 & +0.23$\pm$0.17 & -0.14$\pm$0.15 & -0.31$\pm$0.07 & -0.05$\pm$0.17 & &\\
V455\,Sco           &  8.42$\pm$0.02 &  8.81$\pm$0.07 &  9.16$\pm$0.03 &  4.20$\pm$0.10 &  5.37$\pm$0.05 &  7.83$\pm$0.05 &  6.41$\pm$0.10 & --       & 8.4$\pm$1.0 \\
                    & -0.01$\pm$0.07 & +0.98$\pm$0.12 & +0.47$\pm$0.08 & +1.04$\pm$0.14 & +0.44$\pm$0.10 & +0.36$\pm$0.09 & +0.21$\pm$0.14 & &\\
Hen\,2-247          &  8.28$\pm$0.02 &  8.55$\pm$0.05 &  8.98$\pm$0.01 &  4.43$\pm$0.09 &  5.48$\pm$0.09 &  7.63$\pm$0.05 &  6.36$\pm$0.06 & --       & 10.4$\pm$1.0 \\
                    & -0.15$\pm$0.07 & +0.72$\pm$0.10 & +0.29$\pm$0.06 & +1.27$\pm$0.13 & +0.55$\pm$0.14 & +0.16$\pm$0.09 & +0.16$\pm$0.10 & &\\
RT\,Ser             &  8.01$\pm$0.12 &  7.93$\pm$0.27 &  8.36$^d$      &  3.88$\pm$0.10 &  4.89$\pm$0.08 &  6.96$\pm$0.04 &  5.73$^d$      & --       & 7.8$\pm$1.7 \\
                    & -0.42$\pm$0.17 & +0.10$\pm$0.32 & -0.33          & +0.72$\pm$0.14 & -0.04$\pm$0.13 & -0.51$\pm$0.08 & -0.47          & &\\
AE\,Ara             &  8.25$\pm$0.02 &  8.14$\pm$0.06 &  8.66$\pm$0.02 &  4.20$\pm$0.10 &  5.22$\pm$0.07 &  7.45$\pm$0.05 &  6.23$\pm$0.09 & --       & 10.1$\pm$0.8 \\
                    & -0.18$\pm$0.07 & +0.31$\pm$0.11 & -0.03$\pm$0.07 & +1.04$\pm$0.14 & +0.29$\pm$0.12 & -0.02$\pm$0.09 & +0.03$\pm$0.13 & &\\
SS73\,96            &  8.27$\pm$0.03 &  7.83$\pm$0.08 &  8.58$\pm$0.02 &  3.71$\pm$0.21 &  4.75$\pm$0.14 &  7.23$\pm$0.07 &  5.96$\pm$0.26 & --       & 9.1$\pm$0.4 \\
                    & -0.16$\pm$0.08 & +0.00$\pm$0.13 & -0.11$\pm$0.07 & +0.55$\pm$0.25 & -0.18$\pm$0.19 & -0.24$\pm$0.11 & -0.24$\pm$0.30 & &\\
AS\,270             &  8.26$\pm$0.02 &  8.09$\pm$0.03 &  8.63$\pm$0.01 &  3.89$\pm$0.18 &  4.95$\pm$0.07 &  7.50$\pm$0.03 &  6.20$\pm$0.11 & --       & 10.0$\pm$0.9 \\
                    & -0.17$\pm$0.07 & +0.26$\pm$0.08 & -0.06$\pm$0.06 & +0.73$\pm$0.22 & +0.02$\pm$0.12 & +0.03$\pm$0.07 & ~0.00$\pm$0.15 & &\\
Y\,CrA              &  7.84$\pm$0.01 &  7.86$\pm$0.03 &  8.43$\pm$0.02 &  3.40$\pm$0.06 &  4.83$\pm$0.04 &  7.07$\pm$0.02 &  5.94$\pm$0.05 & --       & 10.4$\pm$2.9 \\
                    & -0.59$\pm$0.06 & +0.03$\pm$0.08 & -0.26$\pm$0.07 & +0.24$\pm$0.10 & -0.10$\pm$0.09 & -0.40$\pm$0.06 & -0.26$\pm$0.09 & &\\
Hen\,2-374          &  7.85$\pm$0.08 &  7.99$\pm$0.12 &  8.36$^d$      &  3.87$\pm$0.04 &  4.90$\pm$0.03 &  6.95$\pm$0.04 &  5.73$^d$      & --       & 6.4$\pm$0.5 \\
                    & -0.58$\pm$0.13 & +0.16$\pm$0.17 & -0.33          & +0.71$\pm$0.08 & -0.03$\pm$0.08 & -0.52$\pm$0.08 & -0.47          & &\\
Hen\,3-1761         &  7.81$\pm$0.02 &  7.79$\pm$0.03 &  8.30$\pm$0.02 &  3.63$\pm$0.09 &  4.77$\pm$0.06 &  7.22$\pm$0.04 &  6.17$\pm$0.06 & --       & 6.9$\pm$0.6 \\
                    & -0.62$\pm$0.07 & -0.04$\pm$0.08 & -0.39$\pm$0.07 & +0.47$\pm$0.13 & -0.16$\pm$0.11 & -0.25$\pm$0.08 & -0.03$\pm$0.10 & &\\
  \hline
Sun                 &  8.43$\pm$0.05 &  7.83$\pm$0.05 &  8.69$\pm$0.05 &  3.16$\pm$0.04 &  4.93$\pm$0.04 &  7.47$\pm$0.04 &  6.20$\pm$0.04 & &\\
  \hline
  \hline
\end{tabular}
\begin{list}{}{}
\item[$Notes.$ $^a$3$\sigma$.]
\item[$^b$The]\,abundance of scandium is based on only one strong \mbox{Sc\,{\sc i}} line at $\lambda \sim 22272.8$\AA\, and it may be less reliable than other abundances. Broadening of the\\ infrared scandium lines by hyperfine structure has not been included in the analysis (see Paper\,I).
\item[$^c$Relative]\,to the Sun [$X$] abundances estimated in relation to the solar composition of \citet{Asp2009} for CNO and \citet{Sco2015} for elements around the iron peak.
\item[$^d$Adopted.]
\end{list}
\end{table*}

\section{Discussion}\label{secdiscussion}

We measured the photospheric chemical abundances (CNO and elements around
the iron peak: Sc, Ti, Fe, and Ni) for the first time in a sample 23
classical S-type symbiotic systems with red giant primary and in one
yellow-type symbiotic system Hen\,3-1213.  The abundances of carbon,
nitrogen, oxygen, iron, and titanium are based on the large number of
absorption features in the spectra and should be relatively well determined. 
As shown above, the uncertainties are typically $\sim$0.2--0.3\,dex ranging
up to $\sim$0.4\,dex for the case of titanium.  Elemental abundances in
symbiotic giants can be used to address number of issues, for example to
investigate the evolutionary status of these systems and to associate the
systems with their parent populations in the Milky Way.  Table\,\ref{T6}
summarizes absolute abundances of carbon and nitrogen, and relative
abundances $[$O$/$Fe$]$, $[$Ti$/$Fe$]$, and $[$Fe$/$H$]$.

The metal abundances in the yellow symbiotic Hen\,3-1213 in our sample were
studied previously by \citet{Per2009}.  They find
$\log{\epsilon(\mbox{Fe\,{\sc i}})}$ = 6.59 $\pm$ 0.16,
$\log{\epsilon(\mbox{Ni\,{\sc i}})}$ = 5.38 $\pm$ 0.15, and
$\log{\epsilon(\mbox{Ti\,{\sc i}})}$ = 4.68 based on analysis of optical
spectra.  Their values are lower by $\sim$0.2--0.3\,dex than ours but
consistent within the uncertainties.  Taking into account that our adopted
$\log{g}=1.5$ is higher than the $\log{g}=1.1$ adopted by \citet{Per2009}
and given the sensitivity of Fe, Ni, and Ti abundances to $\log{g}$
(Table\,\ref{T5}) the agreement between our results is even better, in the
case of iron almost perfect to $\sim$0.1\,dex.

\begin{table}
 \centering
  \caption{Sensitivity of abundances to uncertainties in the stellar
parameters for M-type giants (top) and the yellow symbiotic Hen\,3-1213
(bottom).}
\label{T5}
  \begin{tabular}{@{}lrrrr@{}}
  \hline
  \hline
  $\Delta X$ & $\Delta T_{\rmn{eff}} = +100$\,K & $\Delta \log{g} = +0.5$ & $\Delta  \xi_{\rmn{t}} = +0.25$ & $\Delta^a$\\
  \hline
  \multicolumn{5}{c}{M-type giants}\\
  C              & +0.02 & +0.23 & -0.05 & $\pm$0.23 \\
  N              & +0.03 & +0.02 & -0.05 & $\pm$0.06 \\
  O              & +0.12 & +0.08 & -0.06 & $\pm$0.16 \\
  Sc             & +0.10 & +0.15 & -0.30 & $\pm$0.35 \\
  Ti             & +0.06 & +0.15 & -0.24 & $\pm$0.29 \\
  Fe             & -0.05 & +0.16 & -0.08 & $\pm$0.18 \\
  Ni             & -0.06 & +0.18 & -0.11 & $\pm$0.22 \\
  \hline
  \multicolumn{5}{c}{Hen\,3-1213}\\
  C              & +0.05 & +0.20 &~~0.00 & $\pm$0.21 \\
  N              & +0.10 & -0.05 & -0.04 & $\pm$0.12 \\
  O              & +0.20 &  0.00 & -0.04 & $\pm$0.20 \\
  Sc             & +0.18 & +0.03 & -0.05 & $\pm$0.19 \\
  Ti             & +0.17 & +0.06 & -0.11 & $\pm$0.21 \\
  Fe             & +0.02 & +0.11 & -0.06 & $\pm$0.13 \\
  Ni             & -0.01 & +0.10 & -0.05 & $\pm$0.11 \\
  \hline
\end{tabular}
\begin{list}{}{}
\item[$^a$] $[(\Delta T_{\rmn{eff}})^2 + (\Delta \log{g})^2 + (\Delta \xi_{\rmn{t}})^2]^{0.5}$.
\end{list}
\end{table}

\begin{table}
 \centering
  \caption{Absolute and relative abundances adopted for comparison with
Galactic stellar populations.  Abundances of CH\,Cyg and V2116\,Oph from the
literature are shown for comparison at the bottom.}
  \label{T6}
  \begin{tabular}{@{}lccccc@{}}
  \hline
   Object          & $A$($^{12}$C) & $A$($^{14}$N) & $[$O$/$Fe$]$ & $[$Ti$/$Fe$]$ & $[$Fe$/$H$]$\\
  \hline
BX\,Mon            & 7.64 & 7.79 & -0.09 & +0.18 & -0.40\\
V694\,Mon          & 8.06 & 7.90 & +0.06 & -0.04 & -0.35\\
Hen\,3-461         & 8.26 & 8.35 & -0.07 & +0.11 & +0.12\\
SY\,Mus            & 8.02 & 8.12 & +0.07 & +0.15 & -0.15\\
Hen\,2-87          & 8.59 & 8.30 & +0.13 & -0.36 & +0.17\\
Hen\,3-828         & 8.27 & 8.21 & -0.01 & +0.47 & +0.03\\
CD-36$^{\circ}$8436& 7.69 & 7.99 & -0.03 & +0.08 & -0.30\\
RW\,Hya            & 7.43 & 7.48 & +0.22 & +0.19 & -0.77\\
Hen\,3-916         & 7.92 & 7.90 & +0.07 & +0.20 & -0.46\\
Hen\,3-1092        & 7.36 & 7.47 & +0.14 & +0.07 & -0.79\\
WRAY\,16-202       & 8.06 & 8.24 & -0.20 & +0.27 & +0.17\\
Hen\,3-1213        & 7.95 & 7.76 & +0.87 & +0.73 & -0.68\\
Hen\,2-173         & 8.13 & 8.17 & +0.29 & +0.37 & -0.18\\
KX\,TrA            & 7.97 & 7.90 & +0.27 & +0.40 & -0.34\\
CL\,Sco            & 7.93 & 8.19 & +0.19 & +0.17 & -0.31\\
V455\,Sco          & 8.37 & 8.81 & +0.11 & +0.08 & +0.36\\
Hen\,2-247         & 8.23 & 8.55 & +0.13 & +0.39 & +0.16\\
RT\,Ser            & 7.96 & 7.93 & +0.18 & +0.47 & -0.51\\
AE\,Ara            & 8.20 & 8.14 & -0.01 & +0.31 & -0.02\\
SS73\,96           & 8.22 & 7.83 & +0.13 & +0.06 & -0.24\\
AS\,270            & 8.21 & 8.09 & -0.09 & -0.01 & +0.03\\
Y\,CrA             & 7.79 & 7.86 & +0.14 & +0.30 & -0.40\\
Hen\,2-374         & 7.80 & 7.99 & +0.19 & +0.49 & -0.52\\
Hen\,3-1761        & 7.76 & 7.79 & -0.14 & +0.09 & -0.25\\
CH\,Cyg$^{[1]}$    & 8.35 & 8.08 & +0.04 & +0.28 & +0.03\\
V2116\,Oph$^{[2]}$ & 8.03 & 8.97 & -0.25 & -0.33 & -0.02\\
  \hline
Sun                & 8.43 & 7.83 & ~0.0  & ~0.0  & ~0.0 \\
  \hline
\end{tabular}
\begin{list}{}{}
\item[$Notes.$ References: $^{[1]}$\,\citet{Sch2006}; $^{[2]}$\,\citet{Hin2006}.]
\end{list}
\end{table}

\begin{figure}
  \resizebox{\hsize}{!}{\includegraphics[]{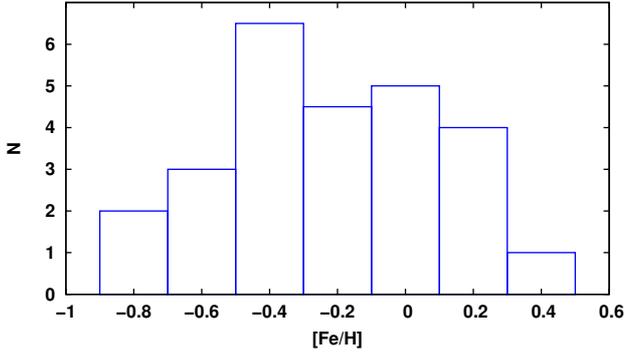}}
  \caption{The distribution for the number ($N$) of objects, counted at
0.2\,dex intervals, as a function of [Fe/H].}
  \label{F8}
\end{figure}

The sample is large enough to benefit from undertaking a statistical
analysis.  The distribution of objects as a function of [Fe/H] is shown in
Fig.\,\ref{F8}.  [Fe/H] is often regarded as a proxy for metallicity.  The
metallicities cover a wide range from significantly subsolar
([Fe$/$H]$=-0.79$\,dex) to slightly supersolar ([Fe$/$H]$=+0.36$\,dex)
with maxima around slightly subsolar ([Fe$/$H]$\sim-0.4$ to $-0.3$\,dex) and
near-solar metallicity.

Based on an analysis of $JHK$ and $IRAS$ photometry of a large sample of
Galactic symbiotic systems \citet{WhMu1992} argued that symbiotic giants
could be related to the metal-rich M stars found in the Galactic bulge and
elsewhere, i.e.~that the symbiotic giants have low masses and higher than
solar metallicity.  They also noted that the mass-loss rates of the
symbiotic giants, derived from the $K-[12]$ and $[25]-[12]$ colours,
although systematically greater than for the local bright giants are similar
to those of the bulge-like stars.  Our abundance results, which is for a
sample of southern symbiotic stars overlapping Whitelock \& Munari's sample,
do not confirm the increased metallicity in symbiotic giants.  On the
contrary, we find subsolar metallicities that suggest efficient mass loss is
needed to explain symbiotic activity.  This is in line with conclusions from
radio studies (\citealt{Sea1993, Mik2002}) that the symbiotic giants tend to
have higher mass-loss rates than single giants of the same spectral type. 
\citet{Gro2013} found that light curves of most symbiotic systems have more
or less regular variations with time-scales of 50--200\,d, most likely due
to stellar pulsations of the cool giant component.  The presence of SRb
variables in these systems can account for relatively high mass-loss rates
from symbiotic giants, about a few times 10$^{-7}$ M$_{\sun}$yr$^{-1}$, as
compared with single field giants \citep{Gro2007b}.

\begin{figure}
\resizebox{\hsize}{!}{\includegraphics[]{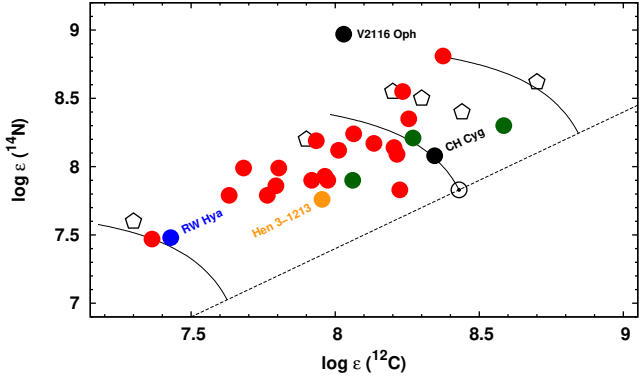}}
  \caption{Nitrogen versus carbon for the symbiotic giants (circles).  The
dashed line represents scaled solar abundances, [$^{12}$C/Fe] = 0 and
[$^{14}$N/Fe] = 0.  The solid curves delineate constant $^{12}$C+$^{14}$N. 
Bulge giants \citep{CuSm2006} are shown as pentagons.  Giants with highest
$^{12}$C/$^{13}$C $\geq 15$ (green circles, see Fig.\,\ref{F10}) are less
elevated in this plane, for instance CH\,Cyg with $^{12}$C$/^{13}$C = 18. 
See online edition for colour version.}
  \label{F9}
\end{figure}

The measured abundances of carbon, nitrogen, and oxygen are similar to
values typical for single Galactic M giants \citep[e.g.][]{SmLa1990}.  In
particular all our giants show an enhancement of nitrogen and a depletion of
carbon.  It is well known that during evolution on the red giant branch the
abundances of carbon and nitrogen are changing because the CN cycle operates
and extensive convection dredges nuclear processed material up to the outer
layers.  Simplifying, it can be assumed that only the CN cycle operates
effectively in the stellar interior during this phase and thus $^{12}$C
nuclei are converted mostly to $^{14}$N.  As a result the $^{12}$C/$^{14}$N
ratio should be reduced with the total number of C+N nuclei conserved.  The
decrease in $^{12}$C combined with some increase in $^{13}$C nuclei results
in a decrease in the $^{12}$C/$^{13}$C ratio.  This picture has been
confirmed by a number of theoretical and observational studies and is called
the 'first dredge-up'.

\begin{figure}
  \resizebox{\hsize}{!}{\includegraphics[]{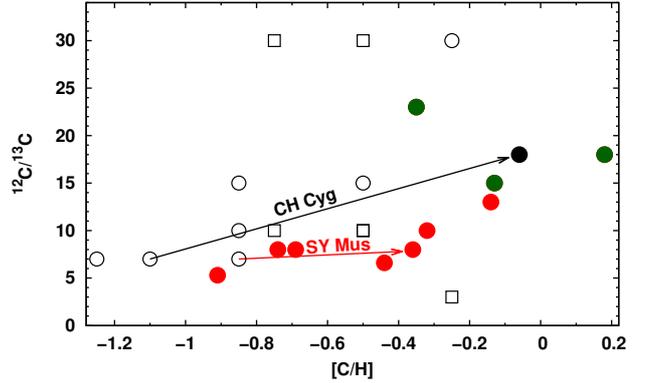}}
  \caption{Chemical abundance ratio of $^{12}$C/$^{13}$C versus [C/H]
measured from $K_{\rm r}$-band spectra for 10 objects in our sample (red and
green filled circles).  CH\,Cyg from high-resolution spectra
\citep{Sch2006}, black circle, from low-resolution spectra
\citep{Sch1992,ScMi2003}, open circles, and squares, respectively.  See
online edition for colour version.}
  \label{F10}
\end{figure}

Evidence of the first dredge-up in our red giants is shown in
Fig.\,\ref{F9}.  The abundances of $^{14}$N versus $^{12}$C for the
symbiotic sample are shown compared to the abundances in Galactic bulge
giants \citep{CuSm2006} .  All our symbiotic giants fall above the scaled
solar line.  This signifies an enhanced $^{14}$N abundance and indicates
that the symbiotic giants have experienced the first dredge-up.  The
occurrence of the first dredge-up is also confirmed by the low
$^{12}$C/$^{13}$C isotopic ratios obtained for 10 objects
(Fig.\,\ref{F10}).  We obtained $^{12}$C/$^{13}$C values in the range
5--23 with average, and median, $^{12}$C/$^{13}$C = $\sim$10.  In
Fig.\,\ref{F9} the position of our giants with highest $^{12}$C/$^{13}$C
ratio is less elevated in $\log{\rmn N}$--$\log{\rmn C}$ plane similar to
CH\,Cyg for which \citet{Sch2006} obtained $^{12}$C$/^{13}$C = 18.

In Fig.\,\ref{F10} our high-resolution determinations of $^{12}$C/$^{13}$C
are compared to [C/H] reported in our papers as well as previous results
derived by \citet{ScMi2003} and \citet{Sch1992}.  The \citet{ScMi2003} and
\citet{Sch1992} results are based on spectra at significantly lower
resolution.  The low-resolution spectra which lack measurable CN molecular
lines could not provide a good measure of carbon abundances.  The
underestimation of the carbon abundance is clearly visible in the case of
CH\,Cyg and SY\,Mus (arrows in Fig.\,\ref{F10}).  The conclusion is that
those abundances from low-resolution spectra are not suitable for comparison
with theoretical models.

\begin{figure}
\resizebox{\hsize}{!}{\includegraphics[]{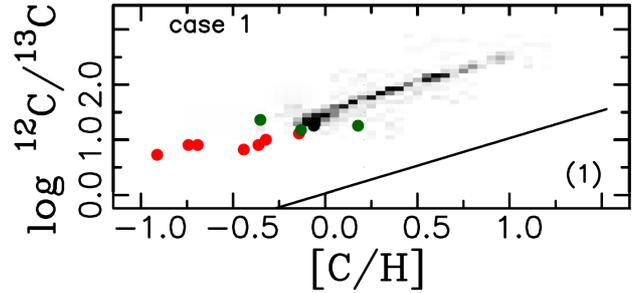}}
  \caption{$^{12}$C/$^{13}$C versus [C/H] from high-resolution spectra
compared with one selected case (fig.\,2, {\sl{left}}, case\,1) derived by
\citet{Lu2008} from theoretical abundances.  See online edition for colour
version.}
  \label{F11}
\end{figure}

\citet{Lu2008} performed the first theoretical study of the chemical
abundances in symbiotic giants.  Among the elements studied are CNO
abundances and $^{12}$C/$^{13}$C isotopic ratio.  The confrontation of
observational $^{12}$C/$^{13}$C and [C/H] with results of theoretical
calculations shows that first dredge-up is insufficient to explain observed
carbon abundances (Fig.\,\ref{F11}).  \citet{Lu2008} suggested that some
additional mixing process, for instance thermohaline mixing applicable to
low-mass giants \citep{ChZa2007}, is required to model the measurements. 
However, \citet{Lu2008} used the solar initial composition in their
calculations and did not consider the effect of metallicity on the carbon
abundances.  Our sample is not homogeneous in respect of [Fe/H] and it is
dominated by objects with subsolar metallicities.

\begin{figure}
\resizebox{\hsize}{!}{\includegraphics[]{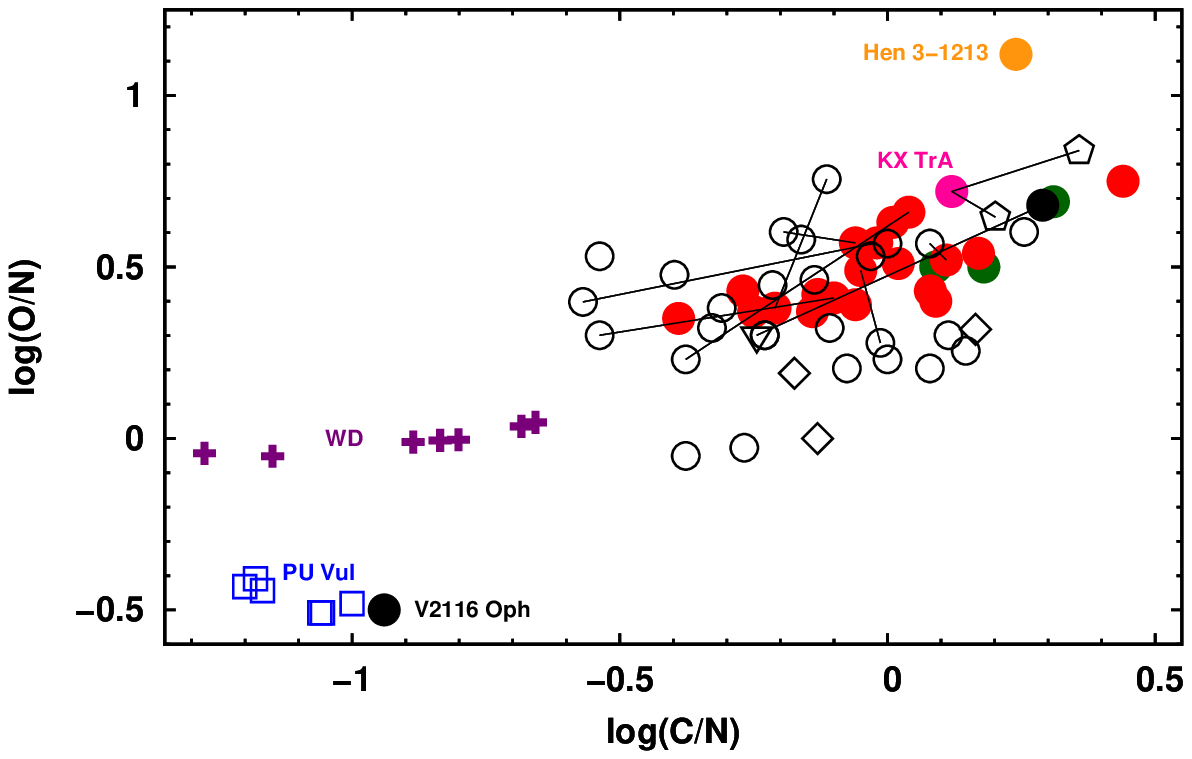}}
\caption{O/N versus C/N from the current photospheric abundances (filled
symbols) compared with values from nebular lines (open symbols),
\citet[][circles]{Nus1988}, \citet[][diamonds]{ScSc1990},
\citet[][pentagons]{Per1995} for KX\,TrA, \citet[][triangle]{Sch2006}
for CH\,Cyg, and \citet[][squares]{VoNu1992} for PU\,Vul during outburst. 
Crosses represent theoretical predictions for nova ejecta from CO WD with
0.65\,M\sun\, \citep{KoPr1997}.  Solid lines link measurements obtained
with two methods (from nebular lines and photospheric) for the same objects. 
See online edition for colour version.}
  \label{F12}
\end{figure}

\begin{figure}
\resizebox{\hsize}{!}{\includegraphics[]{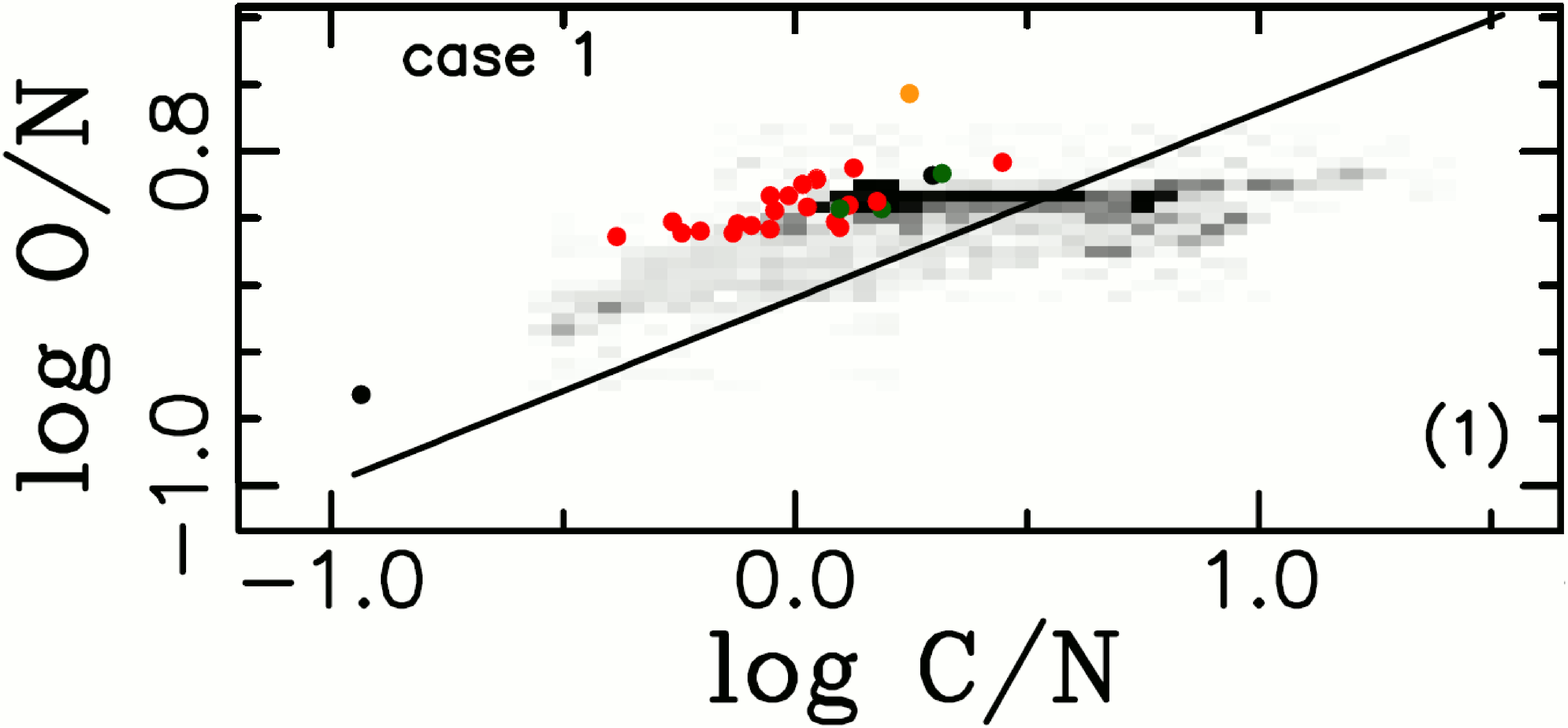}}
  \caption{Comparison of observational $\log{\rm{O/N}}$ versus
$\log{\rm{C/N}}$ (Fig.\,\ref{F12}) with one selected theoretical case
(fig.\,3, {\sl{left}}, case\,1) derived by \citet{Lu2008}.  See online
edition for colour version.}
  \label{F13}
\end{figure}

Most previous studies of C/N/O abundances in symbiotic systems have been
based on nebular emission lines.  In Fig.\,\ref{F12} O/N and C/N ratios
obtained from photospheric abundances are compared with those from nebular
lines (\citealt{Nus1988, ScSc1990, VoNu1992, Per1995, Sch2006}).  The
nebular line results are more scattered in both of the coordinate
directions.  In part this could result from the larger uncertainty in the
abundances derived from the nebular lines.  A larger concern is changes in
the measured abundances resulting from changes of physical conditions in the
nebulae.  In Fig.\,\ref{F12} the 'nebular' points are shifted as a whole
with respect to the 'photospheric' points with the nebular points shifted to
lower O/N and C/N ratios.  The shift is towards the abundance ratios
characterized by the symbiotic nebulae in outburst.  As an example, the
position of PU\,Vul during outburst, and the theoretical O/N and C/N ratios
in ejecta from a 0.65\,M\sun CO white dwarf during nova outburst
\citep{KoPr1997} are shown in Fig.\,\ref{F12}.

The abundances of most symbiotic nebulae are between those of the cool
giants and the materials ejected by the hot companions during symbiotic
novae.  The abundances measured from nebular lines can systematically
underestimate the C and O abundances relative to the N abundance. 
\citet{Nus1988} noted that this effect could reach up to $\sim$30\%. 
Another problem is that the emission spectrum depends significantly on the
spatial orientation of the system.  The abundances measured from emission
lines change with orbital phase (eg.~KX\,TrA in Fig.\,\ref{F12} shows the
different values at orbital phases 0.52 and 0.27, see table\,8 in
Paper\,II).  Reliable values can be obtained only with spectra taken during
superior conjunction of the cool giant (see Paper\,II).  Summarizing, the
behaviour of nebular lines is useful for studying conditions in the
circumbinary environment, for instance the conditions as a function of the
changing projection of the system with orbital phase.  The nebular lines can
also be used to study the heating and supplying a new gas content as a
result of thermonuclear explosions on the white dwarf.

\begin{figure}
\resizebox{\hsize}{!}{\includegraphics[]{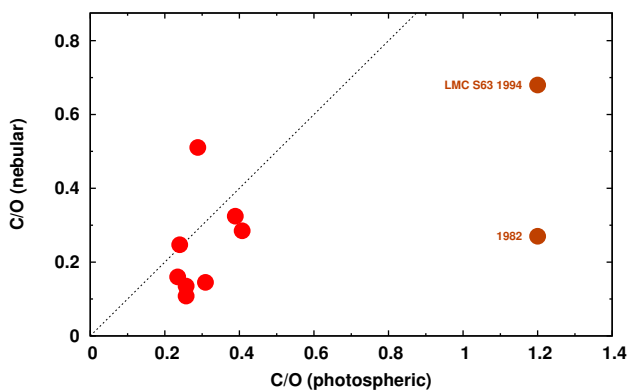}}
  \caption{Nebular versus photospheric C/O ratios for objects with both
values measured.  The changing position of LMC\,S63 is shown.}
  \label{F14}
\end{figure}

\citet{Lu2008} conducted a theoretical analysis of O/N and C/N ratios in
symbiotic giants.  They considered specific regions in the O/N versus C/N
diagram depending on whether or not the cool components have undergone the
third dredge-up.  The occurrence of the third dredge-up in evolved giants is
a function of mass.  Since the third dredge-up enriches envelopes with
carbon and oxygen the relative abundances of O/N and C/N can constrain the
masses of the giants.  The location of our objects in the
$\log{\rm{O/N}}$--$\log{\rm{C/N}}$ spaces defined by \citet{Lu2008}
(Fig.\,\ref{F13}, C/N $<$ O/N) suggests that the cool components of our
symbiotic systems are low-mass giants ($M <$ 4M\sun) that have not undergone
or have undergone only inefficient third dredge-up.  The special location of
V2116\,Oph, at the bottom region of C/N $<$ O/N (Fig.\,\ref{F13}), may
suggest that it could be the only symbiotic giant that passed through third
dredge-up and the hot bottom burning.  However, this is precluded by the
well-established, relatively low mass of this giant $M =$ 1.22M\sun
\citep{Hin2006}.  The high N enrichment clearly visible in
Figs\,\ref{F9},\,\ref{F12},\,and\,\ref{F13}, and peculiar, in general,
chemical composition of the red giant in V2116\,Oph, must be the
manifestation of past mass transfer from the more massive, evolved companion
before it went through the supernova stage and became a neutron star.

\citet{Lu2008} also compared theoretical ratios with those of symbiotic
nebulae, novae, and planetary nebulae.  \citet{Lu2008} did not have
photospheric C/N/O compositions for symbiotic giants.  Now we can compare
the photospheric O/N and C/N ratios with those from symbiotic nebulae and
novae (Figs\,\ref{F12}\,and\,\ref{F13}).  It confirms previous suggestions
that the compositions of the symbiotic nebulae are modified by the material
ejected from the hot components during active phase.  Thus, decreases in
O/N, C/N, and C/O ratios occur.  Following an active phase the abundances
return slowly to the state that existed before activity as material in
nebula becomes dominated by material from the red giant wind.  The wind
material is rich mainly in oxygen and/or carbon.  A good example is
symbiotic S63 in LMC where C/O ratio seems to grow continuously
(Fig.\,\ref{F14}) after active phases \citep{Ilk2015}.

When the evolution of carbon and nitrogen abundances occurs with the total
number of carbon plus nitrogen nuclei conserved it can be presumed that the
oxygen abundance should remain almost unchanged.  In such cases we can
assume, as did \citet{CuSm2006}, that the oxygen abundance is still roughly
the value with which the star was born.  Oxygen is one of the
$\alpha$-elements that are particularly important in studying the evolution
of stars in the context of the formation and chemical evolution of Galactic
populations.  In our sample giants we measured abundances of two
$\alpha$-elements, oxygen and titanium.  The $\alpha$-elements are produced
over relatively short time-scales, originating mostly from massive stars and
Type II supernovae (SNe\,II).  Iron, on the other hand, is produced over
much longer time-scales in the Type Ia supernovae (SNe\,Ia) explosions.  The
contamination of the interstellar medium (ISM) with material originated from
these two sources can lead to significantly different trends for particular
populations.  Thus, clear separations between sequences for various stellar
populations are observed in the [O$/$Fe] versus [Fe$/$H] and [Ti$/$Fe]
versus [Fe$/$H] planes (see e.g.  \citealt{Ben2006,CuSm2006}).\\

\begin{figure}
  \resizebox{\hsize}{!}{\includegraphics[]{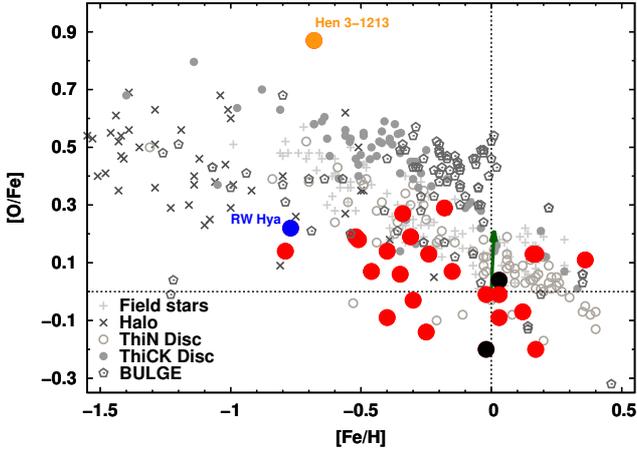}}
  \caption{Oxygen relative to iron for various stellar populations with
positions of our targets denoted with large coloured circles.  Four
populations are distinguished: thin and thick discs, halo, and bulge.  The
thin and thick disc samples contain only those objects with membership
confirmed by their kinematic characteristics.  A large black circles marks
the positions of CH\,Cyg (top) from \citet{Sch2006} and V2116\,Oph (bottom)
the symbiotic neutron star system from \citep{Hin2006}.  The green arrow
denotes the systematic shift from the use of PHOENIX models instead of MARCS
models (see Section\,\ref{secres}).  See online edition for colour version.}
  \label{F15}
\end{figure}

\begin{figure}
  \resizebox{\hsize}{!}{\includegraphics[]{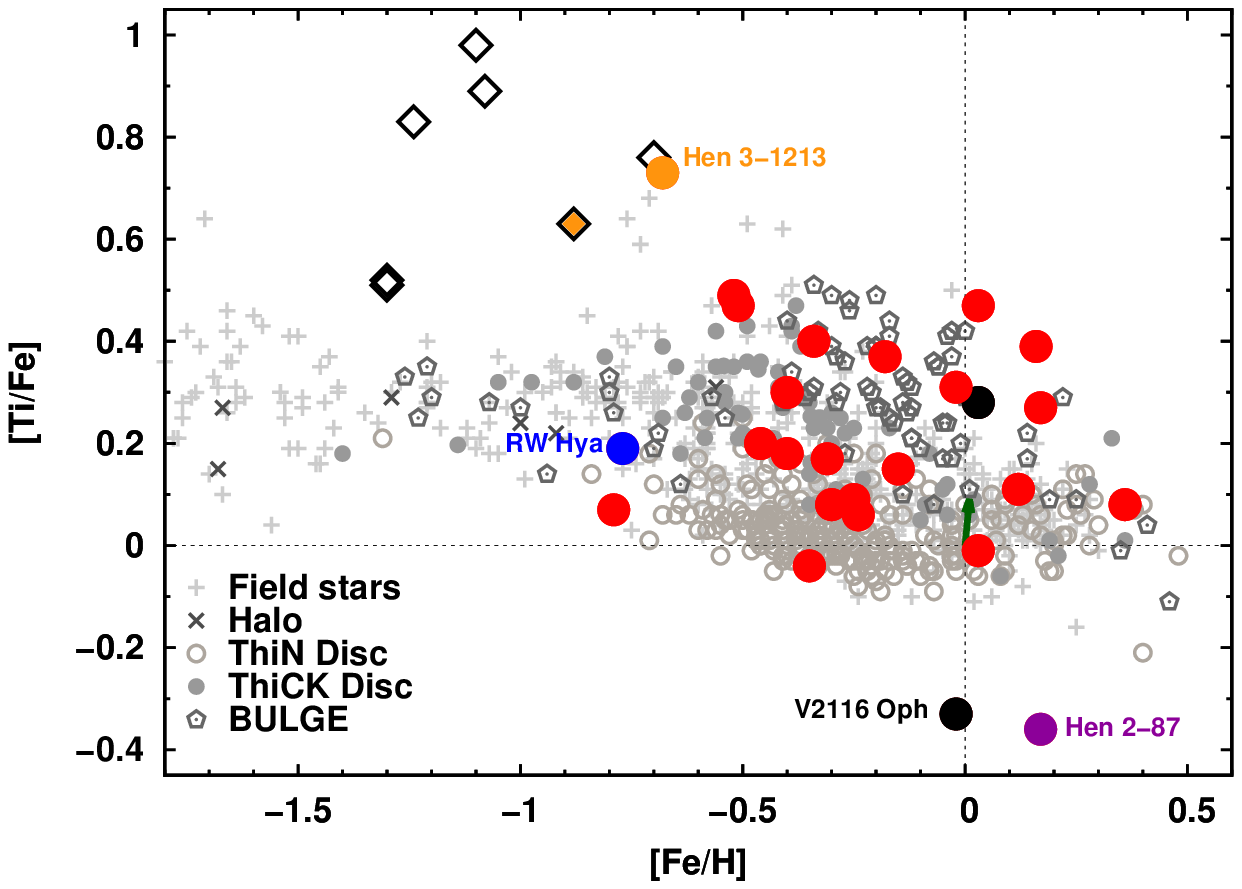}}
  \caption{Titanium relative to iron for various stellar populations with
positions of our targets denoted with large coloured circles.  Symbols
analogous to those in Fig.\,\ref{F15}.  Black open diamonds are giants in
yellow symbiotic systems from \citealt[1997]{Smi1996}, \citealt{Per1998},
and \citealt{Per2009}.  Yellow open diamond are Ti and Fe abundances from
\citet{Per2009} for Hen\,3-1213.  Hen\,2-87 (magenta) has a particularly
very low [Ti$/$Fe] ratio.  See online edition for colour version.}
  \label{F16}
\end{figure}

We use our values for the abundances of oxygen, iron, and titanium in 24
symbiotic giants to investigate the first analyses of symbiotics in term of
chemical evolution and membership in Galactic populations. 
Figs\,\ref{F15} and \ref{F16} show [O$/$Fe] and [Ti$/$Fe], respectively,
as a function of [Fe$/$H] for the symbiotic giants along with values for
various stellar populations (halo, thin and thick disc, and bulge) from the
literature
(\citealt{Gra1988,Edv1993,McW1995,Ful2000,Pro2000,Boy2001,Mel2001,Smi2001a,Joh2002,MeBa2002,Ful2003,Red2003,Ben2005,Ric2005,CuSm2006,Alv2010,Ryd2010,Ben2011,Ric2012,Smi2013}).
The population studies have been scaled to the solar composition of
\citet{Asp2009} for CNO, and \citet{Sco2015} for elements around the
iron peak (Fe and Ti).  The halo, thin- and thick-disc populations are
grouped around clear sequences.  On the contrary, the bulge population seems
be composed of a more or less chemically inhomogeneous groups of stars.  It
is more scattered in the [O$/$Fe] versus [Fe$/$H] (see Paper\,II) and
[Ti$/$Fe] versus [Fe$/$H] planes overlapping partly with all other
populations.  It is, however, in both cases shifted somewhat towards higher
[O$/$Fe] and [Ti$/$Fe] perhaps reflecting more rapid enrichment in metals
from SNe\,II explosions \citep[see][]{CuSm2006}.  In the [O$/$Fe] versus
[Fe$/$H] diagram (Fig.\,\ref{F15}) most of our targets are located on or
somewhat below of area occupied by thin- and thick-disc populations.  The
position of RW\,Hya and Hen\,3-1213 at low metallicity, [Fe$/$H]$\sim$-0.8,
supports their membership in the extended thick-disc/halo population.

\citet{Per2009} from their analysis of four yellow symbiotic systems,
including Hen\,3-1213, concluded that the overall abundance pattern follows
the halo abundances.  In the [Ti$/$Fe]--[Fe$/$H] plane (Fig.\,\ref{F16})
our M-type giants are typically at higher [Ti$/$Fe] in a region occupied
mainly by thick-disc and bulge stars.  The only yellow symbiotic in our
sample, Hen\,3-1213, has even higher [Ti$/$Fe].  Similar enhancement of
[Ti$/$Fe] was found by \citet{Per2009} for yellow symbiotic giants.  They
also noticed that the other $\alpha$-elements to Fe are typical of halo
giant stars of the same metallicity.  We used the published [Ti$/$Fe] values
for AG\,Dra \citep{Smi1996}, BD-21$^{\circ}$3873 \citep{Smi1997}, Hen\,2-467
\citep{Per1998}, CD-43$^{\circ}$14304, Hen\,3-863, StH$\alpha$\,176, and
Hen\,3-1213 \citep{Per2009} to plot their positions in Fig.\,\ref{F16}. 
This plot demonstrates that this Ti anomaly increases with decreasing
metallicity.  The reason for such a high titanium abundance in symbiotic
giants is not known and we are not able to provide any interpretation. 
However, \citet{CuSm2006} noted very similar behaviour for their bulge
giants.  Our present results indicate that this titanium anomaly is present
in both red (i.e.  M-type giants) and yellow symbiotic systems.  The data
available are still relatively scant and noisy but we can speculate that
this anomaly is distributed along a sequence, suggesting this is a genuine
characteristic of S-type symbiotic giants.

\section{Conclusions}

Analysis of the photospheric abundances of CNO and elements around the iron
peak (Fe, Ti, Ni, and Sc) was performed for the giant stars in a sample of
24 southern S-type symbiotic systems.  Our analysis resulted in
metallicities distributed in a wide range from significantly subsolar
([Fe$/$H]$\sim-0.8$\,dex) to slightly supersolar
([Fe$/$H]$\sim+0.35$\,dex), with largest representation around slightly
subsolar ([Fe$/$H]$\sim-0.4$ to $-0.3$\,dex) and near-solar metallicity. 
The enrichment in $^{14}$N isotope, found in all cases, indicates that the
giants have experienced the first dredge-up.  This is confirmed by the low
$^{12}$C/$^{13}$C ratio (5--23) that was measured in a subset of the sample. 
Comparison with abundances from nebular lines shows that the nebulae are
contaminated by activity on the white dwarf and do not provide reliable
abundances for the red giant.  We found that the enhanced [Ti$/$Fe]
abundances previously found for yellow symbiotic systems are also typically
enhanced in red symbiotic giants.  This suggests that enhanced [Ti$/$Fe]
abundance could be a characteristic of the giants in S-type symbiotic
systems.

\section*{Acknowledgements}

This study has been supported in part by the Polish NCN grant no. 
DEC-2011/01/B/ST9/06145.  CG has been also financed by the NCN post-doc
programme FUGA via grant DEC-2013/08/S/ST9/00581.  The observations were
obtained at the Gemini Observatory, which is operated by the Association of
Universities for Research in Astronomy, Inc., under a cooperative agreement
with the NSF on behalf of the Gemini partnership: the National Science
Foundation (USA), the National Research Council (Canada), CONICYT
(Chile), the Australian Research Council (Australia), Minist\'{e}rio da
Ci\^{e}ncia, Tecnologia e Inova\c{c}\~{a}o (Brazil), and Ministerio de
Ciencia, Tecnolog\'{i}a e Innovaci\'{o}n Productiva (Argentina).

                                     
\appendix

\section[]{The full journal of spectroscopic observations.}

\begin{table*}
 \centering
  \caption{Journal of spectroscopic observations. Quadrature sums of the
projected rotational velocities and microturbulence$\,^a$ $(V_{\rmn{rot}}^2
\sin^2{i} + \xi^2_{\rmn{t}})^{0.5}$ shown have been obtained via
cross-correlation technique (CCF) and from measurement of full width at
half-maximum (FWHM) of $K$ band \mbox{Ti\,{\sc i}}, \mbox{Fe\,{\sc i}}, and
\mbox{Sc\,{\sc i}} absorption lines.  Orbital phases have been calculated
according to the referenced literature ephemeris.}
\label{TA1}
  \begin{tabular}{@{}lccccccc@{}}
  \hline
                      & \   Id. num.$^{b}$  & Sp.\,region              & Date         & HJD(mid)  & \multicolumn{2}{c}{$(V_{\rmn{rot}}^2 \sin^2{i} + \xi^2_{\rmn{t}})^{0.5}$} & Orbital phase$^{c}$\\
                      &                     & band($\lambda$[\micron]) & (dd.mm.yyyy) &           & CCF                            & FWHM                                     &                    \\
 \hline
                      &     & $H$ ($\sim$1.56)         & 16.02.2003 & 245 2686.7409 & ~6.08 & --                     & 0.30 \\
 BX\,Mon              &  23 & $K$ ($\sim$2.23)         & 20.04.2003 & 245 2749.5231 & ~7.58 & 8.67 $\pm$ 1.41        & 0.35 \\
                      &     & $K_{\rm r}$ ($\sim$2.36) & 03.04.2006 & 245 3828.5095 & ~8.44 & --                     & 0.20 \\
                      &     &                          &            &               &       & ~8.67 $\pm$ 1.41$^{d}$ &      \\
 \hline
                      &     & $H$ ($\sim$1.56)         & 16.02.2003 & 245 2686.7491 & ~4.19 & --                     & 0.39 \\
 V694\,Mon            &  24 & $K$ ($\sim$2.23)         & 20.04.2003 & 245 2749.5326 & ~6.34 & 8.42 $\pm$ 0.99        & 0.42 \\
                      &     & $K_{\rm r}$ ($\sim$2.36) & 03.04.2006 & 245 3828.5187 & ~7.21 & --                     & 0.98 \\
                      &     & $H_{\rm b}$ ($\sim$1.54) & 12.03.2010 & 245 5267.5052 & ~9.36 & --                     & 0.72 \\
                      &     &                          &            &               &       & ~8.42 $\pm$ 0.99$^{d}$ &      \\
 \hline
                      &     & $H$ ($\sim$1.56)         & 16.02.2003 & 245 2686.7769 & ~3.97 & --                     & 0.98 \\
 Hen\,3-461           &  31 & $K$ ($\sim$2.23)         & 20.04.2003 & 245 2749.5662 & ~7.42 & 8.18 $\pm$ 0.63        & 0.08 \\
                      &     & $K$ ($\sim$2.23)         & 13.12.2003 & 245 2986.7827 & ~7.38 & 7.11 $\pm$ 0.76        & 0.46 \\
                      &     & $K$ ($\sim$2.23)         & 03.04.2004 & 245 3098.6078 & ~6.54 & 7.66 $\pm$ 0.89        & 0.63 \\
                      &     & $K_{\rm r}$ ($\sim$2.36) & 03.04.2006 & 245 3828.5602 & ~6.63 & --                     & 0.78 \\
                      &     & $H$ ($\sim$1.56)         & 02.04.2009 & 245 4923.5570 & ~4.68 & --                     & 0.50 \\
                      &     & $H$ ($\sim$1.56)         & 23.05.2010 & 245 5340.4918 & ~5.24 & --                     & 0.16 \\
                      &     &                          &            &               &       & ~7.68 $\pm$ 0.52$^{d}$ &      \\
 \hline
                      &     & $H$ ($\sim$1.56)         & 17.02.2003 & 245 2687.7566 & ~3.88 & --                     & 0.02 \\
 SY\,Mus              &  33 & $K$ ($\sim$2.23)         & 20.04.2003 & 245 2749.5817 & ~5.98 & 6.91 $\pm$ 0.99        & 0.12 \\
                      &     & $K$ ($\sim$2.23)         & 13.12.2003 & 245 2986.8250 & ~6.84 & 6.97 $\pm$ 0.98        & 0.50 \\
                      &     & $K_{\rm r}$ ($\sim$2.36) & 03.04.2006 & 245 3828.5767 & ~5.01 & --                     & 0.84 \\
                      &     & $H$ ($\sim$1.56)         & 02.04.2009 & 245 4923.5906 & ~5.60 & --                     & 0.60 \\
                      &     & $H_{\rm b}$ ($\sim$1.54) & 23.03.2010 & 245 5278.5907 & ~5.84 & --                     & 0.16 \\
                      &     & $H$ ($\sim$1.56)         & 26.04.2010 & 245 5312.6012 & ~6.53 & --                     & 0.22 \\
                      &     &                          &            &               &       & ~6.94 $\pm$ 0.64$^{d}$ &      \\
 \hline
                      &     & $H$ ($\sim$1.56)         & 17.02.2003 & 245 2687.7725 & ~5.01 & --                     & --   \\
 Hen\,2-87            &  37 & $K$ ($\sim$2.23)         & 20.04.2003 & 245 2749.5942 & ~7.80 & 9.13 $\pm$ 0.76        & --   \\
                      &     & $K$ ($\sim$2.23)         & 13.12.2003 & 245 2986.8410 & 10.62 &10.28 $\pm$ 1.42        & --   \\
                      &     & $K$ ($\sim$2.23)         & 03.04.2004 & 245 3098.6172 & ~8.84 &10.09 $\pm$ 0.82        & --   \\
                      &     & $K_{\rm r}$ ($\sim$2.36) & 03.04.2006 & 245 3828.5853 & ~8.73 & --                     & --   \\
                      &     & $H$ ($\sim$1.56)         & 24.04.2010 & 245 5310.6015 & ~6.53 & --                     & --   \\
                      &     &                          &            &               &       & ~9.79 $\pm$ 0.66$^{d}$ &      \\
 \hline
                      &     & $H$ ($\sim$1.56)         & 16.02.2003 & 245 2686.8053 & ~3.64 & --                     & --   \\
 Hen\,3-828           &  38 & $K$ ($\sim$2.23)         & 20.04.2003 & 245 2749.6031 & ~8.03 & 8.29 $\pm$ 1.07        & --   \\
                      &     & $K$ ($\sim$2.23)         & 13.12.2003 & 245 2986.8478 & ~7.66 & 7.75 $\pm$ 0.82        & --   \\
                      &     & $K$ ($\sim$2.23)         & 03.04.2004 & 245 3098.6278 & ~6.62 & 8.43 $\pm$ 0.72        & --   \\
                      &     & $K_{\rm r}$ ($\sim$2.36) & 03.04.2006 & 245 3828.6050 & ~7.39 & --                     & --   \\
                      &     &                          &            &               &       & ~8.17 $\pm$ 0.54$^{d}$ &      \\
 \hline
                      &     & $H$ ($\sim$1.56)         & 16.02.2003 & 245 2686.8181 & ~6.94 & --                     & --   \\
 CD-36$^{\circ}$8436  &  42 & $K$ ($\sim$2.23)         & 20.04.2003 & 245 2749.6156 & ~6.96 & 8.37 $\pm$ 1.11        & --   \\
                      &     & $K_{\rm r}$ ($\sim$2.36) & 03.04.2006 & 245 3828.6211 & ~6.45 & --                     & --   \\
                      &     & $H_{\rm b}$ ($\sim$1.54) & 15.03.2010 & 245 5270.8938 & ~7.91 & --                     & --   \\
                      &     & $H$ ($\sim$1.56)         & 24.04.2010 & 245 5310.6144 & ~8.81 & --                     & --   \\
                      &     &                          &            &               &       & ~8.37 $\pm$ 1.11$^{d}$ &      \\
 \hline
\end{tabular}   
\begin{list}{}{}
\item[$Notes.$ $^{a}$Units $\rmn{km}\,\rmn{s}^{-1}$.]
\item[$^{b}$Identification number according to \citet{Bel2000}.]
\item[$^{c}$Orbital]\,phases are calculated from the following ephemerides:
BX\,Mon 2449796+1259$\times$E \citep{Fek2000}, V694\,Mon
2448080+1931$\times$E \citep{Gro2007a}, Hen\,3-461 2452063+635$\times$E
\citep{Gro2013}, SY\,Mus 2450176+625$\times$E \citep{Dum1999}, RW\,Hya
2445071.6+370.2$\times$E \citep{KeMi1995} or 2449512+370.4$\times$E
\citep{Sch1996}, Hen\,3-916 2452410+803$\times$E \citep{Gro2013},
Hen\,3-1213 2451806+514$\times$E \citep{Gro2013}, Hen\,2-173
2452625+911$\times$E \citep{Fek2007}.  KX\,TrA 2453053+1350$\times$E
\citep{Fer2003}, CL\,Sco 2452018+625$\times$E \citep{Fek2007}, V455\,Sco
2452641.5+1398$\times$E \citep{Fek2008}, Hen\,2-247 2452355+898$\times$E
\citep{Fek2008}, AE\,Ara 2453449+803.4$\times$E \citep{Fek2010}, AS\,270
2451633+671$\times$E \citep{Fek2007}, Y\,CrA 2454126+1619$\times$E
\citep{Fek2010}, Hen\,2-374 2453173+820$\times$E \citep{Fek2010}.
\item[$^{d}$Values]\,$(V_{\rmn{rot}}^2 \sin^2{i} + \xi^2_{\rmn{t}})^{0.5}$ obtained from all $K$-band spectra jointly -- used for synthetic spectra calculations.
\end{list} 
\end{table*}

\begin{table*}
 \centering
  \contcaption{}
  \begin{tabular}{@{}lccccccc@{}}
  \hline
                      &     Id. num.$^{b}$  & Sp.\,region              & Date         & HJD (mid)  & \multicolumn{2}{c}{$(V_{\rmn{rot}}^2 \sin^2{i} + \xi^2_{\rmn{t}})^{0.5}$} & Orbital phase$^{c}$\\
                      &                     & band($\lambda$[\micron]) & (dd.mm.yyyy) &            & CCF                            & FWHM                                     &                    \\
 \hline
                      &     & $H$ ($\sim$1.56)         & 16.02.2003 & 245 2686.8380 & ~5.74 & --                     & 0.57 \\
 RW\,Hya              &  45 & $K$ ($\sim$2.23)         & 20.04.2003 & 245 2749.6295 & ~6.63 & 6.73 $\pm$ 1.80        & 0.74 \\
                      &     & $K$ ($\sim$2.23)         & 13.12.2003 & 245 2986.8656 & ~6.25 & 6.35 $\pm$ 0.72        & 0.38 \\
                      &     & $K_{\rm r}$ ($\sim$2.36) & 03.04.2006 & 245 3828.6308 & ~5.57 & --                     & 0.65 \\
                      &     & $H$ ($\sim$1.56)         & 24.04.2010 & 245 5310.5915 & ~8.80 & --                     & 0.66 \\
                      &     &                          &            &               &       & ~6.54 $\pm$ 0.94$^{d}$ &      \\
 \hline
                      &     & $K$ ($\sim$2.23)         & 20.04.2003 & 245 2749.6619 & ~8.19 & 8.34 $\pm$ 1.17        & 0.42 \\
 Hen\,3-916           &  46 & $K$ ($\sim$2.23)         & 03.04.2004 & 245 3098.6407 & ~8.78 & 9.11 $\pm$ 1.30        & 0.86 \\
                      &     & $K_{\rm r}$ ($\sim$2.36) & 03.04.2006 & 245 3828.6583 & ~8.38 & --                     & 0.77 \\
                      &     & $H$ ($\sim$1.56)         & 26.04.2010 & 245 5312.5667 & ~6.73 & --                     & 0.62 \\
                      &     & $H$ ($\sim$1.56)         & 22.05.2010 & 245 5338.5624 & ~7.98 & --                     & 0.65 \\
                      &     &                          &            &               &       & ~8.69 $\pm$ 0.90$^{d}$ &      \\
 \hline
                      &     & $H$ ($\sim$1.56)         & 17.02.2003 & 245 2687.7791 & ~5.29 & --                     & --   \\
 Hen\,3-1092          &  53 & $K$ ($\sim$2.23)         & 03.04.2004 & 245 3098.6669 & ~5.80 & 6.16 $\pm$ 0.79        & --   \\
                      &     & $H$ ($\sim$1.56)         & 26.04.2010 & 245 5312.6269 & ~5.26 & --                     & --   \\
                      &     &                          &            &               &       & ~6.16 $\pm$ 0.79$^{d}$ &      \\
 \hline
                      &     & $H$ ($\sim$1.56)         & 17.02.2003 & 245 2687.7936 & ~5.97 & --                     & --   \\
 WRAY\,16-202         &  59 & $K$ ($\sim$2.23)         & 20.04.2003 & 245 2749.7301 & ~6.60 & 8.54 $\pm$ 1.67        & --   \\
                      &     & $K_{\rm r}$ ($\sim$2.36) & 03.04.2006 & 245 3828.6812 & ~8.39 & --                     & --   \\
                      &     & $H$ ($\sim$1.56)         & 24.04.2010 & 245 5310.6498 & ~6.17 & --                     & --   \\
                      &     &                          &            &               &       & ~8.54 $\pm$ 1.67$^{d}$ &      \\
 \hline
                      &     & $H$ ($\sim$1.56)         & 16.02.2003 & 245 2686.8581 & ~9.89 & --                     & 0.71 \\
 Hen\,3-1213          &  65 & $K$ ($\sim$2.23)         & 20.04.2003 & 245 2749.7403 & ~7.24 & 7.82 $\pm$ 0.59        & 0.84 \\
                      &     & $K$ ($\sim$2.23)         & 14.08.2003 & 245 2866.4846 & ~8.38 & 7.60 $\pm$ 0.69        & 0.06 \\
                      &     & $K$ ($\sim$2.23)         & 03.04.2004 & 245 3098.6892 & ~9.43 & 7.68 $\pm$ 0.41        & 0.52 \\
                      &     & $H$ ($\sim$1.56)         & 24.05.2010 & 245 5340.5865 & 11.76 & --                     & 0.88 \\
                      &     &                          &            &               &       & ~7.70 $\pm$ 0.30$^{d}$ &      \\
 \hline
                      &     & $H$ ($\sim$1.56)         & 16.02.2003 & 245 2686.8989 & ~6.36 & --                     & 0.07 \\
 Hen\,2-173           &  66 & $K$ ($\sim$2.23)         & 20.04.2003 & 245 2749.7497 & ~7.38 & 8.31 $\pm$ 1.29        & 0.14 \\
                      &     & $K$ ($\sim$2.23)         & 14.08.2003 & 245 2866.4998 & ~9.06 & 8.27 $\pm$ 0.57        & 0.27 \\
                      &     & $K$ ($\sim$2.23)         & 03.04.2004 & 245 3098.7025 & ~8.15 & 9.34 $\pm$ 0.81        & 0.52 \\
                      &     & $H$ ($\sim$1.56)         & 24.05.2010 & 245 5340.5508 & ~7.50 & --                     & 0.98 \\
                      &     &                          &            &               &       & ~8.64 $\pm$ 0.65$^{d}$ &      \\
\hline
                      &     & $H$ ($\sim$1.56)         & 17.02.2003 & 245 2687.8230 & ~6.05 & --                     & 0.73 \\
 KX\,TrA              &  68 & $K$ ($\sim$2.23)         & 20.04.2003 & 245 2749.7670 & ~6.29 & 8.48 $\pm$ 1.77        & 0.78 \\
                      &     & $K$ ($\sim$2.23)         & 03.04.2004 & 245 3098.7314 & ~5.74 & 8.94 $\pm$ 2.12        & 0.03 \\
                      &     & $H$ ($\sim$1.56)         & 24.05.2010 & 245 5340.6039 & ~6.58 & --                     & 0.69 \\
                      &     &                          &            &               &       & ~8.71 $\pm$ 1.32$^{d}$ &      \\
 \hline
                      &     & $H$ ($\sim$1.56)         & 17.02.2003 & 245 2687.8341 & ~7.02 & --                     & 0.07 \\
 CL\,Sco              &  71 & $K$ ($\sim$2.23)         & 20.04.2003 & 245 2749.7780 & ~6.99 & 7.84 $\pm$ 1.78        & 0.17 \\
                      &     & $K$ ($\sim$2.23)         & 15.08.2003 & 245 2866.5367 & ~8.52 & 8.02 $\pm$ 1.62        & 0.36 \\
                      &     & $K$ ($\sim$2.23)         & 03.04.2004 & 245 3098.7794 & ~8.83 & 8.42 $\pm$ 1.48        & 0.73 \\
                      &     &                          &            &               &       & ~8.09 $\pm$ 0.87$^{d}$ &      \\
 \hline
                      &     & $H$ ($\sim$1.56)         & 17.02.2003 & 245 2687.8626 & ~5.23 & --                     & 0.03 \\
 V455\,Sco            &  73 & $K$ ($\sim$2.23)         & 20.04.2003 & 245 2749.7874 & ~7.78 & 8.62 $\pm$ 1.63        & 0.08 \\
                      &     & $K$ ($\sim$2.23)         & 03.04.2004 & 245 3098.7959 & ~7.40 & 8.65 $\pm$ 1.30        & 0.33 \\
                      &     & $H$ ($\sim$1.56)         & 24.05.2010 & 245 5340.6339 & ~7.25 & --                     & 0.93 \\
                      &     &                          &            &               &       & ~8.63 $\pm$ 0.98$^{d}$ &      \\
 \hline
                      &     & $H$ ($\sim$1.56)         & 17.02.2003 & 245 2687.8745 & ~7.28 & --                     & 0.37 \\
 Hen\,2-247           &  88 & $K$ ($\sim$2.23)         & 20.04.2003 & 245 2749.8453 & ~9.53 & 11.47 $\pm$ 1.07       & 0.44 \\
                      &     & $K$ ($\sim$2.23)         & 15.08.2003 & 245 2866.5144 & 10.95 & 11.29 $\pm$ 0.68       & 0.57 \\
                      &     & $K$ ($\sim$2.23)         & 03.04.2004 & 245 3098.8329 & ~7.69 & ~9.21 $\pm$ 1.55       & 0.83 \\
                      &     & $H$ ($\sim$1.56)         & 27.06.2010 & 245 5374.7414 & ~7.97 & --                     & 0.36 \\
                      &     &                          &            &               &       & ~10.58 $\pm$ 1.05$^{d}$&      \\
 \hline
 RT\,Ser              &  92 & $K$ ($\sim$2.23)         & 20.04.2003 & 245 2749.8550 & ~6.22 & 8.10 $\pm$ 1.78        & --   \\
                      &     &                          &            &               &       & ~8.10 $\pm$ 1.78$^{d}$ &      \\
\hline
\end{tabular}   
\end{table*}

\begin{table*}
 \centering
  \contcaption{}
  \begin{tabular}{@{}lccccccc@{}}
  \hline
                      &     Id. num.$^{b}$  & Sp.\,region              & Date         & HJD (mid)  & \multicolumn{2}{c}{$(V_{\rmn{rot}}^2 \sin^2{i} + \xi^2_{\rmn{t}})^{0.5}$} & Orbital phase$^{c}$\\
                      &                     & band($\lambda$[\micron]) & (dd.mm.yyyy) &            & CCF                            & FWHM                                     &                    \\
 \hline
                      &     & $H$ ($\sim$1.56)         & 17.02.2003 & 245 2687.8830 & ~7.76 & --                      & 0.05 \\
 AE\,Ara              &  93 & $K$ ($\sim$2.23)         & 20.04.2003 & 245 2749.8669 & ~9.16 & 10.06 $\pm$ 0.96        & 0.13 \\
                      &     & $K$ ($\sim$2.23)         & 03.04.2004 & 245 3098.8487 & 10.35 & 10.58 $\pm$ 1.42        & 0.56 \\
                      &     &                          &            &               &       & ~10.30 $\pm$ 0.83$^{d}$ &      \\
 \hline
                      &     & $K$ ($\sim$2.23)         & 20.04.2003 & 245 2749.9407 & 10.68 & ~9.25 $\pm$ 0.71        & --   \\
 SS73\,96             &  94 & $K$ ($\sim$2.23)         & 03.04.2004 & 245 3098.8573 & ~8.47 & ~9.30 $\pm$ 0.66        & --   \\
                      &     & $H$ ($\sim$1.56)         & 02.06.2010 & 245 5349.6673 & ~7.07 & --                      & --   \\
                      &     &                          &            &               &       & ~9.28 $\pm$ 0.46$^{d}$  &      \\
 \hline
                      &     & $H$ ($\sim$1.56)         & 17.02.2003 & 245 2687.8917 & ~6.66 & --                      & 0.57 \\
 AS\,270              & 119 & $K$ ($\sim$2.23)         & 03.04.2004 & 245 3098.8687 & ~8.77 & 10.23 $\pm$ 0.90        & 0.19 \\
                      &     & $H$ ($\sim$1.56)         & 24.05.2010 & 245 5340.9087 & ~8.28 & --                      & 0.53 \\
                      &     &                          &            &               &       & ~10.23 $\pm$ 0.90$^{d}$ &      \\
 \hline
                      &     & $K$ ($\sim$2.23)         & 20.04.2003 & 245 2749.8760 & ~9.56 & 10.57 $\pm$ 2.92        & 0.15 \\
 Y\,CrA               & 131 & $H$ ($\sim$1.56)         & 31.07.2009 & 245 5043.7660 & ~5.67 & --                      & 0.57 \\
                      &     &                          &            &               &       & ~10.57 $\pm$ 2.92$^{d}$ &      \\
 \hline
 Hen\,2-374           & 136 & $K$ ($\sim$2.23)         & 03.04.2004 & 245 3098.8781 & ~5.98 & 6.66 $\pm$ 0.56         & 0.91 \\
                      &     &                          &            &               &       & ~6.66 $\pm$ 0.56$^{d}$  &      \\
 \hline
                      &     & $K$ ($\sim$2.23)         & 15.08.2003 & 245 2866.5289 & ~7.57 & 7.18 $\pm$ 1.02         & --   \\
 Hen\,3-1761          & 170 & $K$ ($\sim$2.23)         & 03.04.2004 & 245 3098.9345 & ~6.38 & 7.27 $\pm$ 0.59         & --   \\
                      &     & $H$ ($\sim$1.56)         & 06.06.2009 & 245 4988.9189 & ~6.68 & --                      & --   \\
                      &     & $H$ ($\sim$1.56)         & 03.06.2010 & 245 5350.7945 & ~5.25 & --                      & --   \\
                      &     &                          &            &               &       & ~7.22 $\pm$ 0.63$^{d}$  &      \\
\hline
\end{tabular}   
\end{table*}

\clearpage

\section[]{Spectra of 24 symbiotic giants observed in $H$-, and/or $H_{\rm
b}$-, $K$-, $K_{\rm r}$-band regions, compared with synthetic fits}

\newpage

\begin{figure}
  \includegraphics[width=84mm]{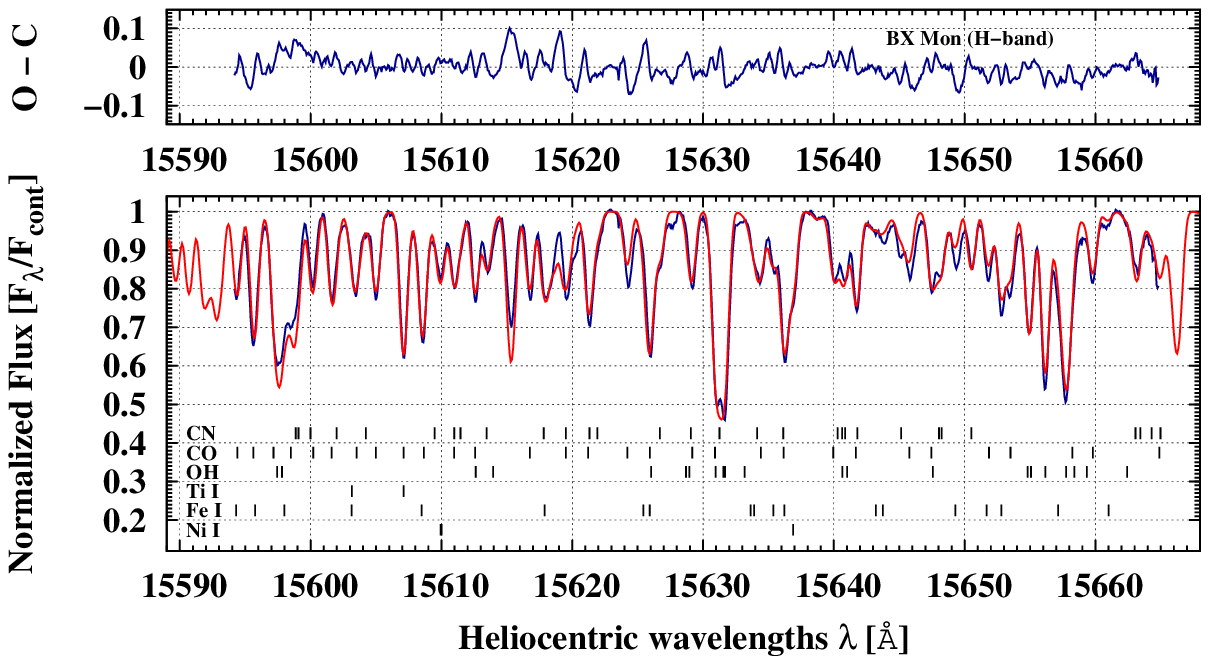}
  \caption{The $H$-band spectrum of BX\,Mon observed 2003 February  (blue
line) and a synthetic spectrum (red line) calculated using the final
abundances and $^{12}$C/$^{13}$C isotopic ratio (Table\,\ref{T4}).}
  \label{FB1}
\end{figure}

\begin{figure}
  \includegraphics[width=84mm]{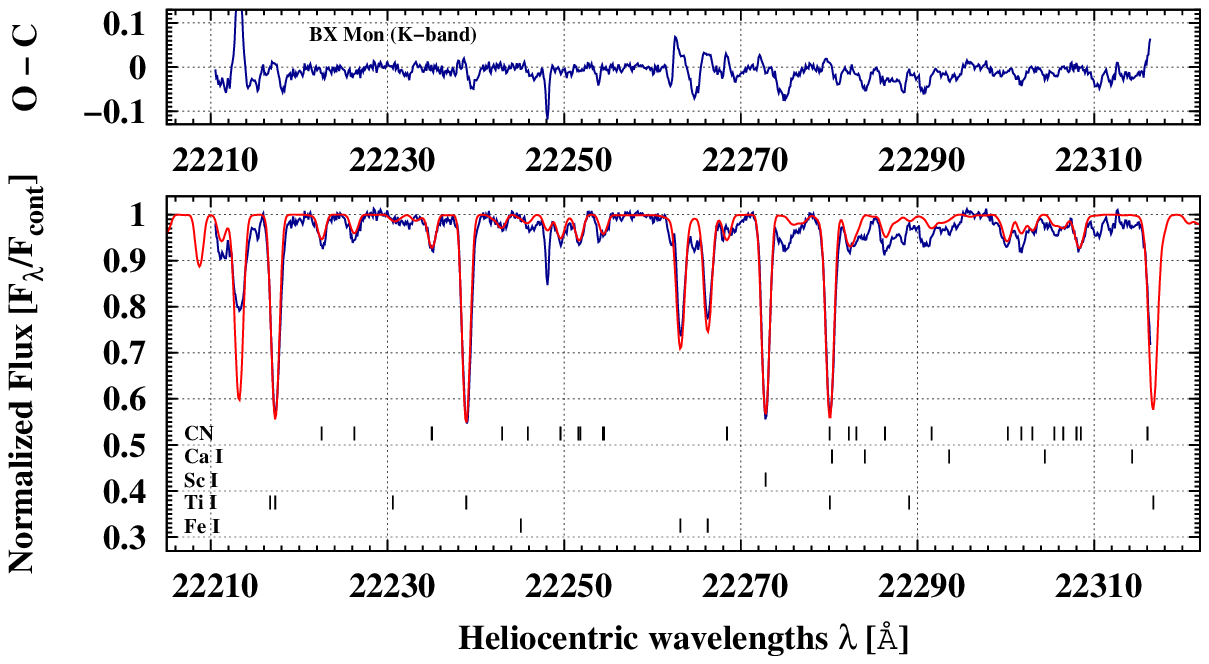} 
  \caption{The $K$-band spectrum of BX\,Mon observed 2003 April (blue line)
and a synthetic spectrum (red line) calculated using the final abundances
and $^{12}$C/$^{13}$C isotopic ratio (Table\,\ref{T4}).}
  \label{FB2}
\end{figure}

\begin{figure}
  \includegraphics[width=84mm]{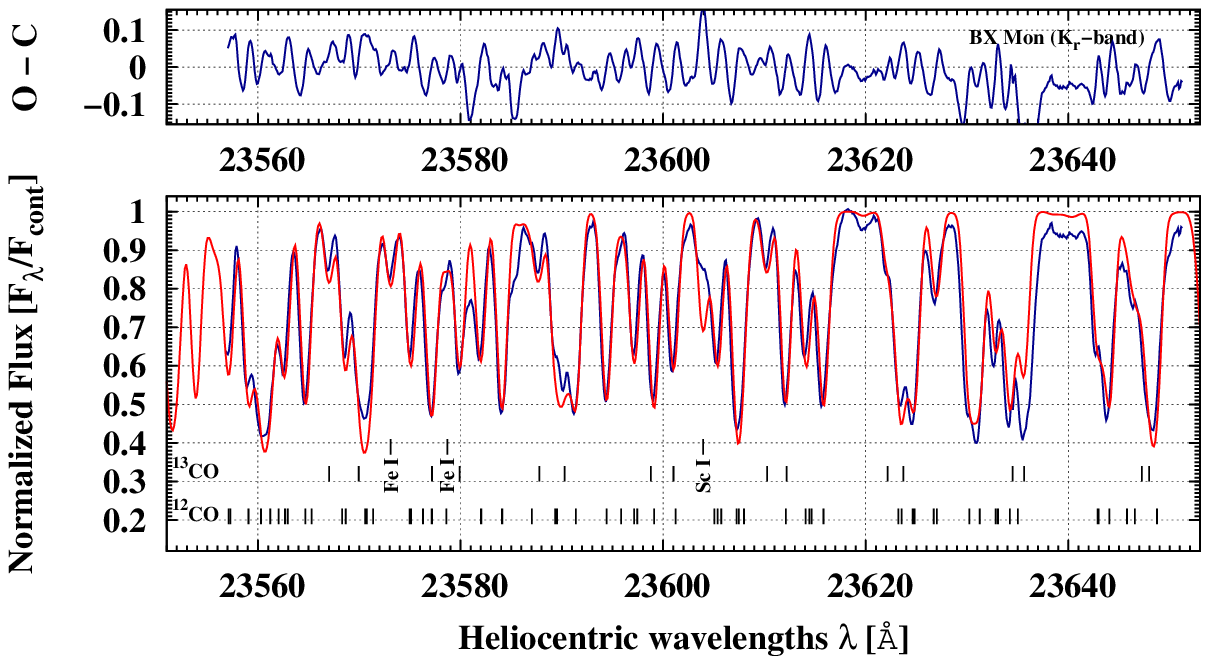} 
  \caption{The $K_{\rm r}$-band spectrum of BX\,Mon observed 2006 April
(blue line) and a synthetic spectrum (red line) calculated using the final
abundances and $^{12}$C/$^{13}$C isotopic ratio (Table\,\ref{T4}).}
  \label{FB3}
\end{figure}

\begin{figure}
  \includegraphics[width=84mm]{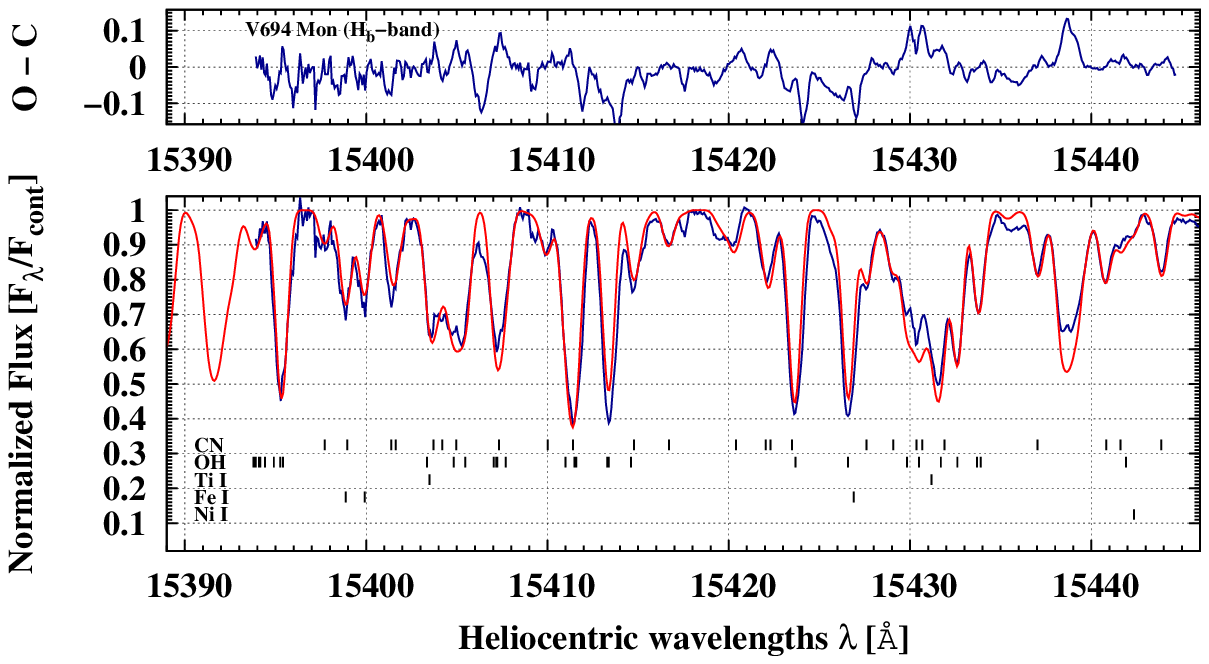}
  \caption{The $H_{\rm b}$-band spectrum of V694\,Mon observed 2010 March
(blue line) and a synthetic spectrum (red line) calculated using the final
abundances and $^{12}$C/$^{13}$C isotopic ratio (Table\,\ref{T4}).}
  \label{FB4}
\end{figure}

\begin{figure}
  \includegraphics[width=84mm]{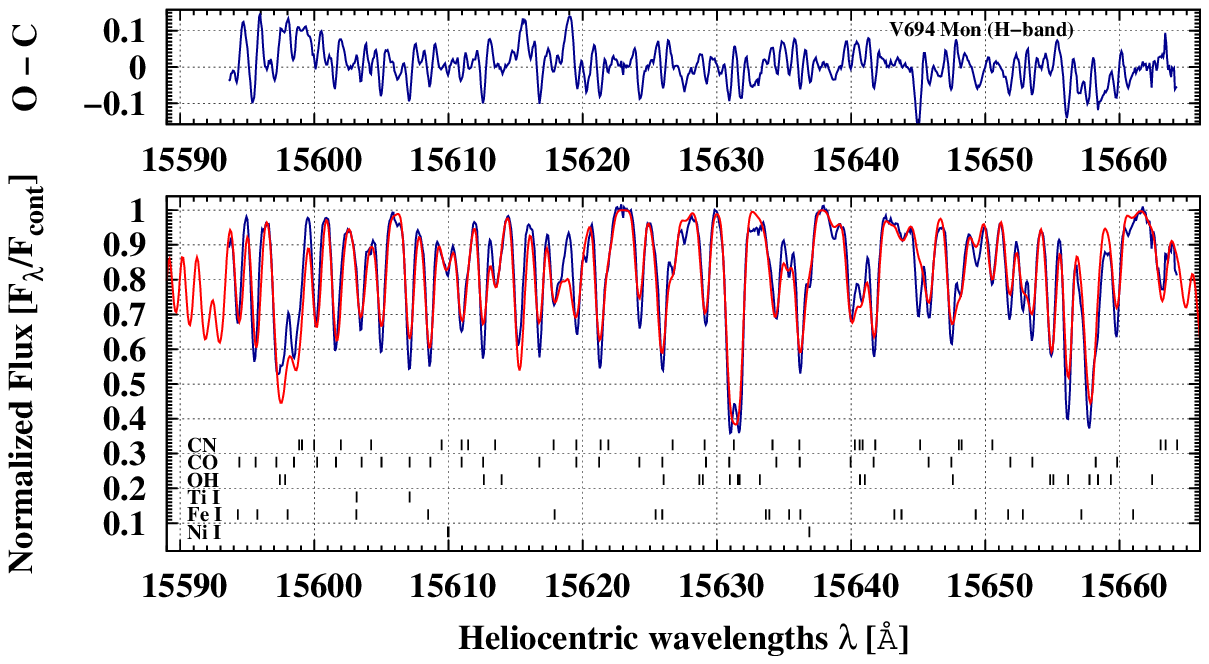}
  \caption{The $H$-band spectrum of V694\,Mon observed 2003 February  (blue
line) and a synthetic spectrum (red line) calculated using the final
abundances and $^{12}$C/$^{13}$C isotopic ratio (Table\,\ref{T4}).}
  \label{FB5}
\end{figure}

\begin{figure}
  \includegraphics[width=84mm]{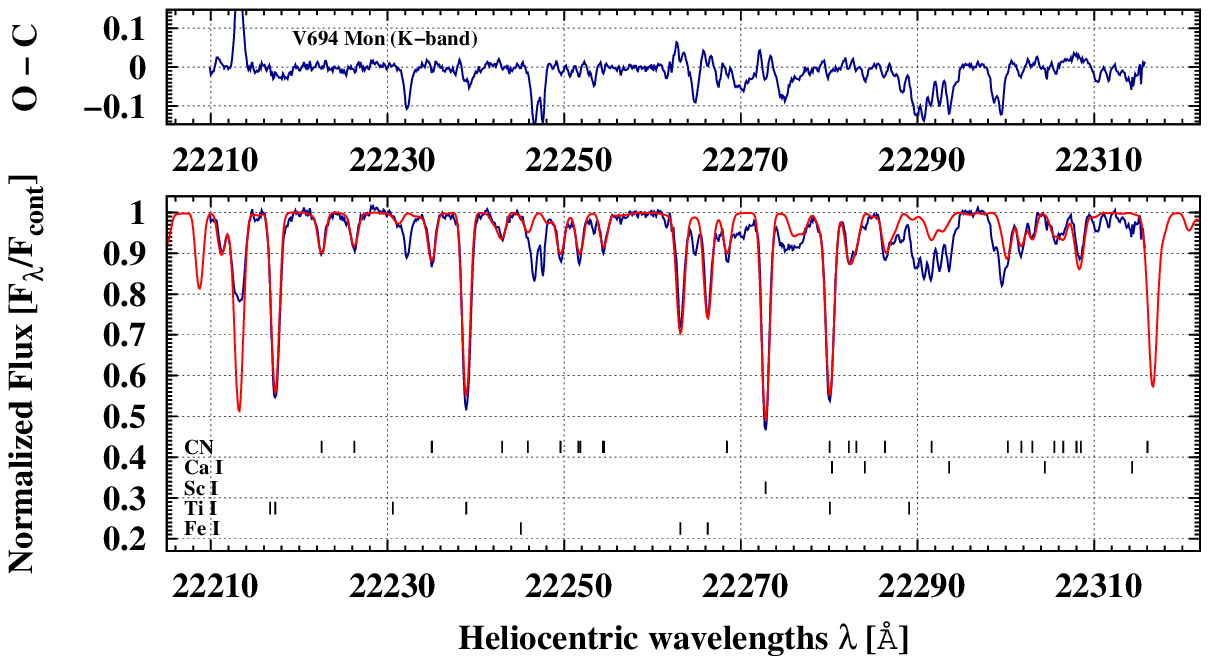} 
  \caption{The $K$-band spectrum of V694\,Mon observed 2003 April (blue
line) and a synthetic spectrum (red line) calculated using the final
abundances and $^{12}$C/$^{13}$C isotopic ratio (Table\,\ref{T4}).}
  \label{FB6}
\end{figure}

\begin{figure}
  \includegraphics[width=84mm]{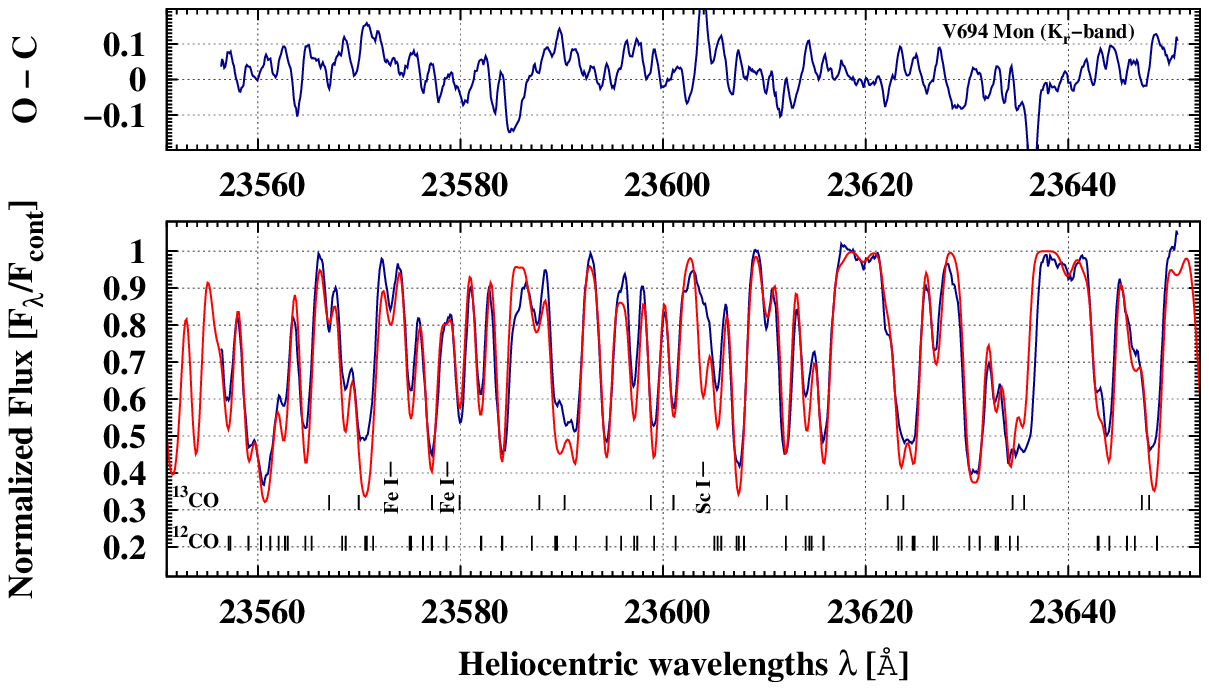} 
  \caption{The $K_{\rm r}$-band spectrum of V694\,Mon observed 2006 April
(blue line) and a synthetic spectrum (red line) calculated using the final
abundances and $^{12}$C/$^{13}$C isotopic ratio (Table\,\ref{T4}).}
  \label{FB7}
\end{figure}

\begin{figure}
  \includegraphics[width=84mm]{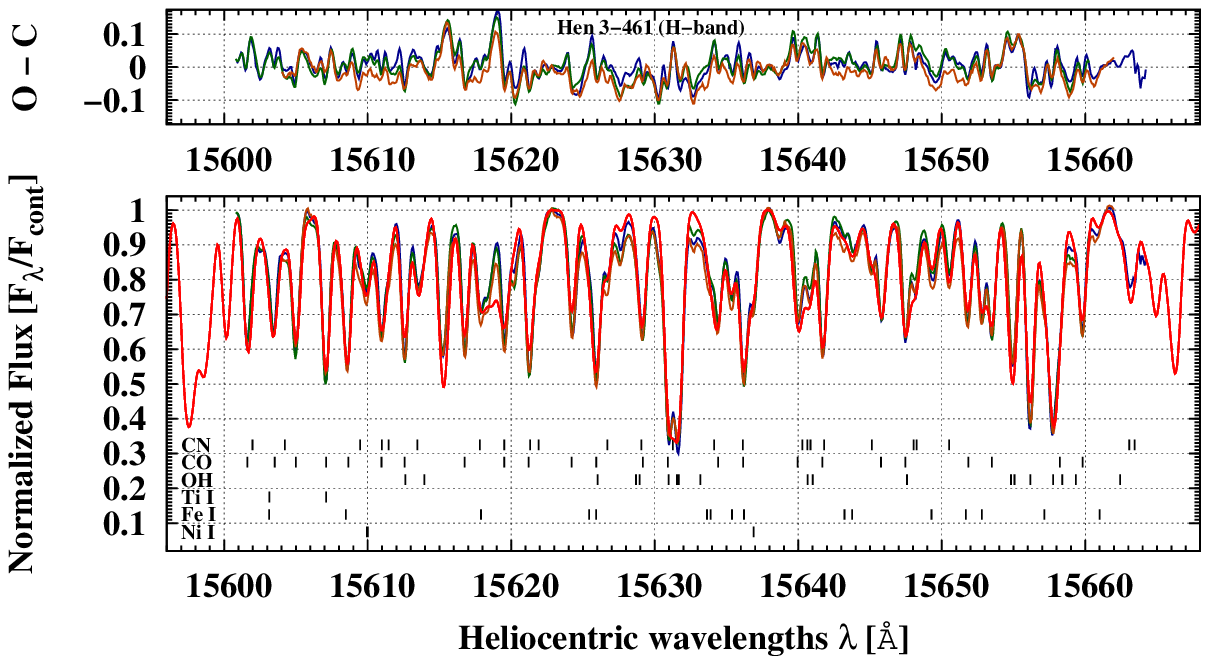}
  \caption{$H$-band spectra of Hen\,3-461 observed 2003 February  (blue
line), 2009 April (green line), 2010 May (dark-orange line), and a synthetic
spectrum (red line) calculated using the final abundances and
$^{12}$C/$^{13}$C isotopic ratio (Table\,\ref{T4}).}
  \label{FB8}
\end{figure}

\begin{figure}
  \includegraphics[width=84mm]{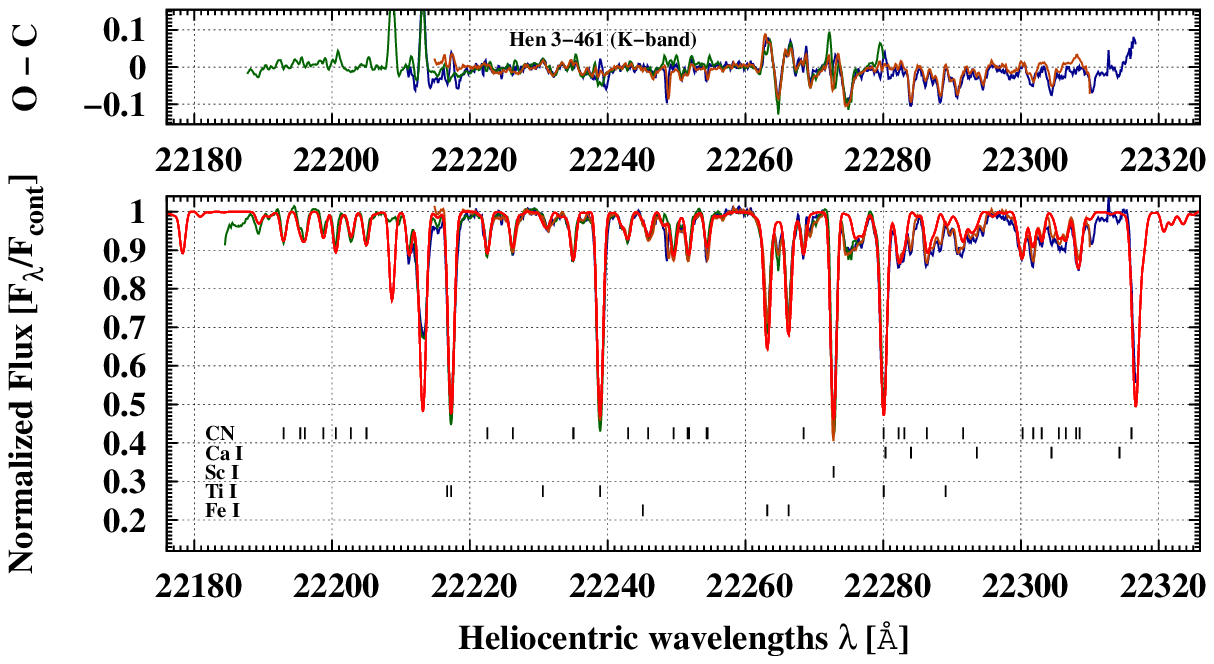} 
  \caption{$K$-band spectra of Hen\,3-461 observed 2003 April (blue line),
2003 December (green line), 2004 April (dark-orange line), and a synthetic
spectrum (red line) calculated using the final abundances
(Table\,\ref{T4}).}
  \label{FB9}
\end{figure}

\begin{figure}
  \includegraphics[width=84mm]{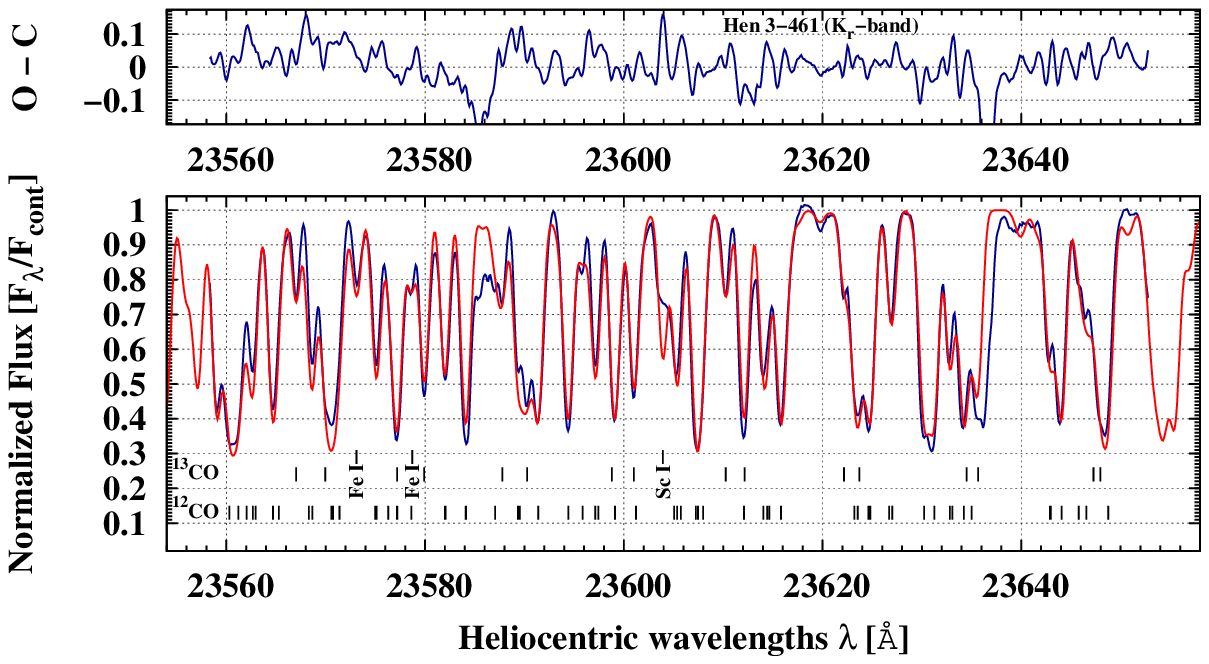} 
  \caption{The $K_{\rm r}$-band spectrum of Hen\,3-461 observed 2006 April
(blue line) and a synthetic spectrum (red line) calculated using the final
abundances and $^{12}$C/$^{13}$C isotopic ratio (Table\,\ref{T4}).}
  \label{FB10}
\end{figure}

\begin{figure}
  \includegraphics[width=84mm]{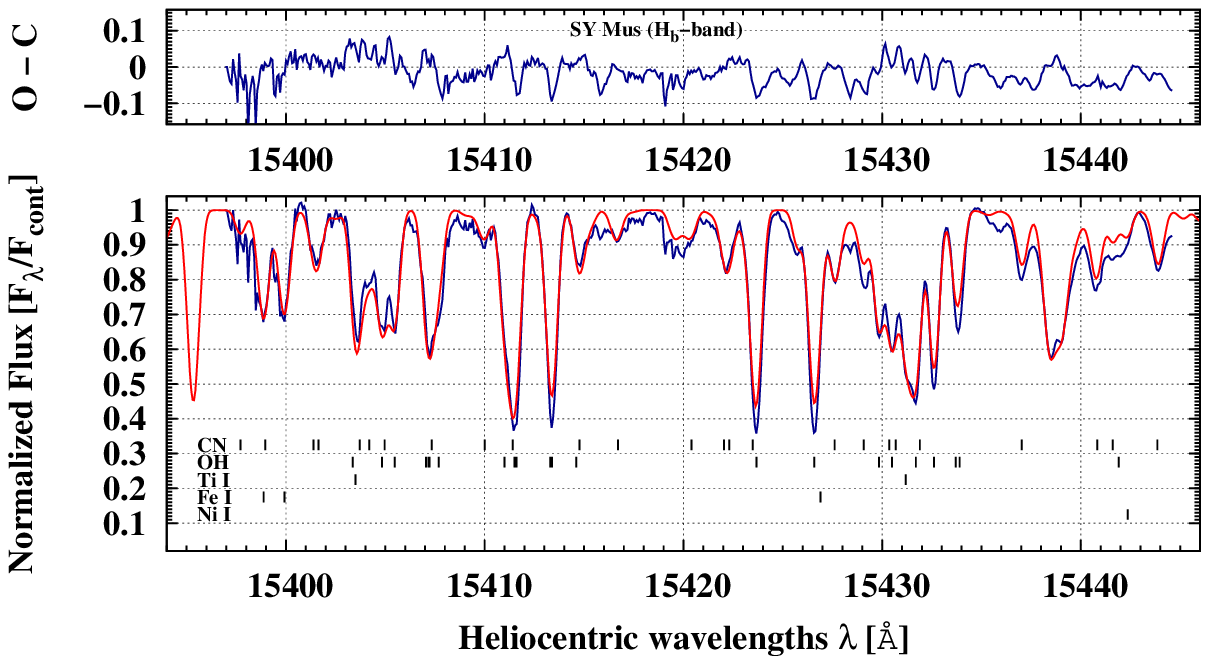}
  \caption{The $H_{\rm b}$-band spectrum of SY\,Mus observed 2010 March
(blue line) and a synthetic spectrum (red line) calculated using the final
abundances and $^{12}$C/$^{13}$C isotopic ratio (Table\,\ref{T4}).}
  \label{FB11}
\end{figure}

\begin{figure}
  \includegraphics[width=84mm]{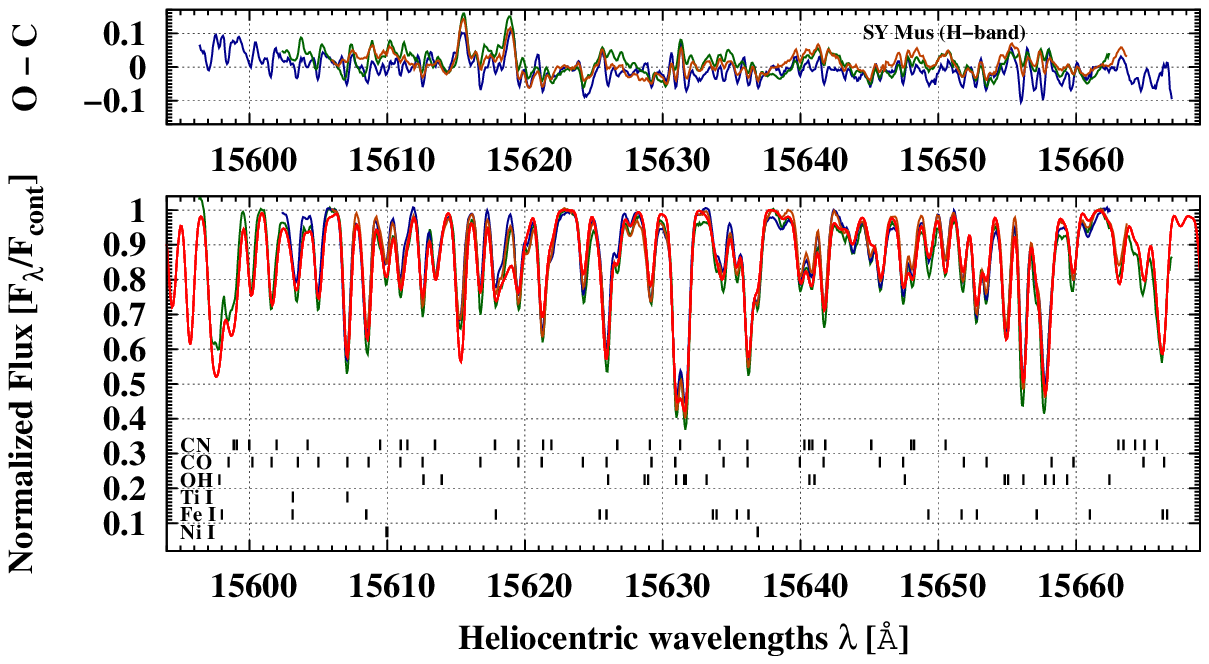}
  \caption{$H$-band spectra of SY\,Mus observed 2003 February  (blue line),
2009 April (green line), 2010 April (dark-orange line), and a synthetic
spectrum (red line) calculated using the final abundances and
$^{12}$C/$^{13}$C isotopic ratio (Table\,\ref{T4}).}
  \label{FB12}
\end{figure}

\begin{figure}
  \includegraphics[width=84mm]{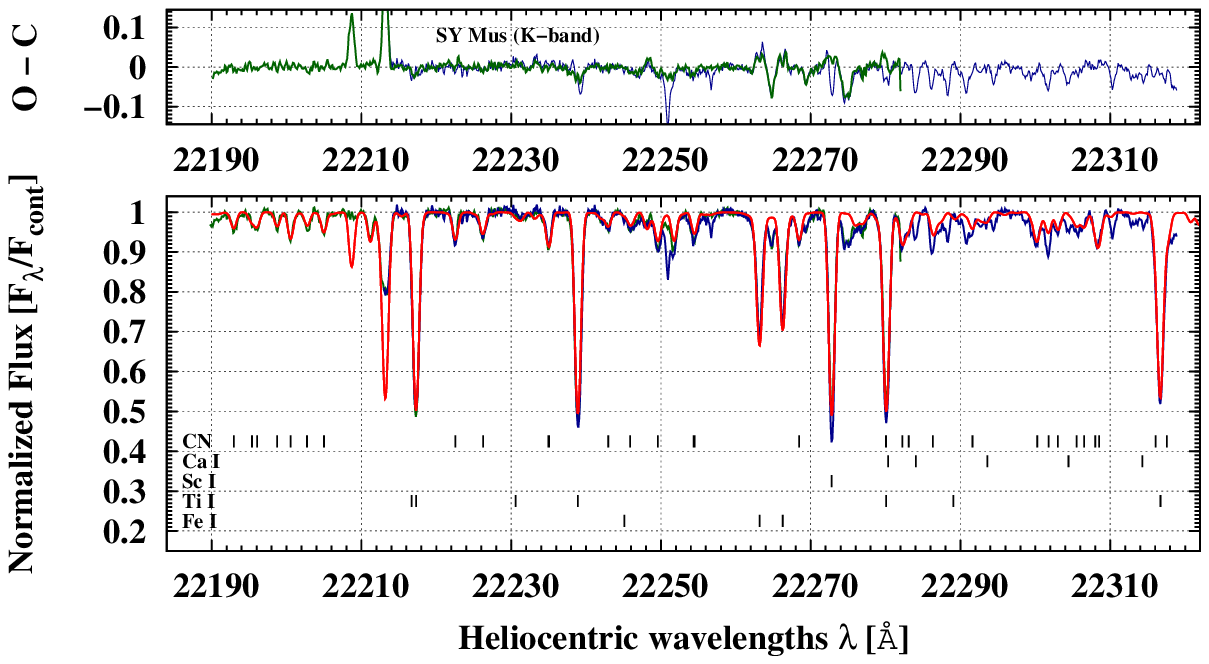} 
  \caption{$K$-band spectra of SY\,Mus observed 2003 April (blue line), 2003
December (green line), and a synthetic spectrum (red line) calculated using
the final abundances (Table\,\ref{T4}).}
  \label{FB13}
\end{figure}

\begin{figure}
  \includegraphics[width=84mm]{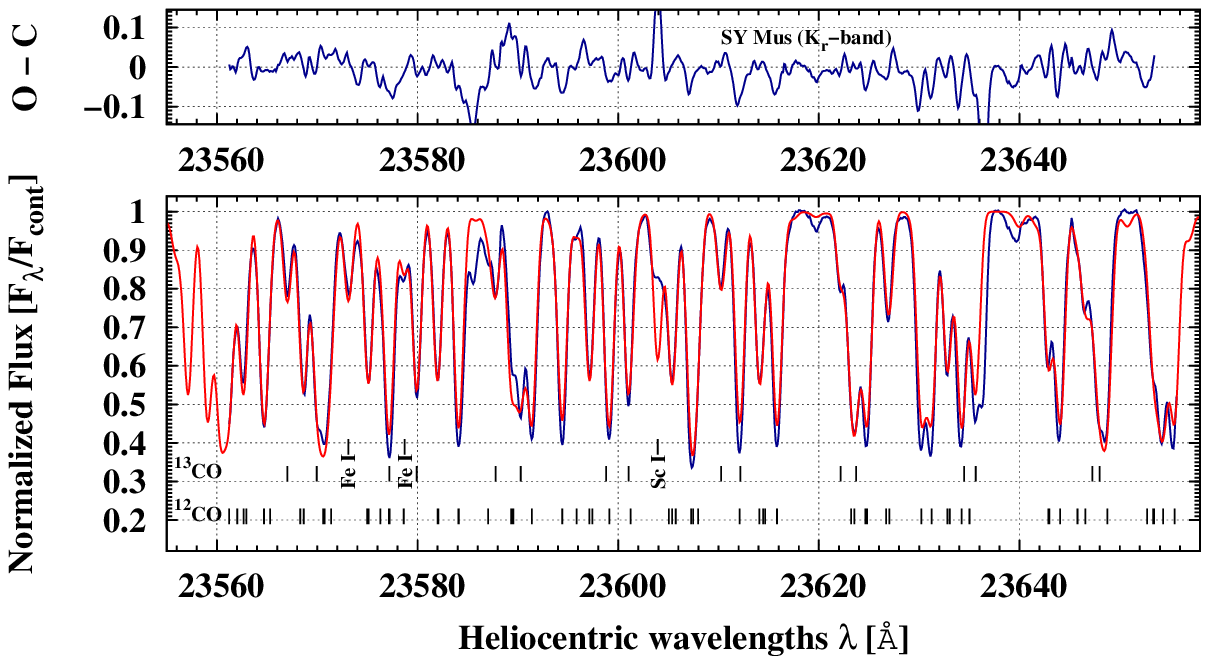} 
  \caption{The $K_{\rm r}$-band spectrum of SY\,Mus observed 2006 April
(blue line) and a synthetic spectrum (red line) calculated using the final
abundances and $^{12}$C/$^{13}$C isotopic ratio (Table\,\ref{T4}).}
  \label{FB14}
\end{figure}

\begin{figure}
  \includegraphics[width=84mm]{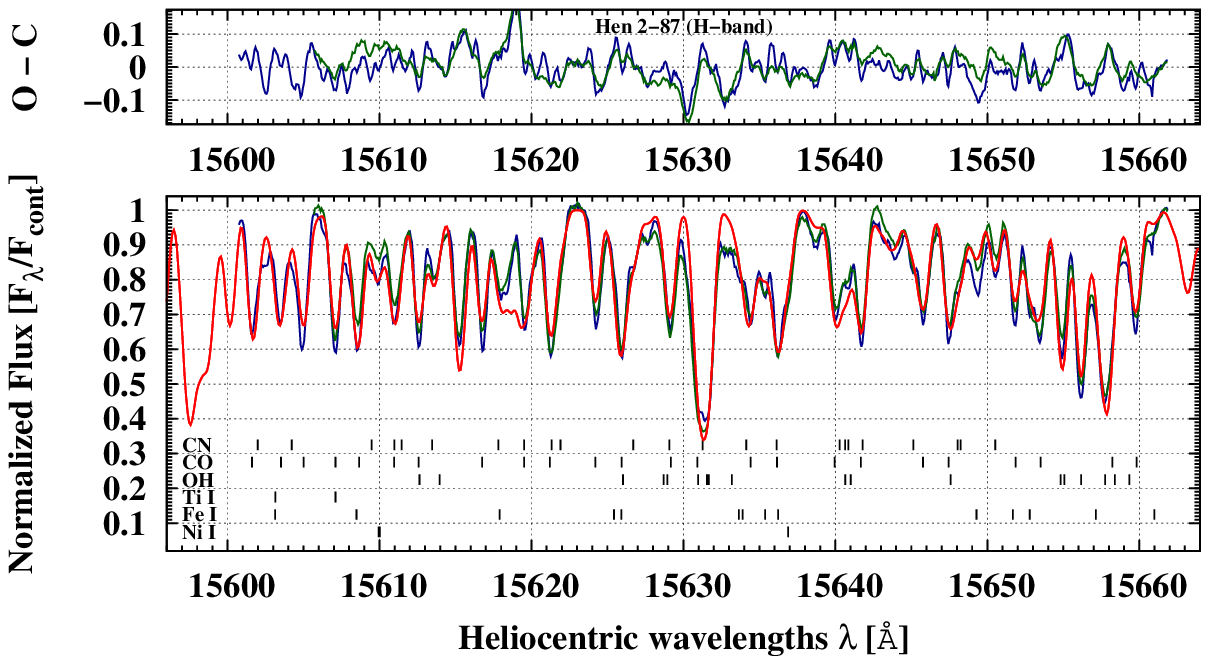}
  \caption{$H$-band spectra of Hen\,2-87 observed 2003 February  (blue
line), 2010 April (green line), and a synthetic spectrum (red line)
calculated using the final abundances and $^{12}$C/$^{13}$C isotopic ratio
(Table\,\ref{T4}).}
  \label{FB15}
\end{figure}

\begin{figure}
  \includegraphics[width=84mm]{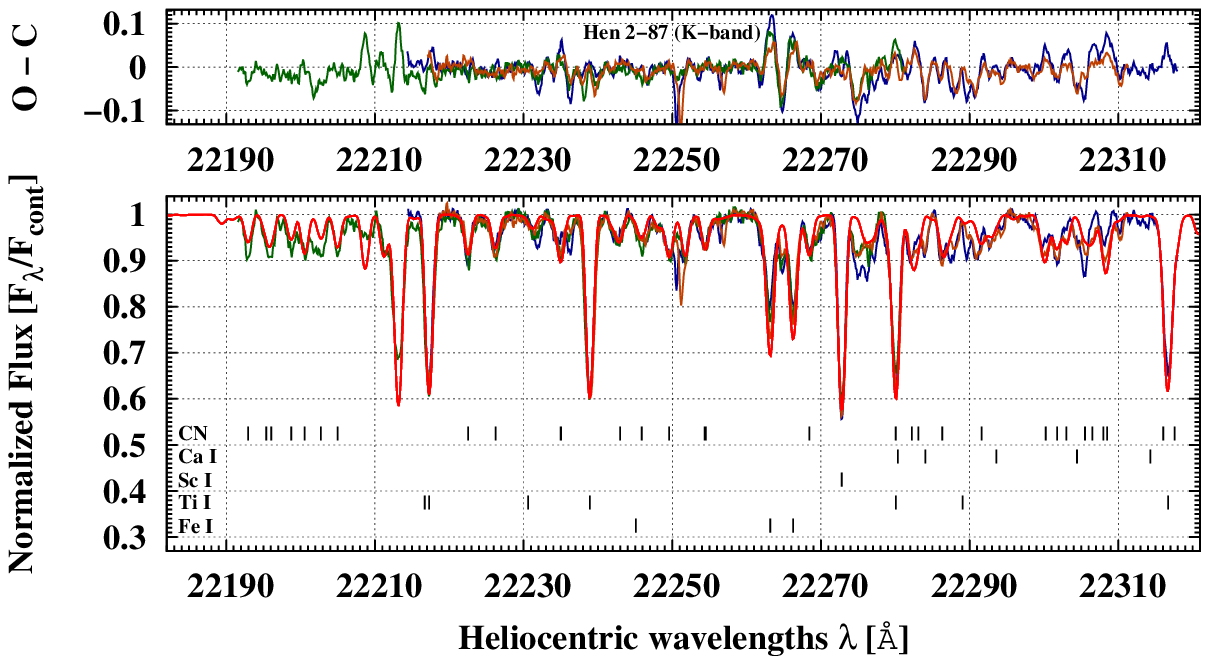} 
  \caption{$K$-band spectra of Hen\,2-87 observed 2003 April (blue line),
2003 December (green line), 2004 April (dark-orange line), and a synthetic
spectrum (red line) calculated using the final abundances
(Table\,\ref{T4}).}
  \label{FB16}
\end{figure}

\begin{figure}
  \includegraphics[width=84mm]{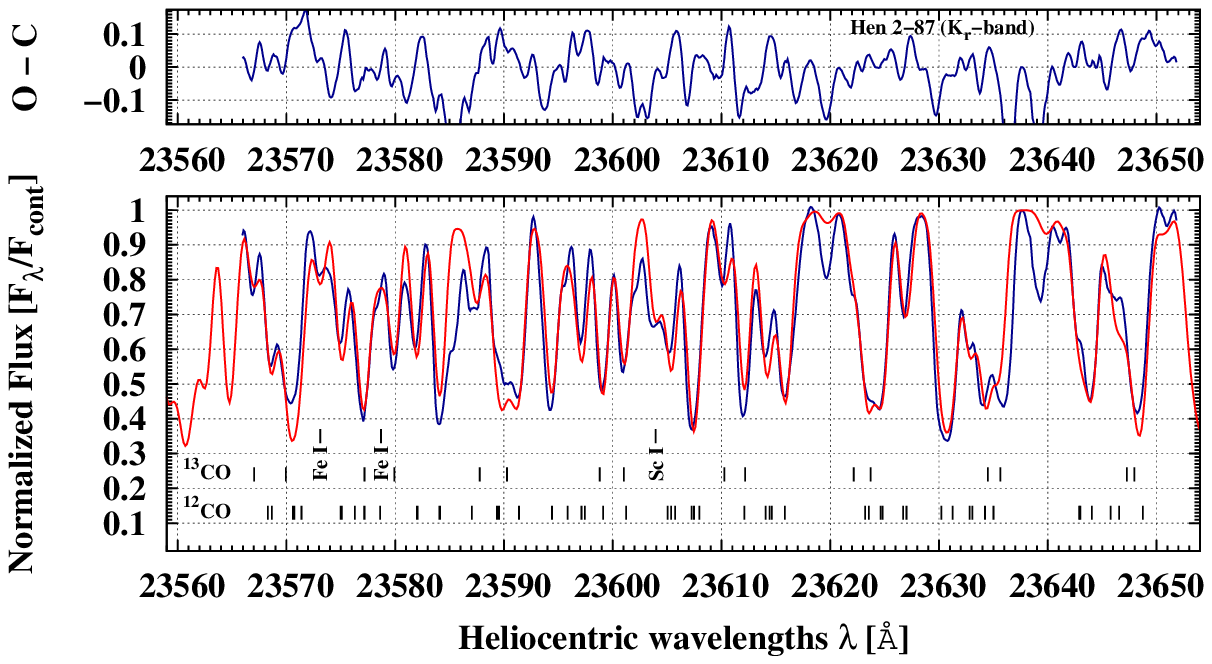} 
  \caption{The $K_{\rm r}$-band spectrum of Hen\,2-87 observed 2006 April
(blue line) and a synthetic spectrum (red line) calculated using the final
abundances and $^{12}$C/$^{13}$C isotopic ratio (Table\,\ref{T4}).}
  \label{FB17}
\end{figure}

\begin{figure}
  \includegraphics[width=84mm]{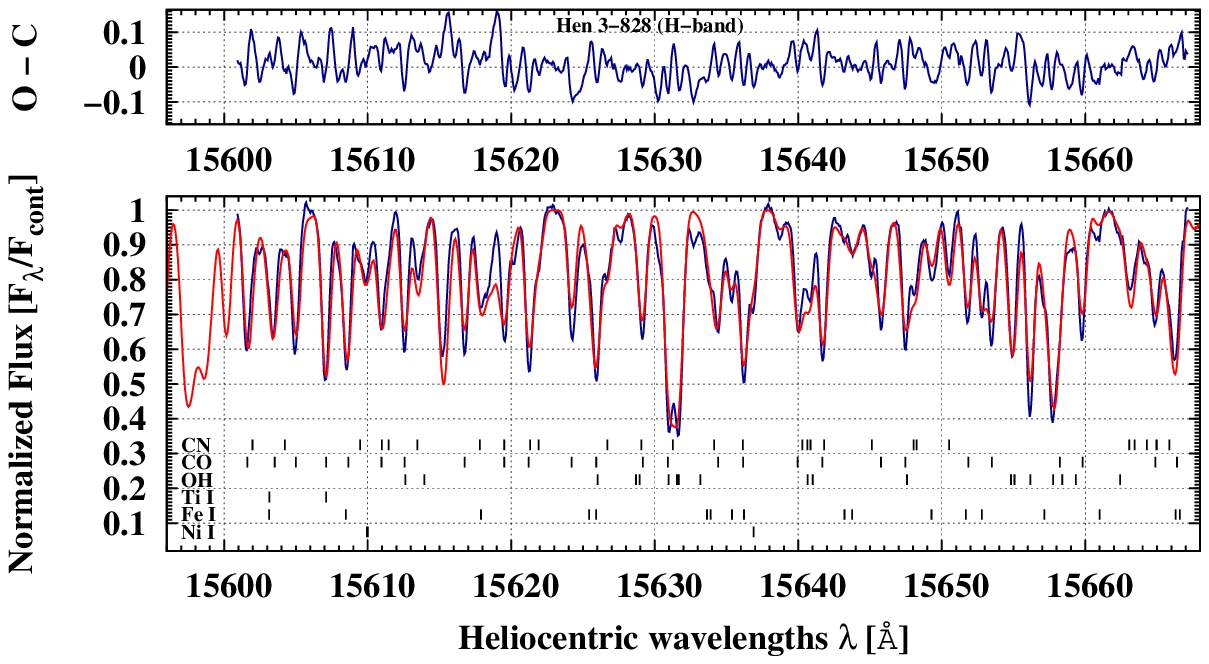}
  \caption{The $H$-band spectrum of Hen\,3-828 observed 2003 February  (blue
line) and a synthetic spectrum (red line) calculated using the final
abundances and $^{12}$C/$^{13}$C isotopic ratio (Table\,\ref{T4}).}
  \label{FB18}
\end{figure}


\begin{figure}
  \includegraphics[width=84mm]{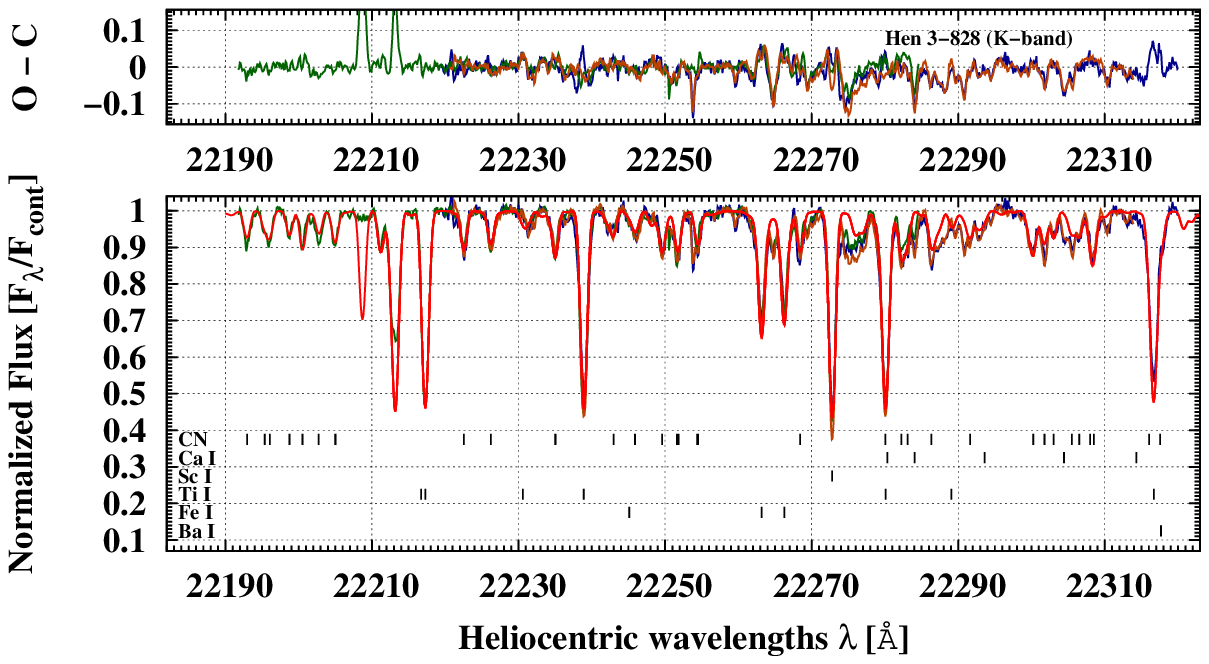} 
  \caption{$K$-band spectra of Hen\,3-828 observed 2003 April (blue line),
2003 December (green line), 2004 April (dark-orange line), and a synthetic
spectrum (red line) calculated using the final abundances
(Table\,\ref{T4}).}
  \label{FB19}
\end{figure}

\begin{figure}
  \includegraphics[width=84mm]{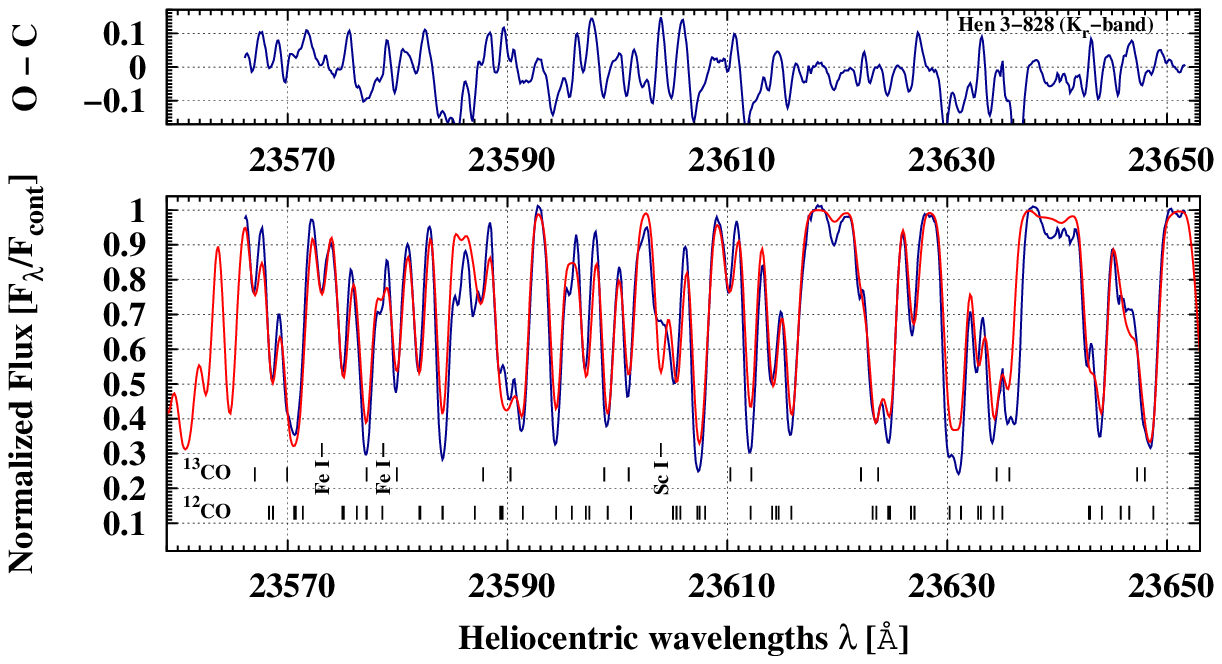} 
  \caption{The $K_{\rm r}$-band spectrum of Hen\,3-828 observed 2006 April
(blue line) and a synthetic spectrum (red line) calculated using the final
abundances and $^{12}$C/$^{13}$C isotopic ratio (Table\,\ref{T4}).}
  \label{FB20}
\end{figure}

\begin{figure}
  \includegraphics[width=84mm]{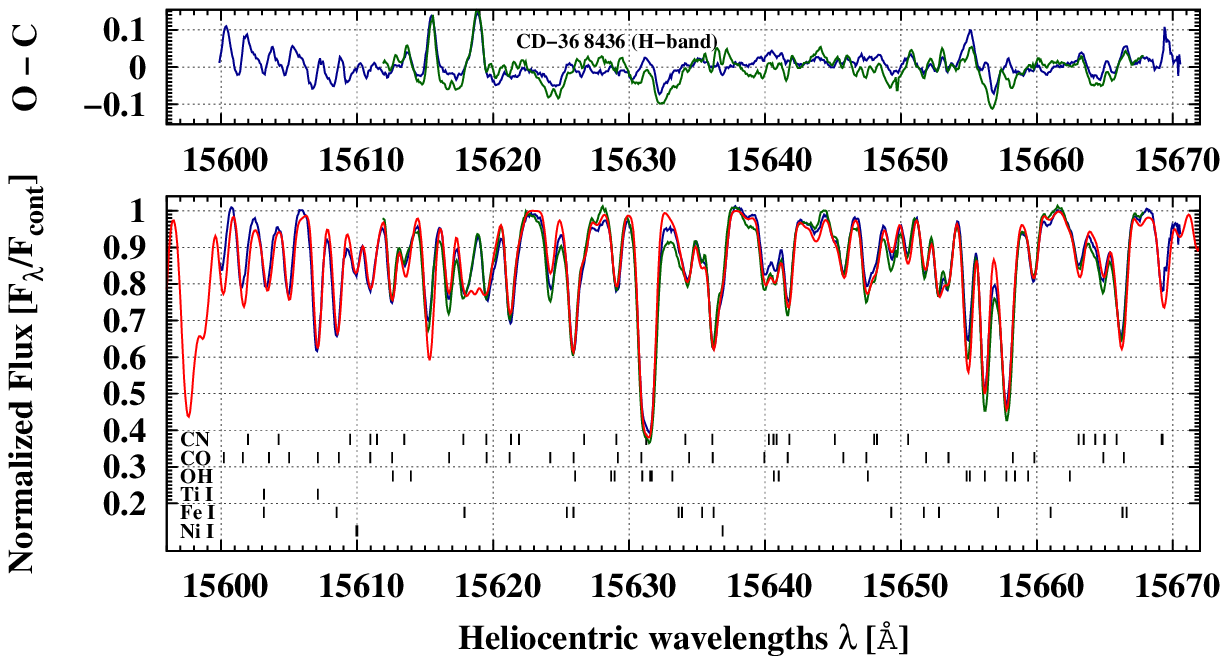}
  \caption{$H$-band spectra of CD-36$^{\circ}$8436 observed 2003 February
(blue line), 2010 April (green line), and a synthetic spectrum (red line)
calculated using the final abundances and $^{12}$C/$^{13}$C isotopic ratio
(Table\,\ref{T4}).}
  \label{FB21}
\end{figure}

\begin{figure}
  \includegraphics[width=84mm]{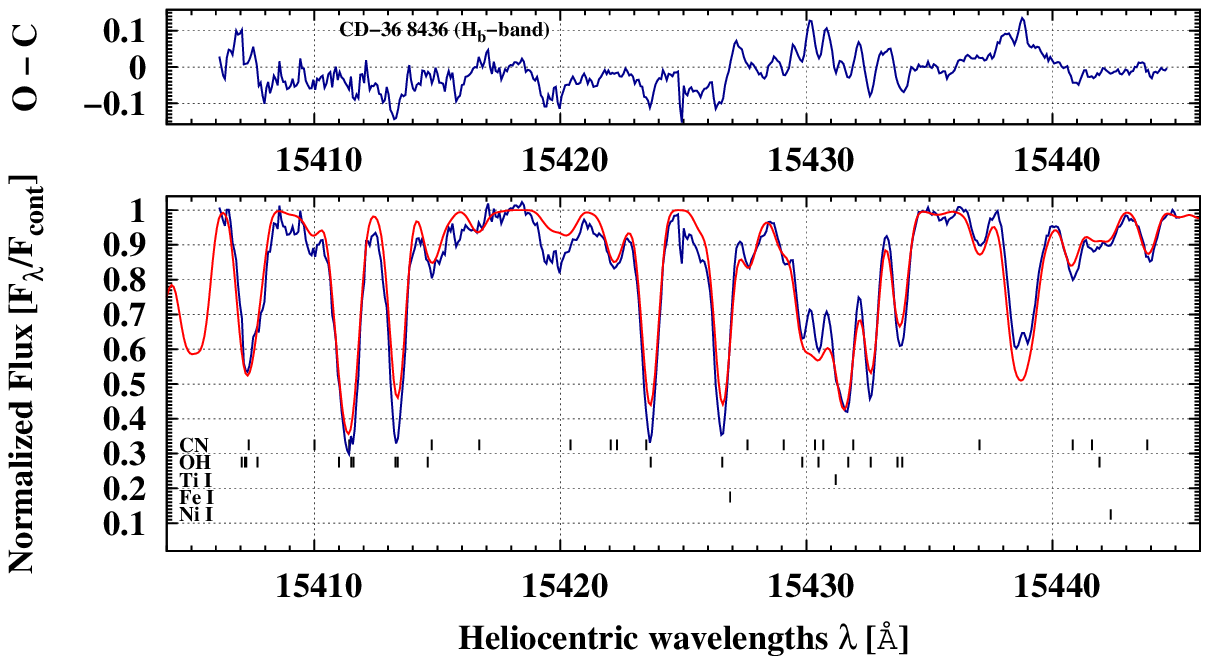}
  \caption{The $H_{\rm b}$-band spectrum of CD-36$^{\circ}$8436 observed
2010 March (blue line) and a synthetic spectrum (red line) calculated using
the final abundances and $^{12}$C/$^{13}$C isotopic ratio
(Table\,\ref{T4}).}
  \label{FB22}
\end{figure}

\begin{figure}
  \includegraphics[width=84mm]{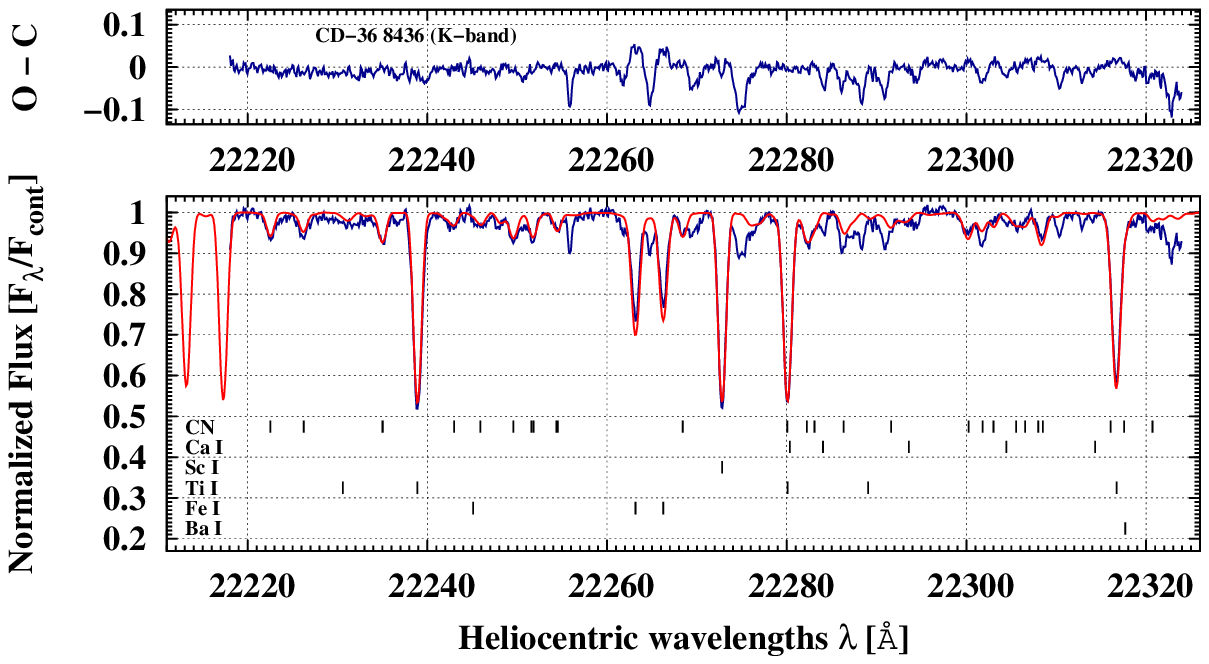} 
  \caption{The $K$-band spectrum of CD-36$^{\circ}$8436 observed 2003 April
(blue line) and a synthetic spectrum (red line) calculated using the final
abundances and $^{12}$C/$^{13}$C isotopic ratio (Table\,\ref{T4}).}
  \label{FB23}
\end{figure}

\begin{figure}
  \includegraphics[width=84mm]{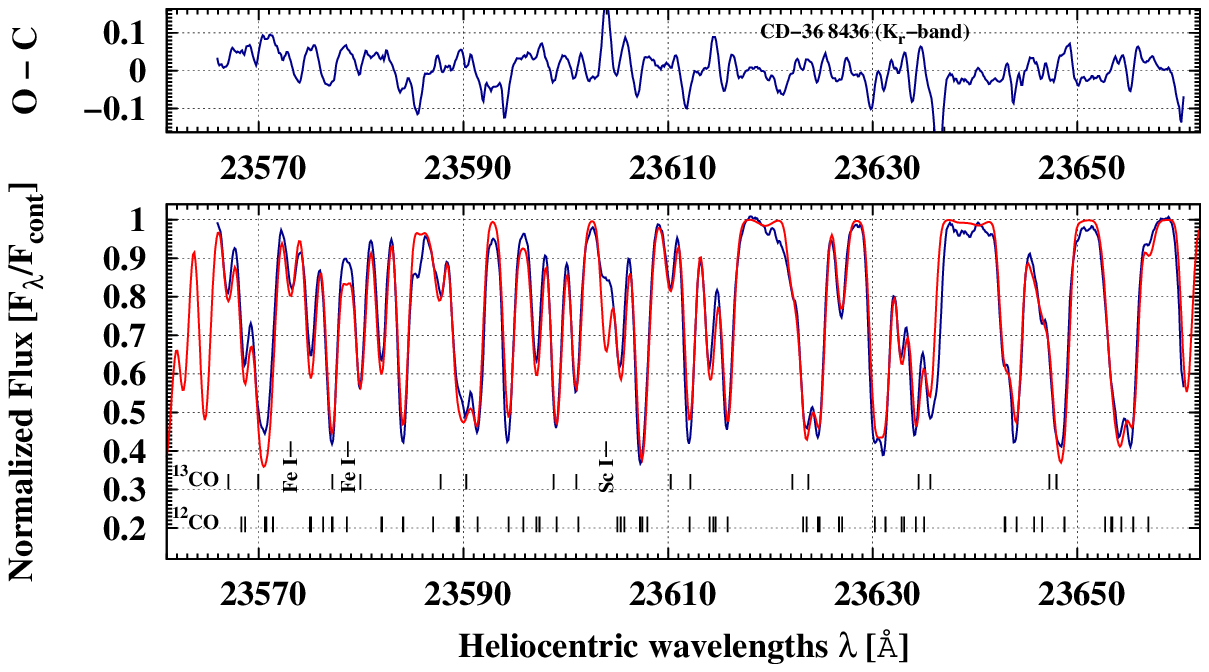} 
  \caption{The $K_{\rm r}$-band spectrum of CD-36$^{\circ}$8436 observed
2006 April (blue line) and a synthetic spectrum (red line) calculated using
the final abundances and $^{12}$C/$^{13}$C isotopic ratio
(Table\,\ref{T4}).}
  \label{FB24}
\end{figure}

\begin{figure}
  \includegraphics[width=84mm]{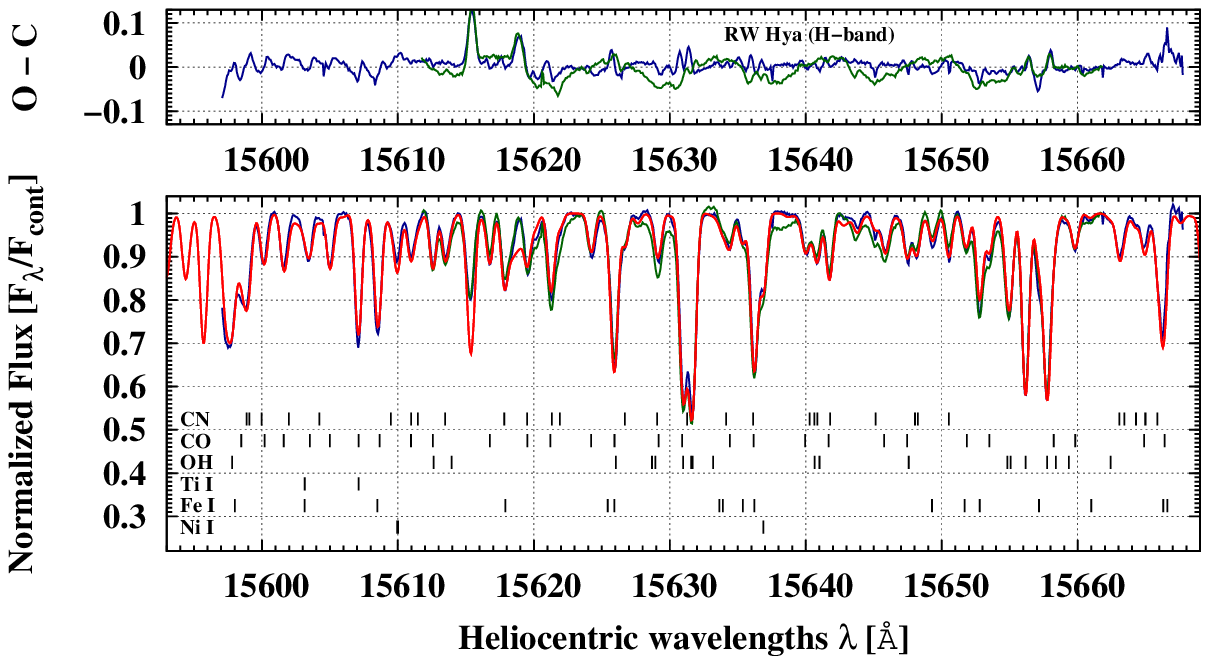}
  \caption{$H$-band spectra of RW\,Hya observed 2003 February  (blue line),
2010 April (green line), and a synthetic spectrum (red line) calculated
using the final abundances and $^{12}$C/$^{13}$C isotopic ratio
(Table\,\ref{T4}).}
  \label{FB25}
\end{figure}

\begin{figure}
  \includegraphics[width=84mm]{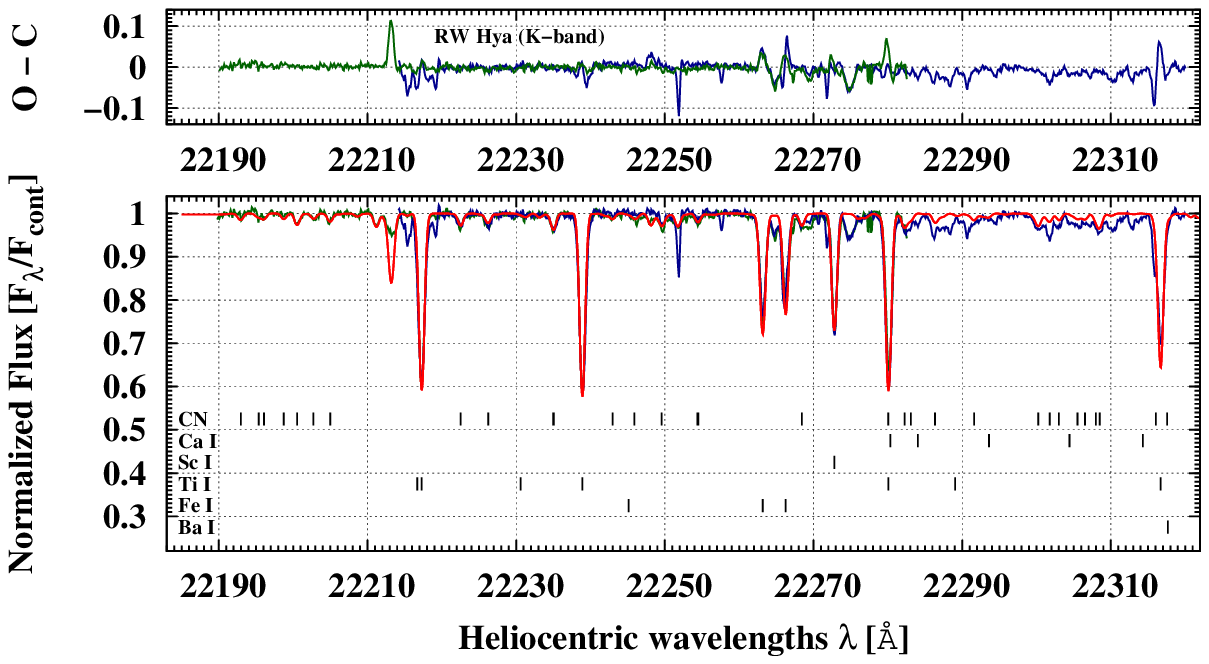} 
  \caption{$K$-band spectra of RW\,Hya observed 2003 April (blue line), 2003
December (green line), and a synthetic spectrum (red line) calculated using
the final abundances (Table\,\ref{T4}).}
  \label{FB26}
\end{figure}

\begin{figure}
  \includegraphics[width=84mm]{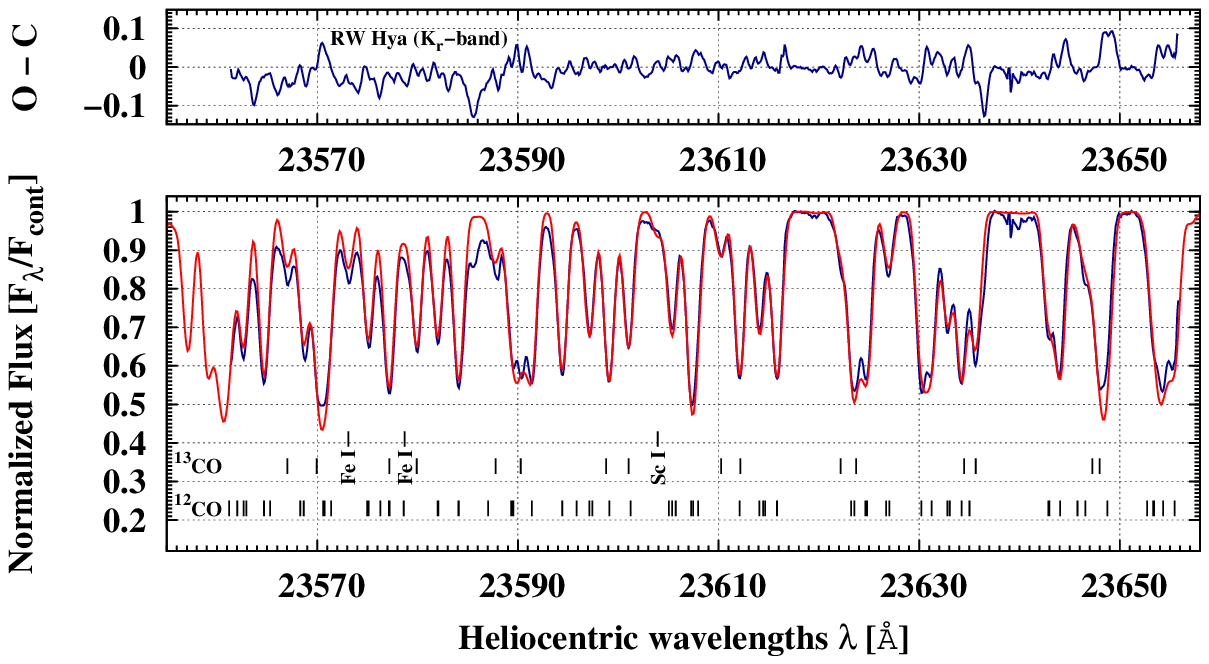} 
  \caption{The $K_{\rm r}$-band spectrum of RW\,Hya observed 2006 April
(blue line) and a synthetic spectrum (red line) calculated using the final
abundances and $^{12}$C/$^{13}$C isotopic ratio (Table\,\ref{T4}).}
  \label{FB27}
\end{figure}

\begin{figure}
  \includegraphics[width=84mm]{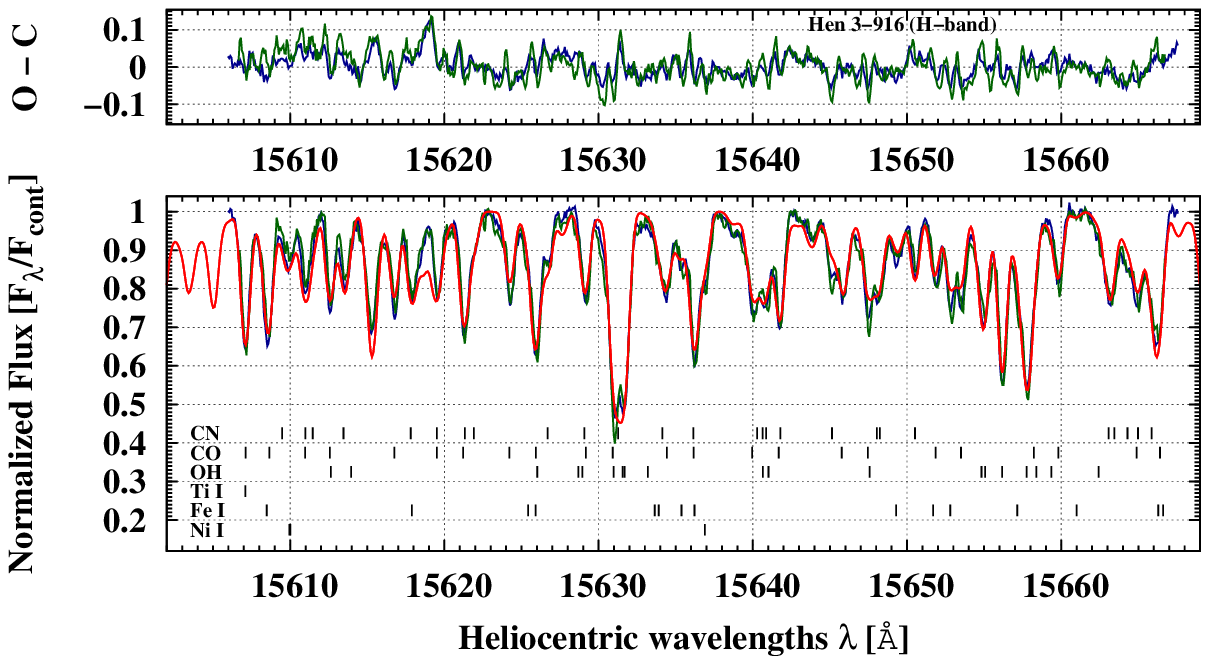}
  \caption{$H$-band spectra of Hen\,3-916 observed 2010 April  (blue line),
2010 May (green line), and a synthetic spectrum (red line) calculated using
the final abundances and $^{12}$C/$^{13}$C isotopic ratio
(Table\,\ref{T4}).}
  \label{FB28}
\end{figure}

\begin{figure}
  \includegraphics[width=84mm]{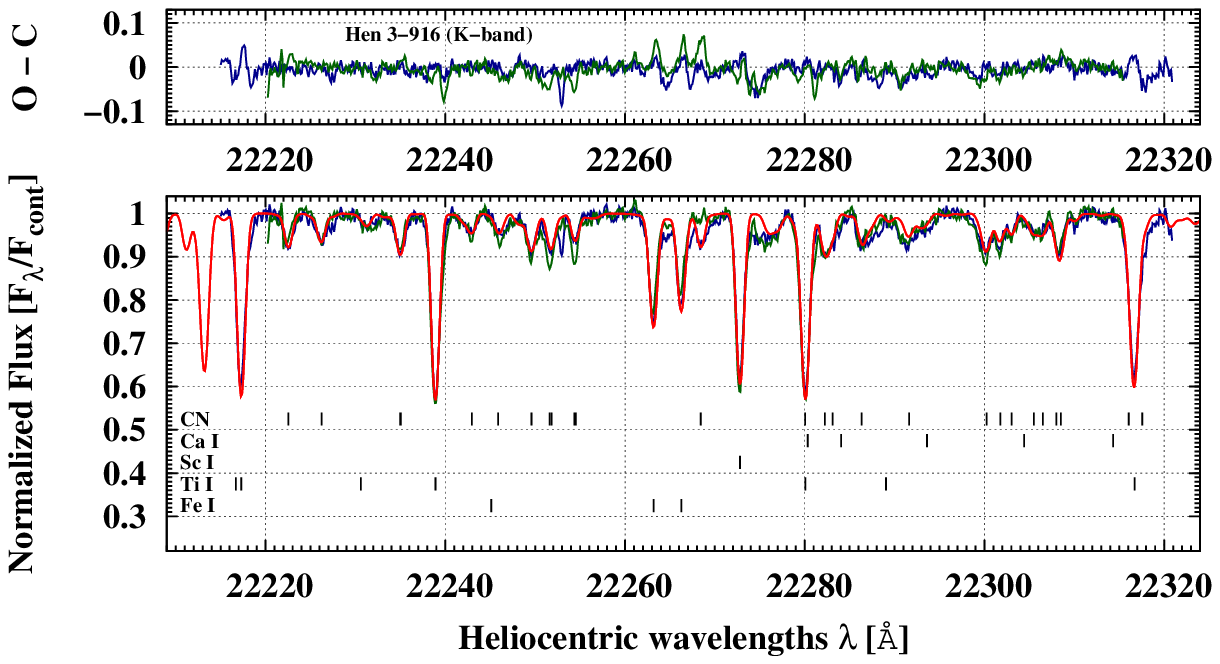} 
  \caption{$K$-band spectra of Hen\,3-916 observed 2003 April (blue line),
2004 April (green line), and a synthetic spectrum (red line) calculated
using the final abundances (Table\,\ref{T4}).}
  \label{FB29}
\end{figure}

\begin{figure}
  \includegraphics[width=84mm]{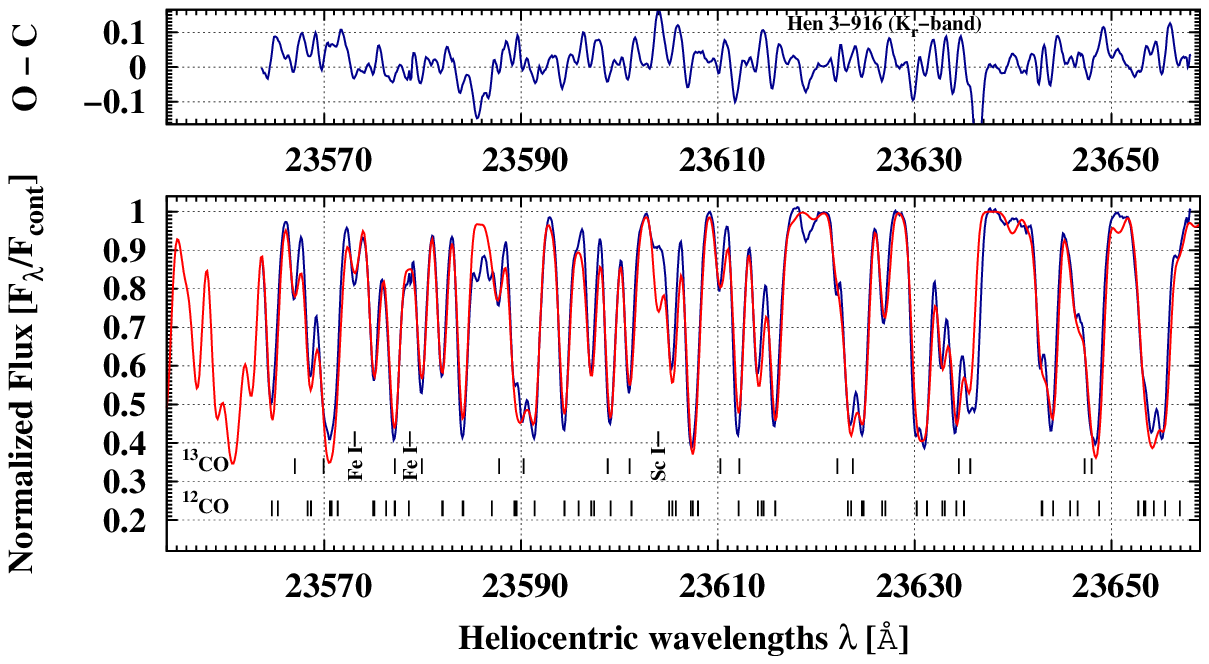} 
  \caption{The $K_{\rm r}$-band spectrum of Hen\,3-916 observed 2006 April
(blue line) and a synthetic spectrum (red line) calculated using the final
abundances and $^{12}$C/$^{13}$C isotopic ratio (Table\,\ref{T4}).}
  \label{FB30}
\end{figure}

\begin{figure}
  \includegraphics[width=84mm]{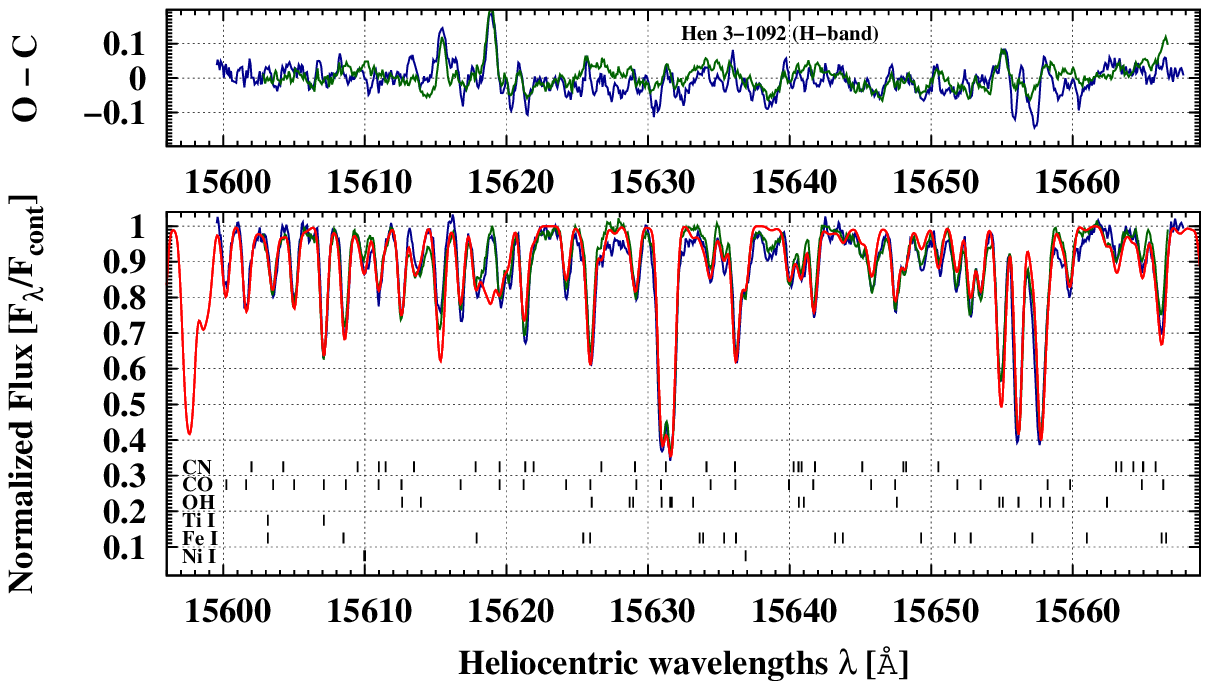}
  \caption{$H$-band spectra of Hen\,3-1092 observed 2003 February  (blue
line), 2010 April (green line), and a synthetic spectrum (red line)
calculated using the final abundances (Table\,\ref{T4}).}
  \label{FB31}
\end{figure}

\begin{figure}
  \includegraphics[width=84mm]{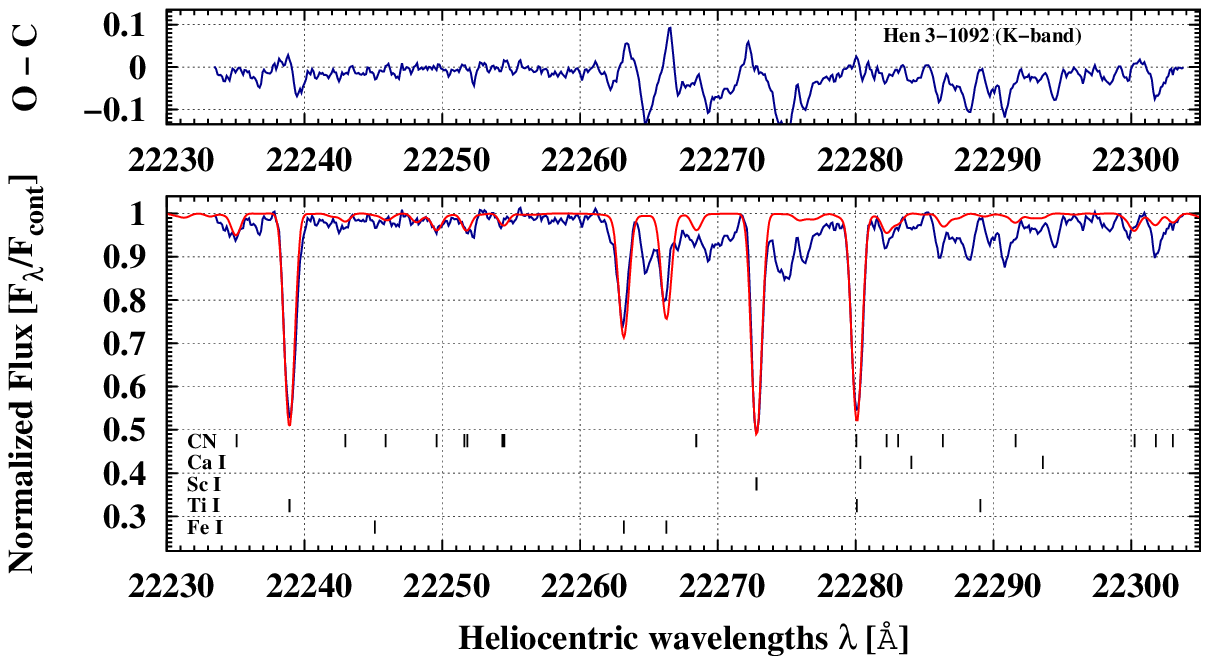} 
  \caption{The $K$-band spectrum of Hen\,3-1092 observed 2004 April (blue
line) and a synthetic spectrum (red line) calculated using the final
abundances (Table\,\ref{T4}).}
  \label{FB32}
\end{figure}

\begin{figure}
  \includegraphics[width=84mm]{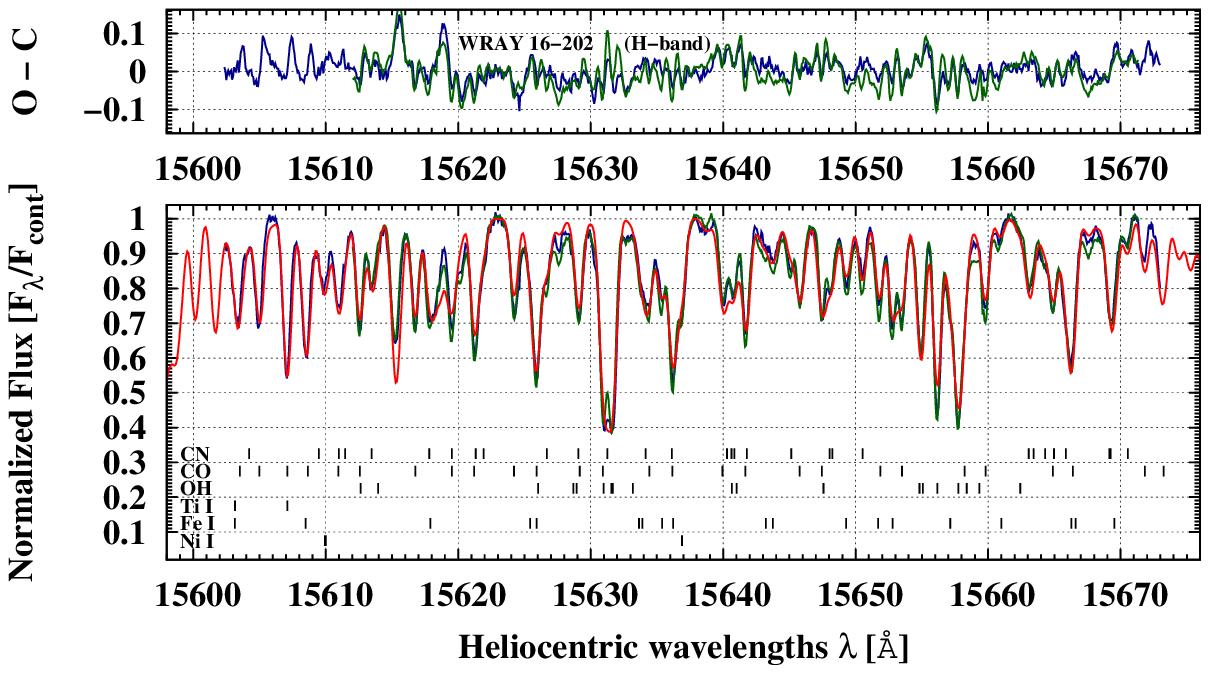}
  \caption{$H$-band spectra of WRAY\,16-202 observed 2003 February (blue
line), 2010 April (green line), and a synthetic spectrum (red line)
calculated using the final abundances and $^{12}$C/$^{13}$C isotopic ratio
(Table\,\ref{T4}).}
  \label{FB33}
\end{figure}

\clearpage

\begin{figure}
  \includegraphics[width=84mm]{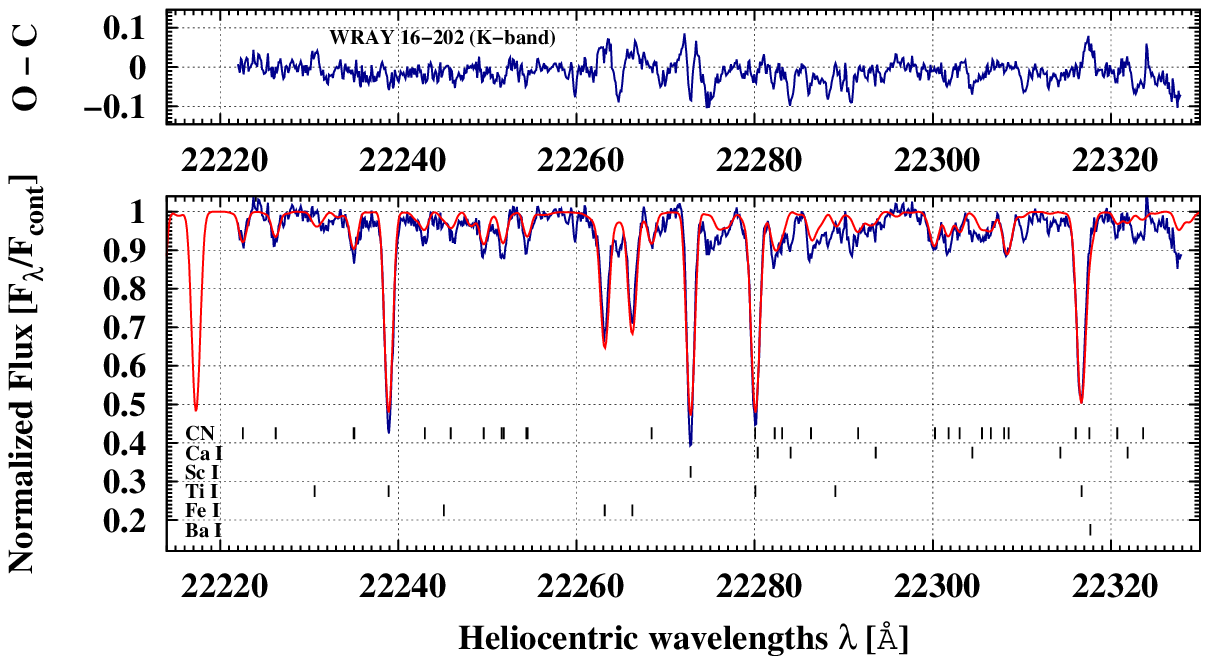} 
  \caption{The $K$-band spectrum of WRAY\,16-202 observed 2003 April (blue
line) and a synthetic spectrum (red line) calculated using the final
abundances and $^{12}$C/$^{13}$C isotopic ratio (Table\,\ref{T4}).}
  \label{FB34}
\end{figure}

\begin{figure}
  \includegraphics[width=84mm]{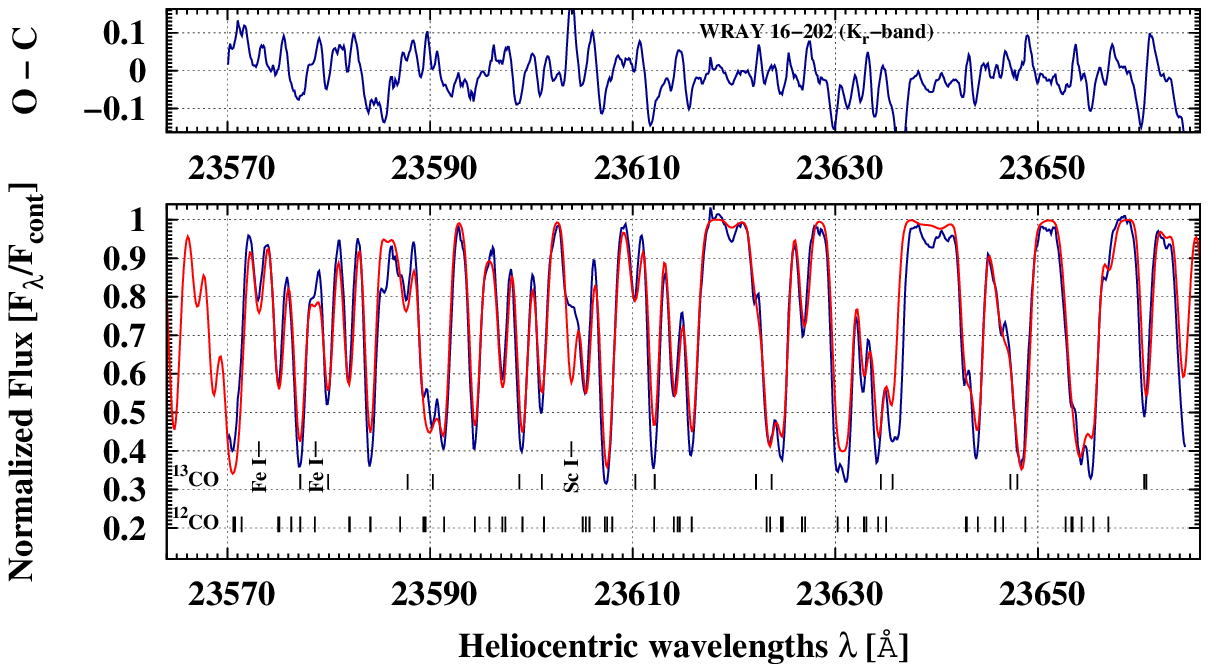} 
  \caption{The $K_{\rm r}$-band spectrum of WRAY\,16-202 observed 2006 April
(blue line) and a synthetic spectrum (red line) calculated using the final
abundances and $^{12}$C/$^{13}$C isotopic ratio (Table\,\ref{T4}).}
  \label{FB35}
\end{figure}

\begin{figure}
  \includegraphics[width=84mm]{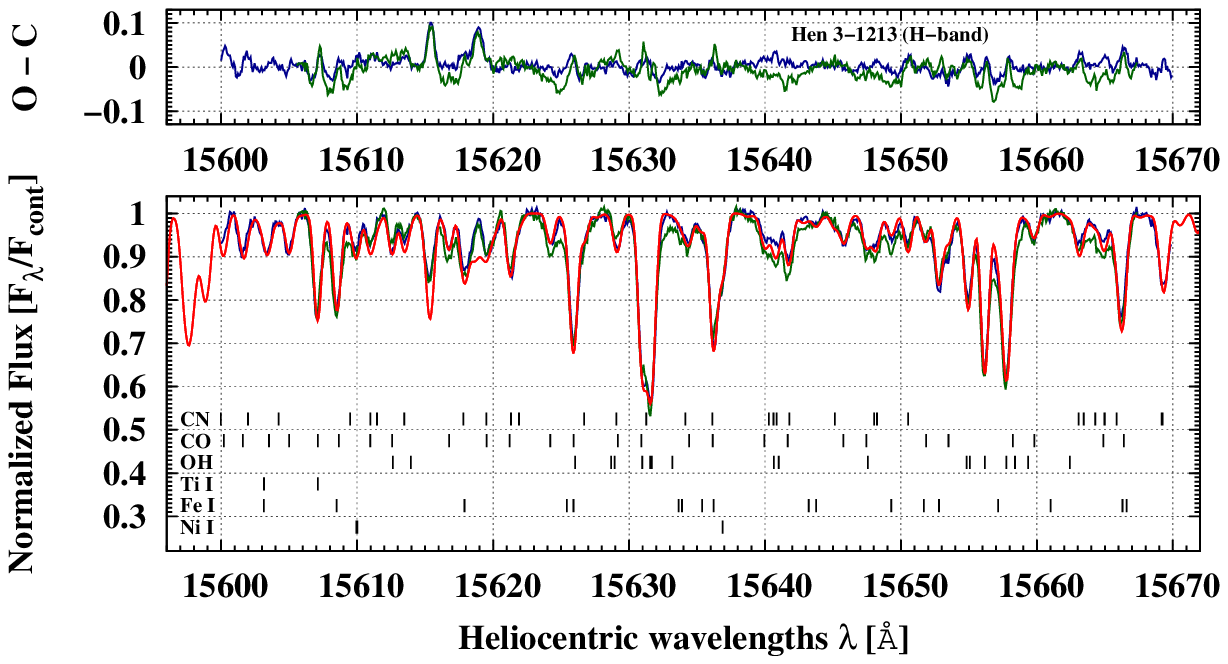}
  \caption{$H$-band spectra of Hen\,3-1213 observed 2003 February (blue
line), 2010 May (green line), and a synthetic spectrum (red line) calculated
using the final abundances (Table\,\ref{T4}).}
  \label{FB36}
\end{figure}

\begin{figure}
  \includegraphics[width=84mm]{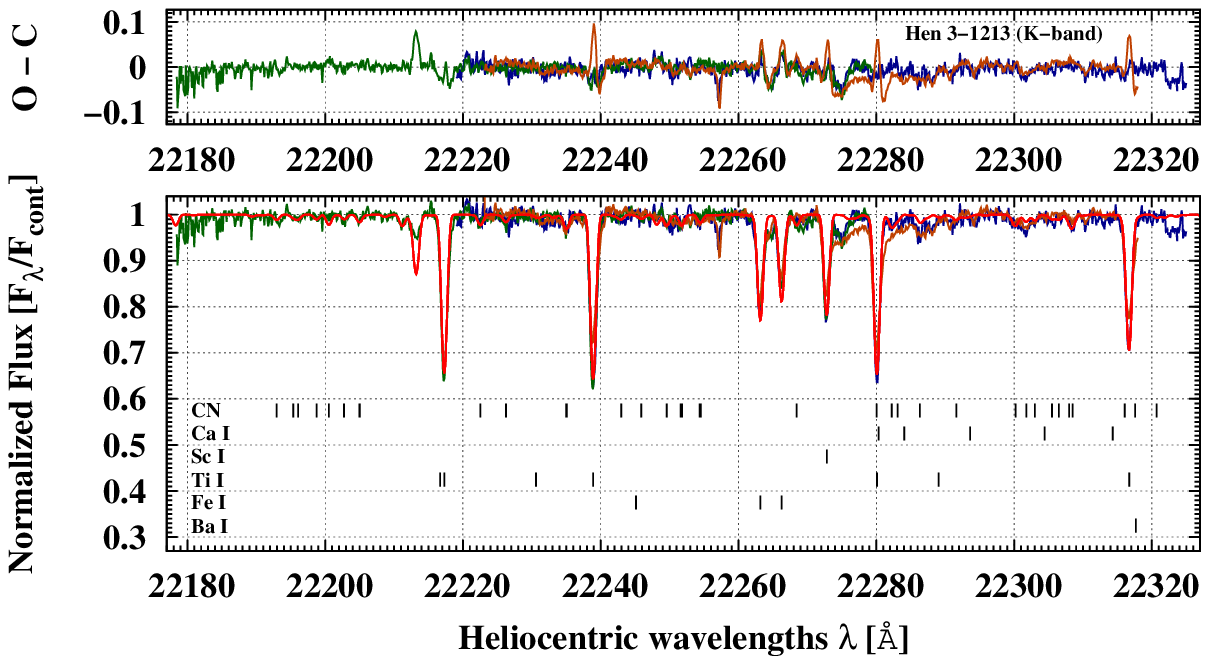} 
  \caption{$K$-band spectra of Hen\,3-1213 observed 2003 April (blue line),
2003 August (green line), 2004 April (dark-orange line), and a synthetic
spectrum (red line) calculated using the final abundances
(Table\,\ref{T4}).}
  \label{FB37}
\end{figure}

\begin{figure}
  \includegraphics[width=84mm]{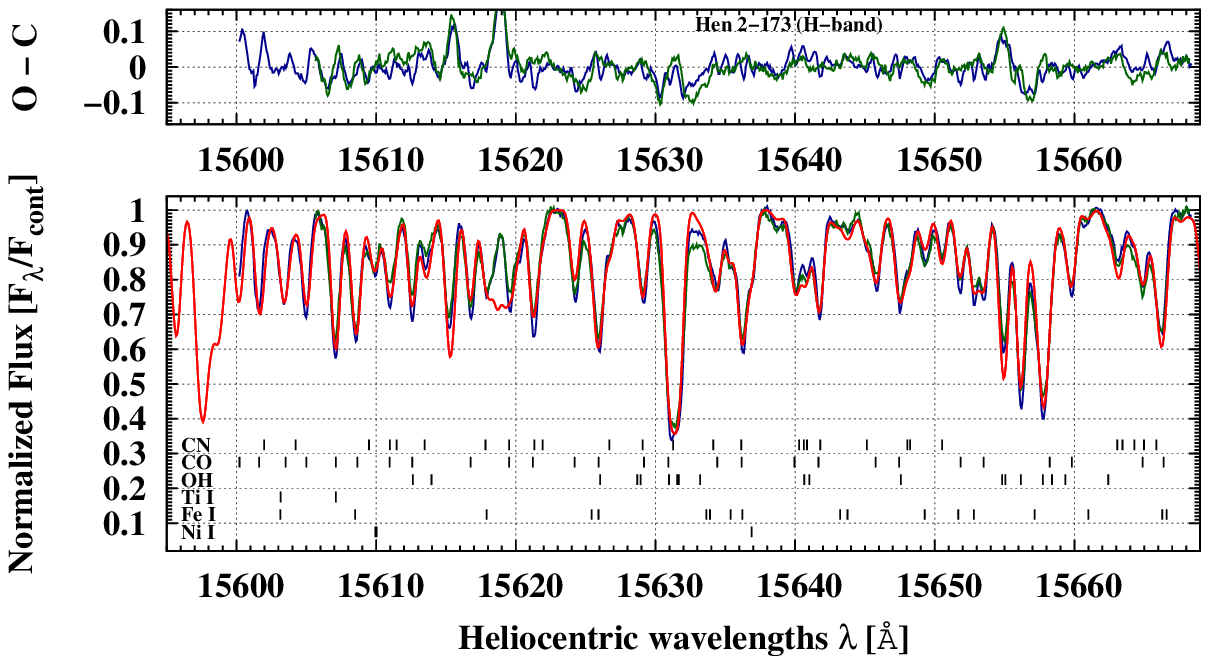}
  \caption{$H$-band spectra of Hen\,2-173 observed 2003 February (blue
line), 2010 May (green line), and a synthetic spectrum (red line) calculated
using the final abundances (Table\,\ref{T4}).}
  \label{FB38}
\end{figure}

\begin{figure}
  \includegraphics[width=84mm]{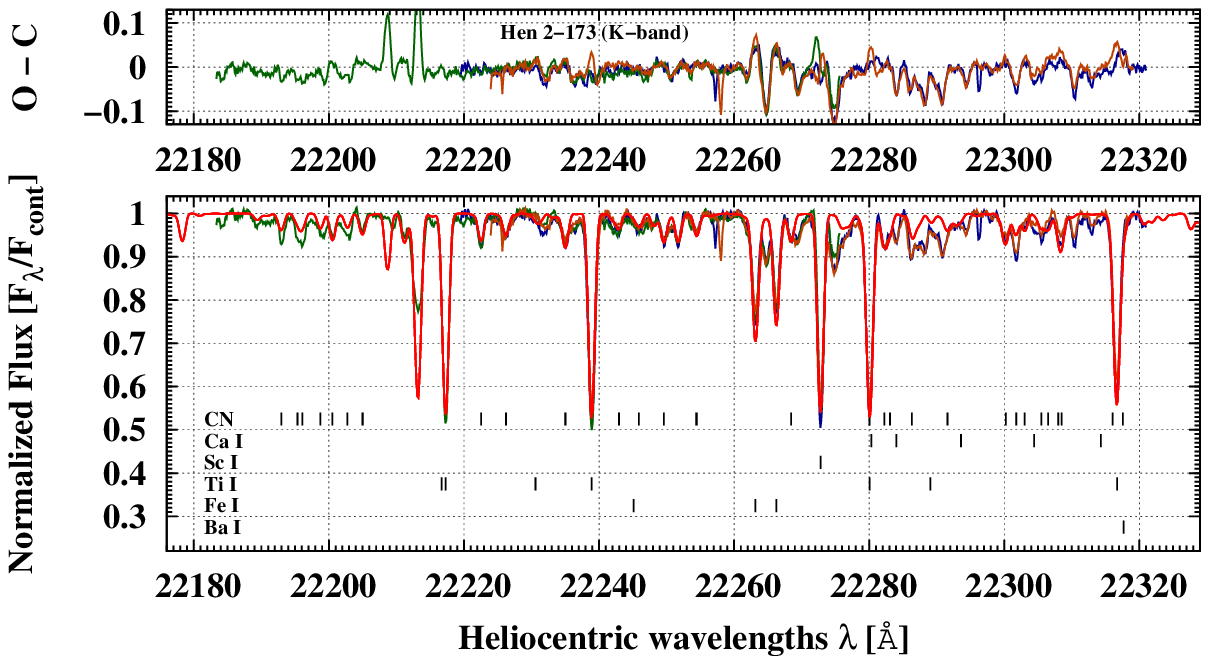} 
  \caption{$K$-band spectra of Hen\,2-173 observed 2003 April (blue line),
2003 August (green line), 2004 April (dark-orange line), and a synthetic
spectrum (red line) calculated using the final abundances
(Table\,\ref{T4}).}
  \label{FB39}
\end{figure}

\begin{figure}
  \includegraphics[width=84mm]{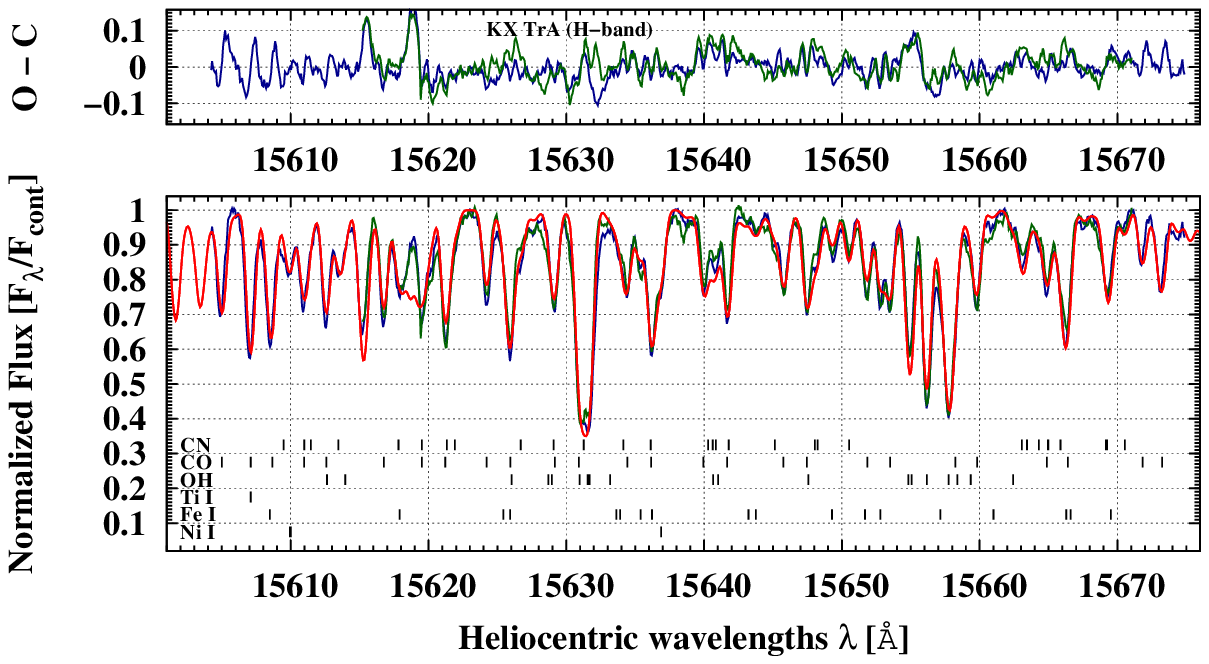}
  \caption{$H$-band spectra of KX\,TrA observed 2003 February  (blue line),
2010 May (green line), and a synthetic spectrum (red line) calculated using
the final abundances (Table\,\ref{T4}).}
  \label{FB40}
\end{figure}

\begin{figure}
  \includegraphics[width=84mm]{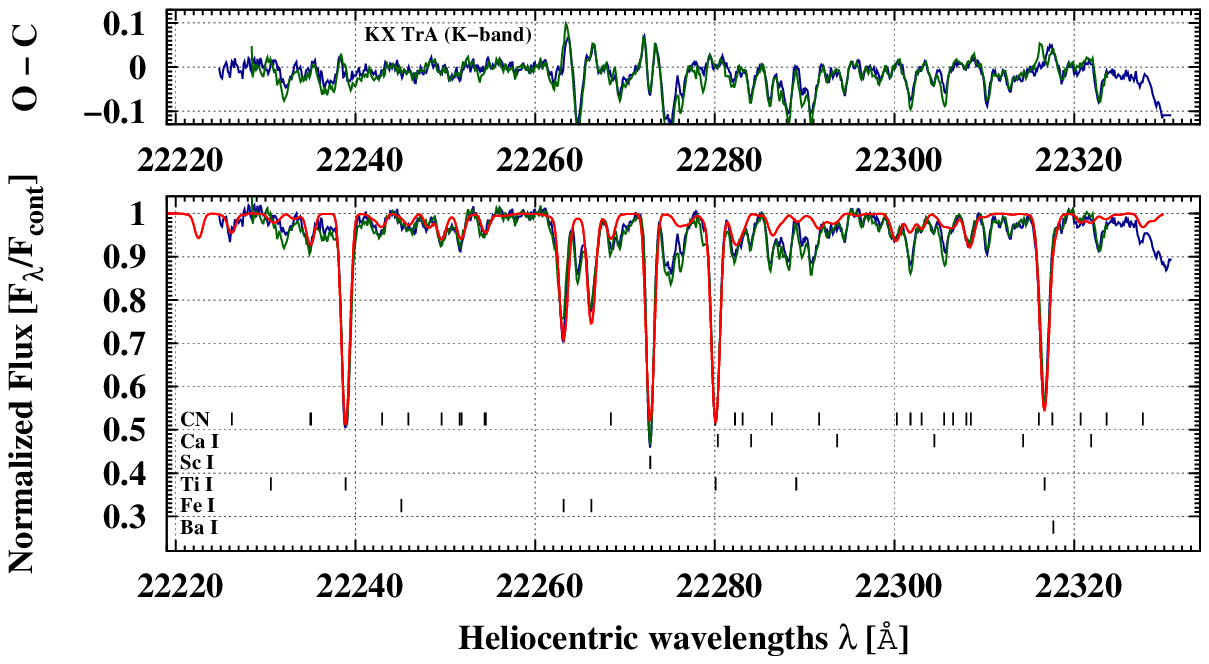} 
  \caption{$K$-band spectra of KX\,TrA observed 2003 April (blue line), 2004
April (green line), and a synthetic spectrum (red line) calculated using the
final abundances (Table\,\ref{T4}).}
  \label{FB41}
\end{figure}

\begin{figure}
  \includegraphics[width=84mm]{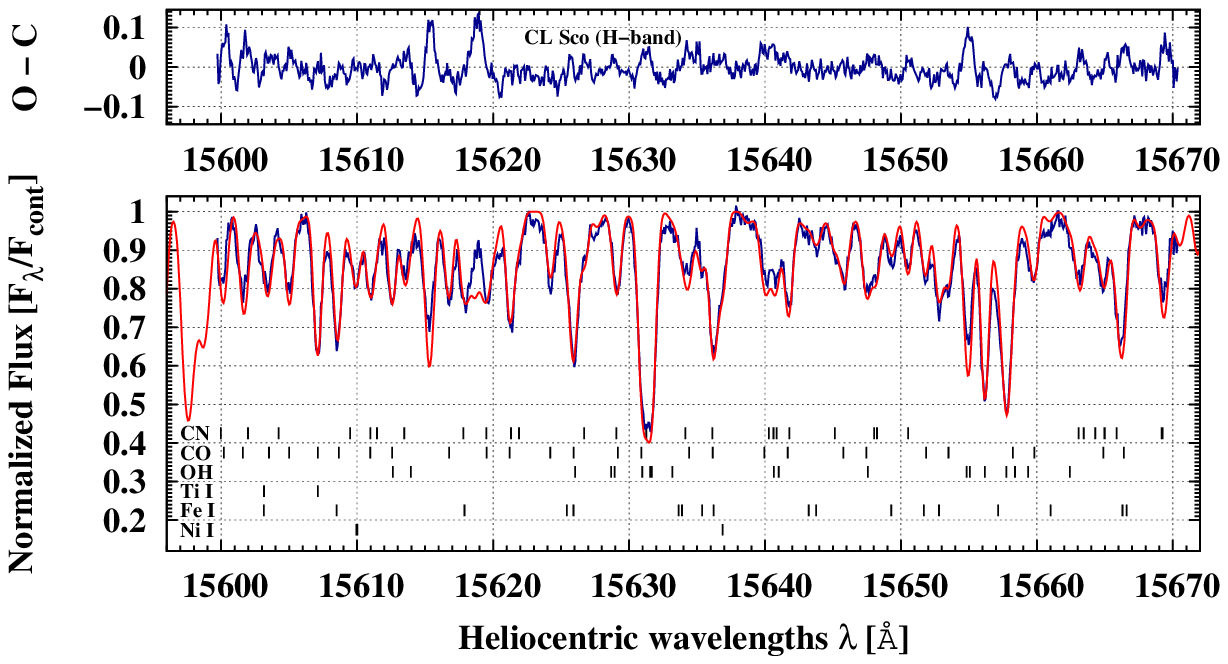}
  \caption{The $H$-band spectrum of CL\,Sco observed 2003 February  (blue
line) and a synthetic spectrum (red line) calculated using the final
abundances (Table\,\ref{T4}).}
  \label{FB42}
\end{figure}

\begin{figure}
  \includegraphics[width=84mm]{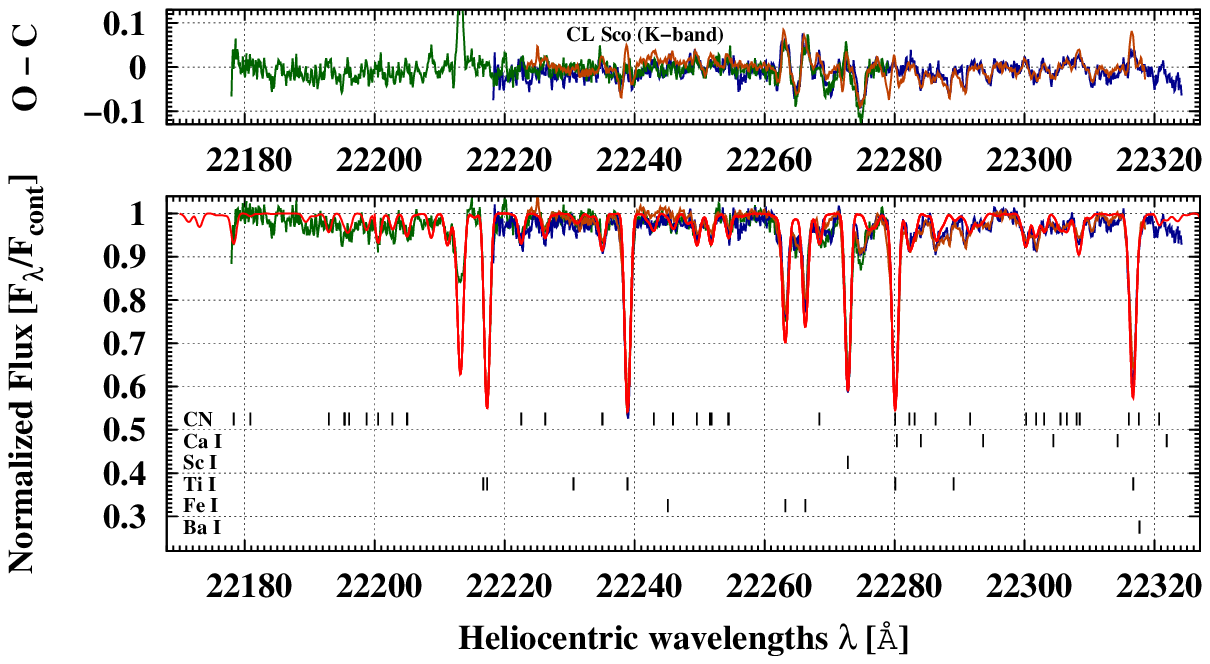} 
  \caption{$K$-band spectra of CL\,Sco observed 2003 April (blue line), 2003
August (green line), 2004 April (dark-orange line), and a synthetic spectrum
(red line) calculated using the final abundances (Table\,\ref{T4}).}
  \label{FB43}
\end{figure}

\begin{figure}
  \includegraphics[width=84mm]{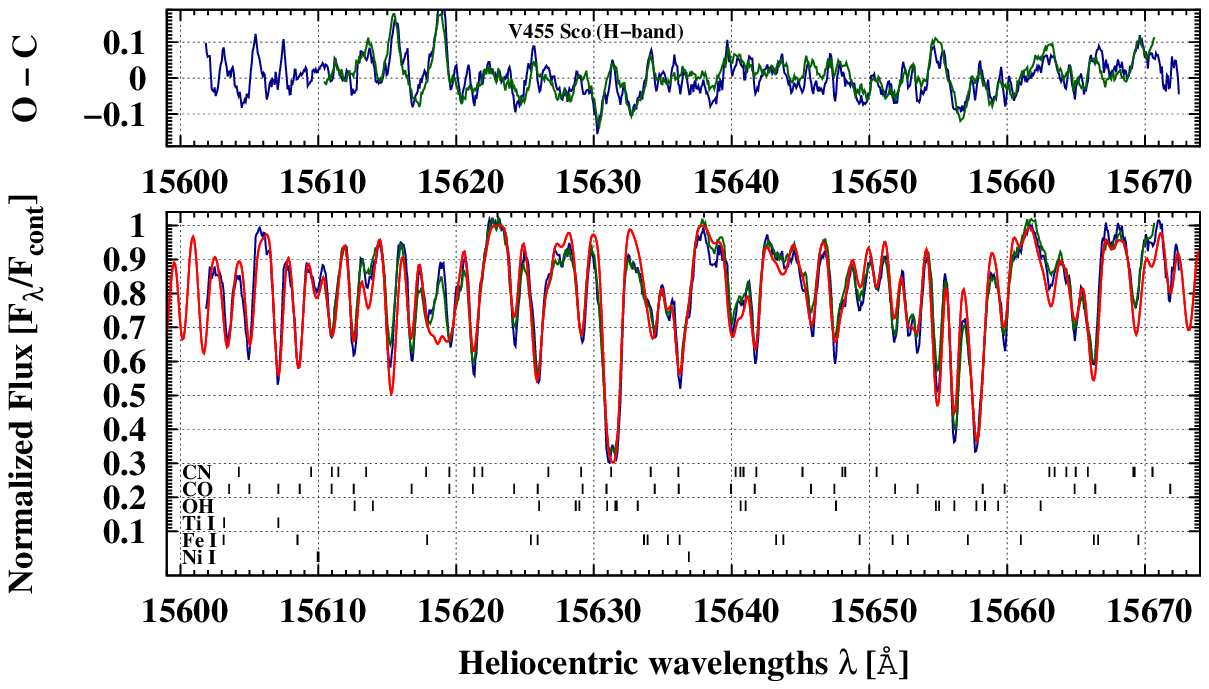}
  \caption{$H$-band spectra of V455\,Sco observed 2003 February (blue line),
2010 May (green line), and a synthetic spectrum (red line) calculated using
the final abundances (Table\,\ref{T4}).}
  \label{FB44}
\end{figure}

\begin{figure}
  \includegraphics[width=84mm]{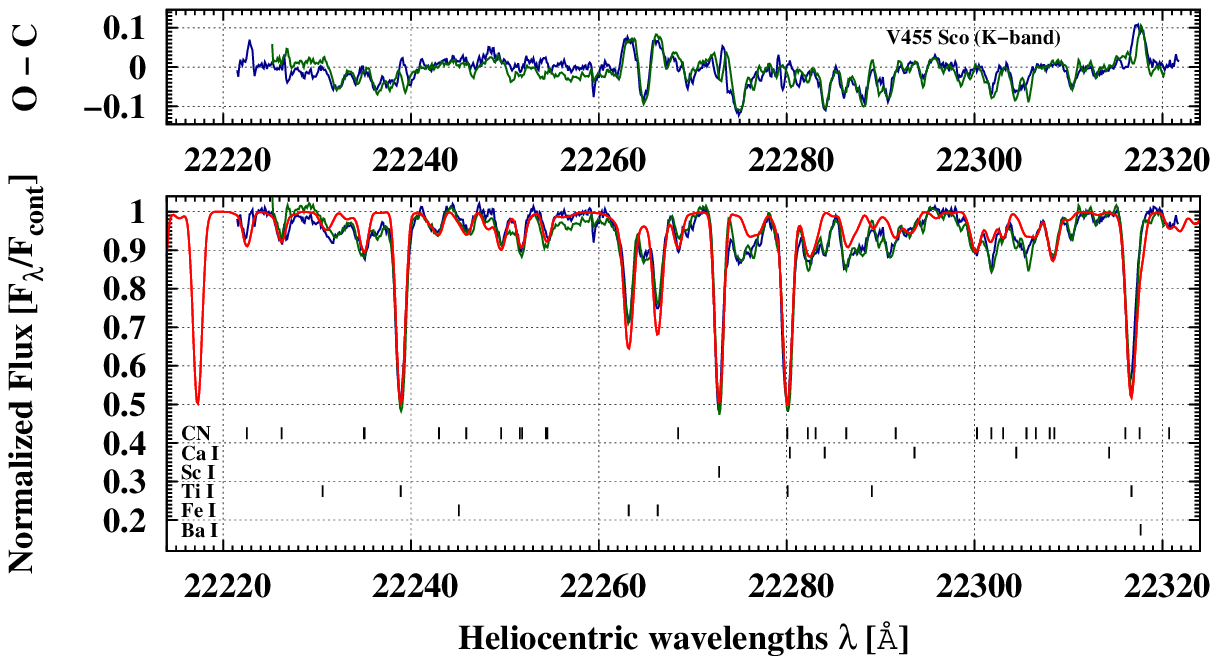} 
  \caption{$K$-band spectra of V455\,Sco observed 2003 April (blue line),
2004 April (green line), and a synthetic spectrum (red line) calculated
using the final abundances (Table\,\ref{T4}).}
  \label{FB45}
\end{figure}

\begin{figure}
  \includegraphics[width=84mm]{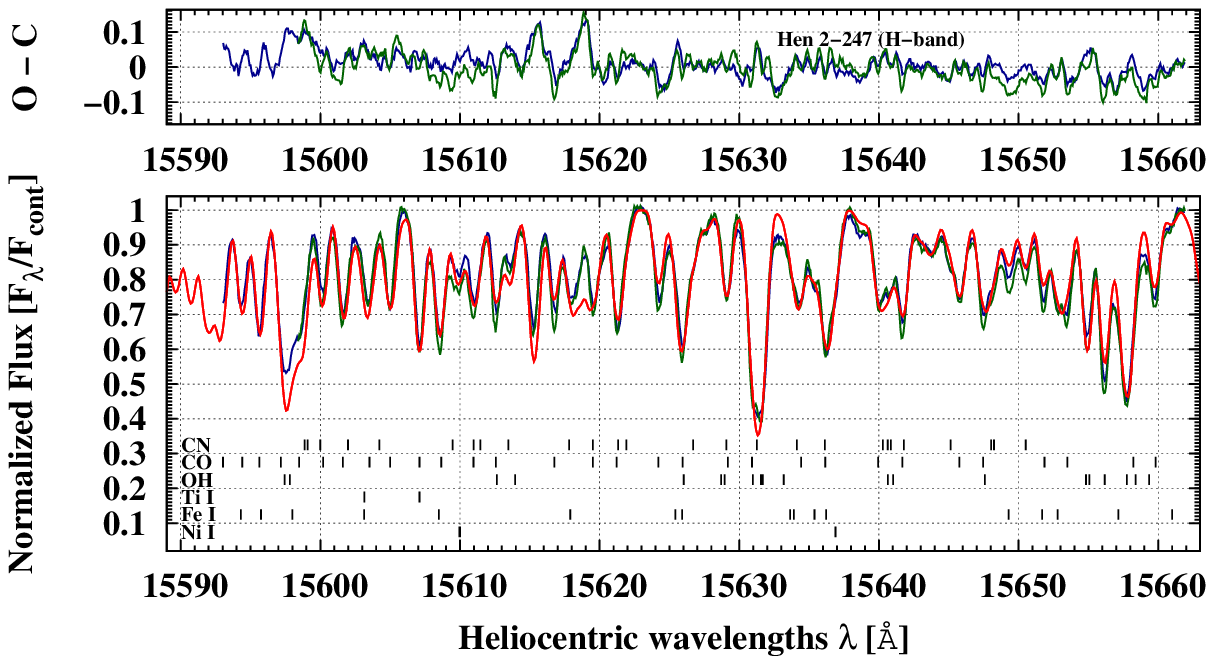}
  \caption{$H$-band spectra of Hen\,2-247 observed 2003 February (blue
line), 2010 June (green line), and a synthetic spectrum (red line)
calculated using the final abundances (Table\,\ref{T4}).}
  \label{FB46}
\end{figure}

\begin{figure}
  \includegraphics[width=84mm]{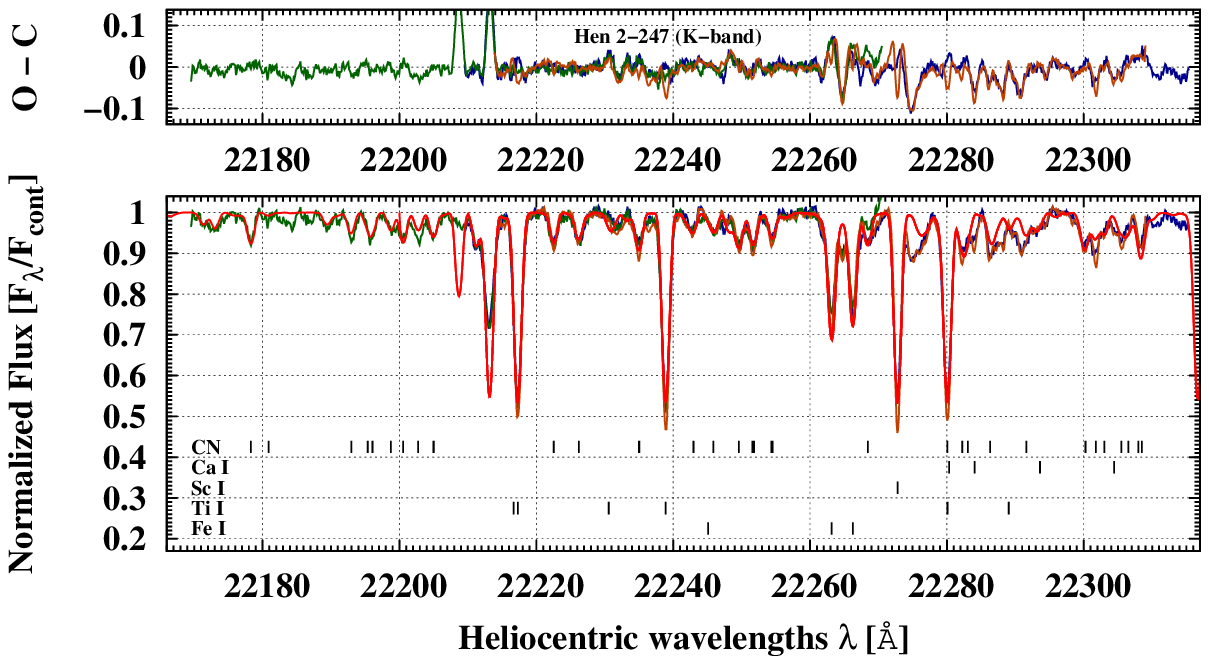} 
  \caption{$K$-band spectra of Hen\,2-247 observed 2003 April (blue line),
2003 August (green line), 2004 April (dark-orange line), and a synthetic
spectrum (red line) calculated using the final abundances
(Table\,\ref{T4}).}
  \label{FB47}
\end{figure}

\begin{figure}
  \includegraphics[width=84mm]{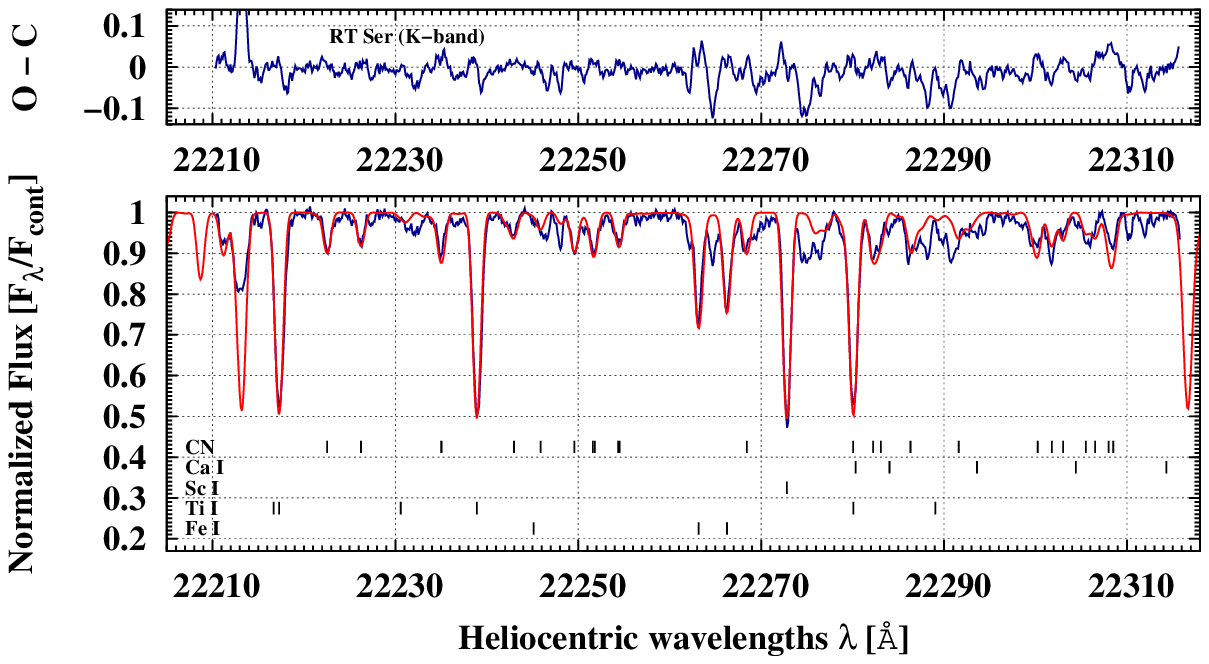} 
  \caption{The $K$-band spectrum of RT\,Ser observed 2003 April (blue line)
and a synthetic spectrum (red line) calculated using the final abundances
(Table\,\ref{T4}).}
  \label{FB48}
\end{figure}


\begin{figure}
  \includegraphics[width=84mm]{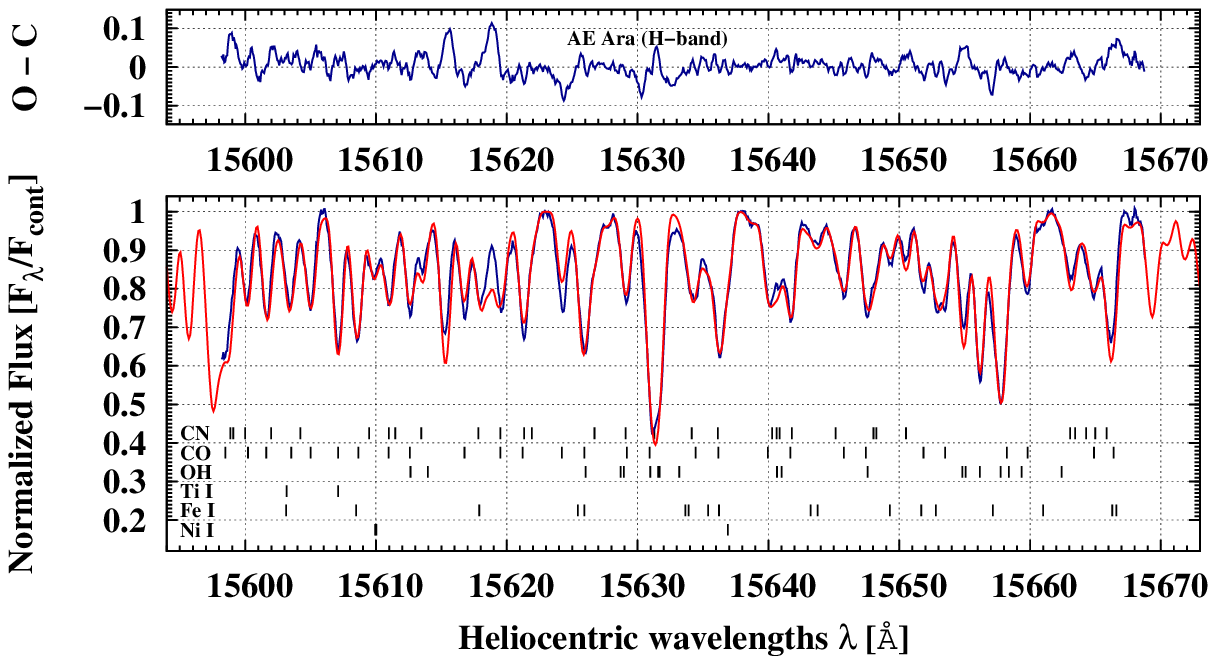}
  \caption{The $H$-band spectrum of AE\,Ara observed 2003 February (blue
line) and a synthetic spectrum (red line) calculated using the final
abundances (Table\,\ref{T4}).}
  \label{FB49}
\end{figure}

\begin{figure}
  \includegraphics[width=84mm]{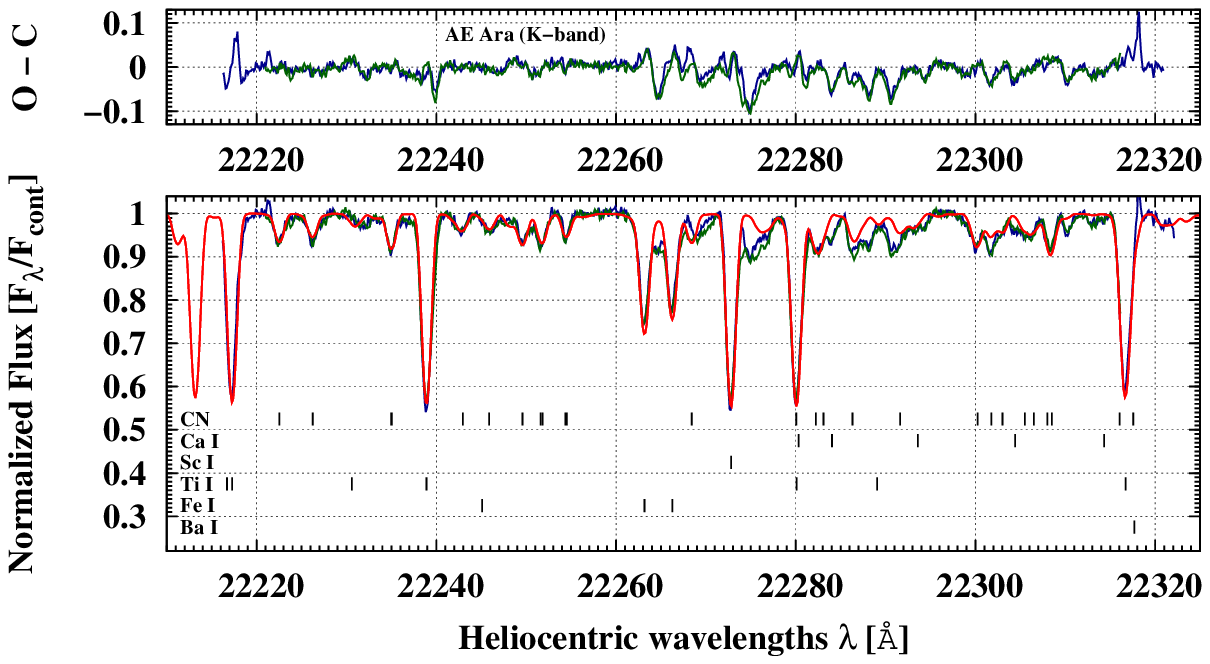} 
  \caption{$K$-band spectra of AE\,Ara observed 2003 April (blue line), 2004
April (green line), and a synthetic spectrum (red line) calculated using the
final abundances (Table\,\ref{T4}).}
  \label{FB50}
\end{figure}

\begin{figure}
  \includegraphics[width=84mm]{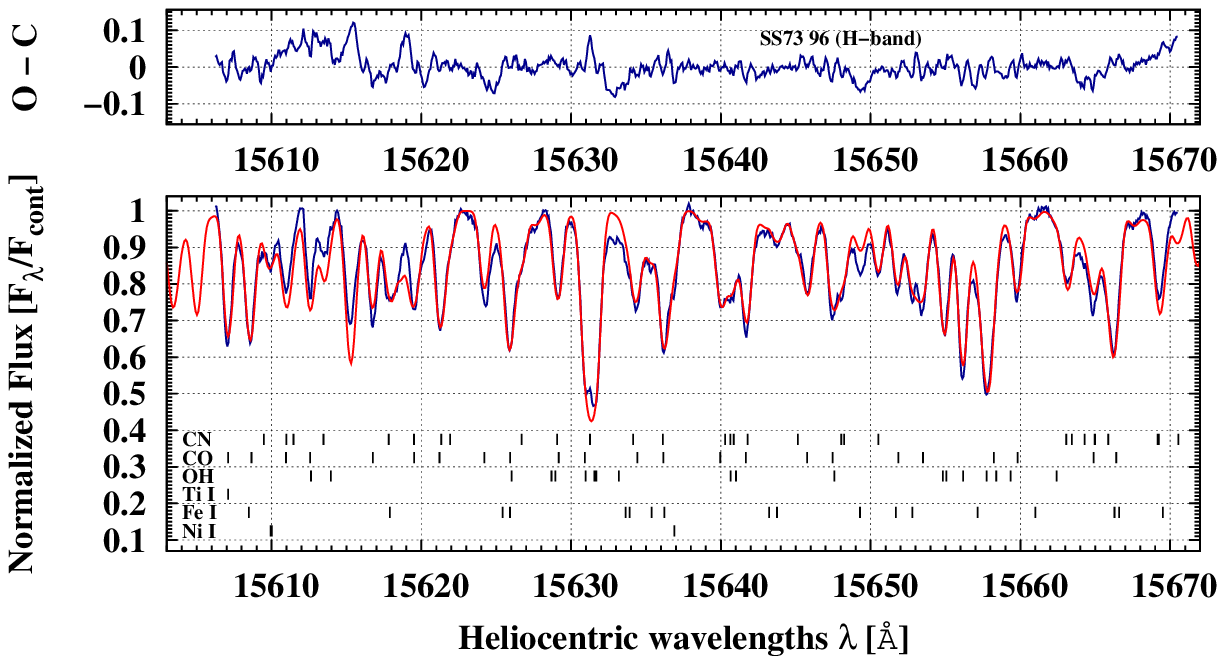}
  \caption{The $H$-band spectrum of SS73\,96 observed 2010 June (blue line)
and a synthetic spectrum (red line) calculated using the final abundances
(Table\,\ref{T4}).}
  \label{FB51}
\end{figure}

\begin{figure}
  \includegraphics[width=84mm]{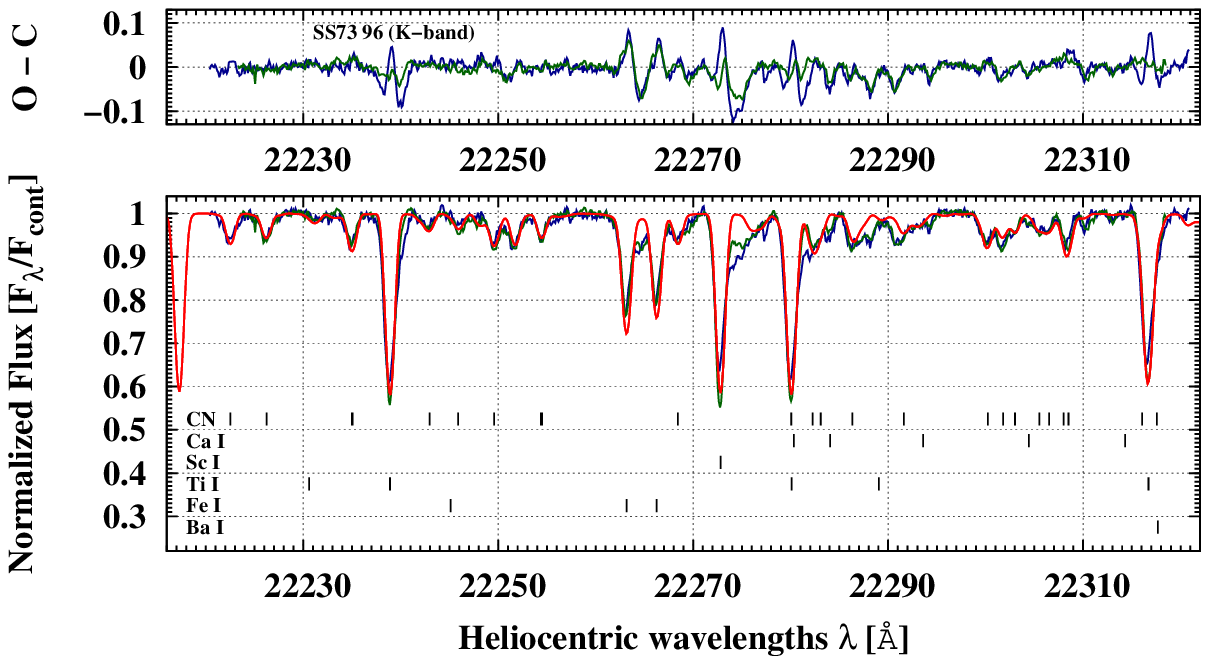} 
  \caption{$K$-band spectra of SS73\,96 observed 2003 April (blue line),
2004 April (green line), and a synthetic spectrum (red line) calculated
using the final abundances (Table\,\ref{T4}).}
  \label{FB52}
\end{figure}

\begin{figure}
  \includegraphics[width=84mm]{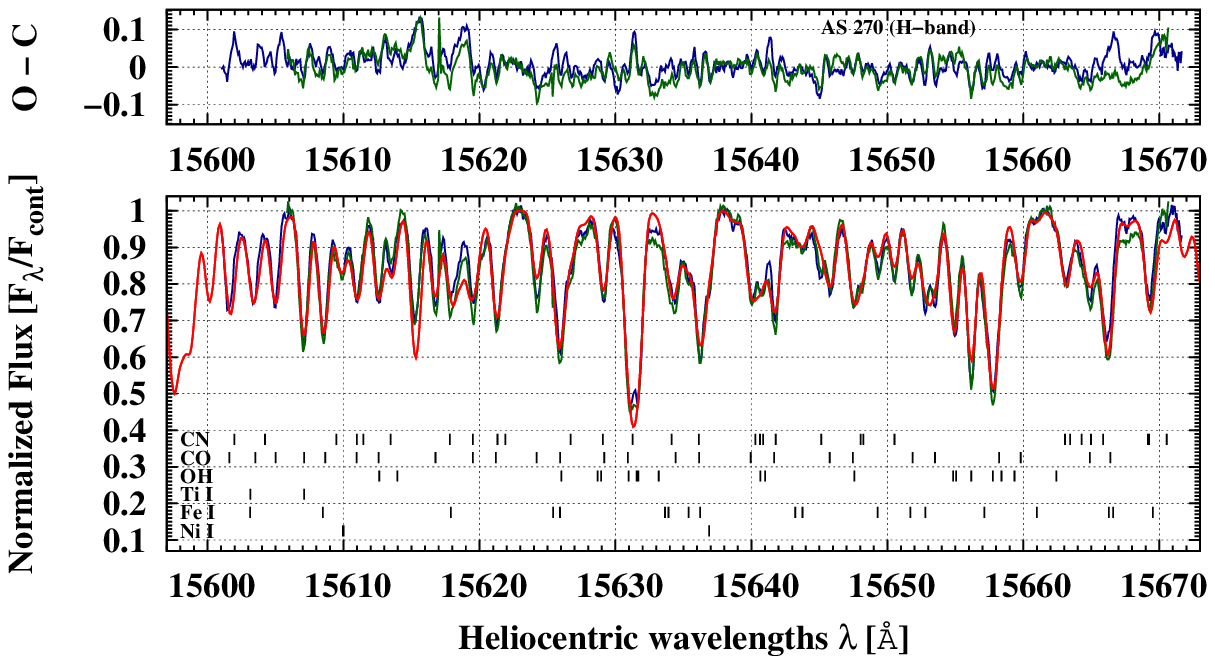}
  \caption{$H$-band spectrum of AS\,270 observed 2003 February (blue line),
2010 May (green line), and a synthetic spectrum (red line) calculated using
the final abundances (Table\,\ref{T4}).}
  \label{FB53}
\end{figure}

\begin{figure}
  \includegraphics[width=84mm]{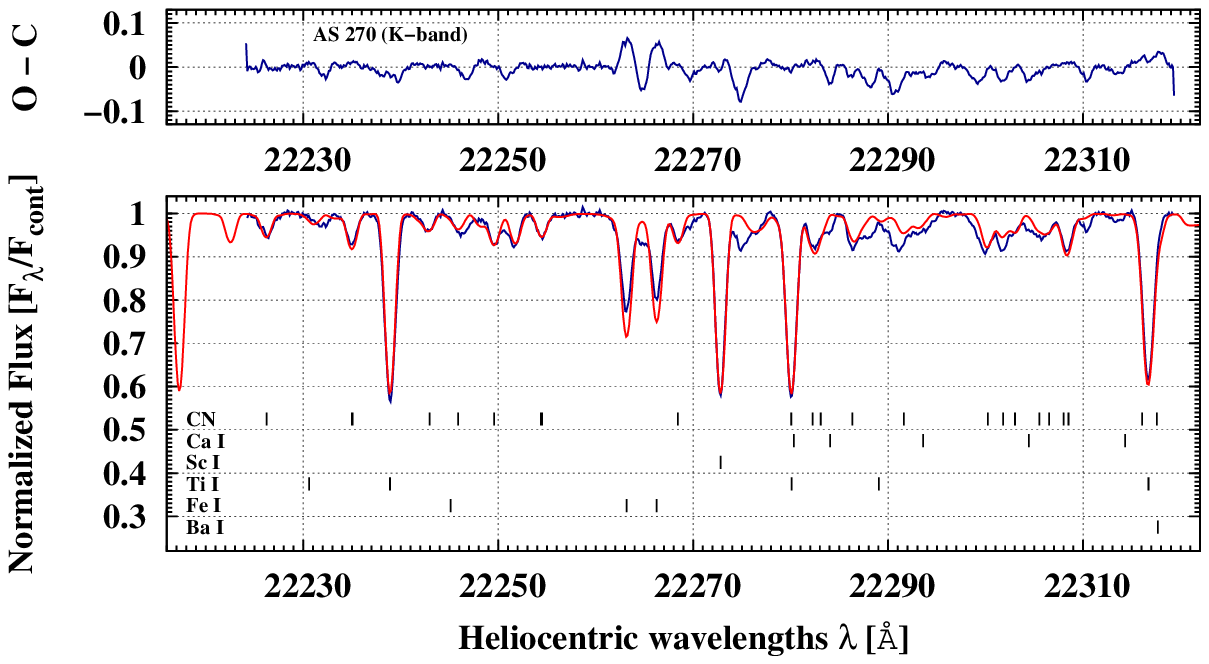} 
  \caption{The $K$-band spectrum of AS\,270 observed 2004 April (blue line)
and a synthetic spectrum (red line) calculated using the final abundances
(Table\,\ref{T4}).}
  \label{FB54}
\end{figure}

\begin{figure}
  \includegraphics[width=84mm]{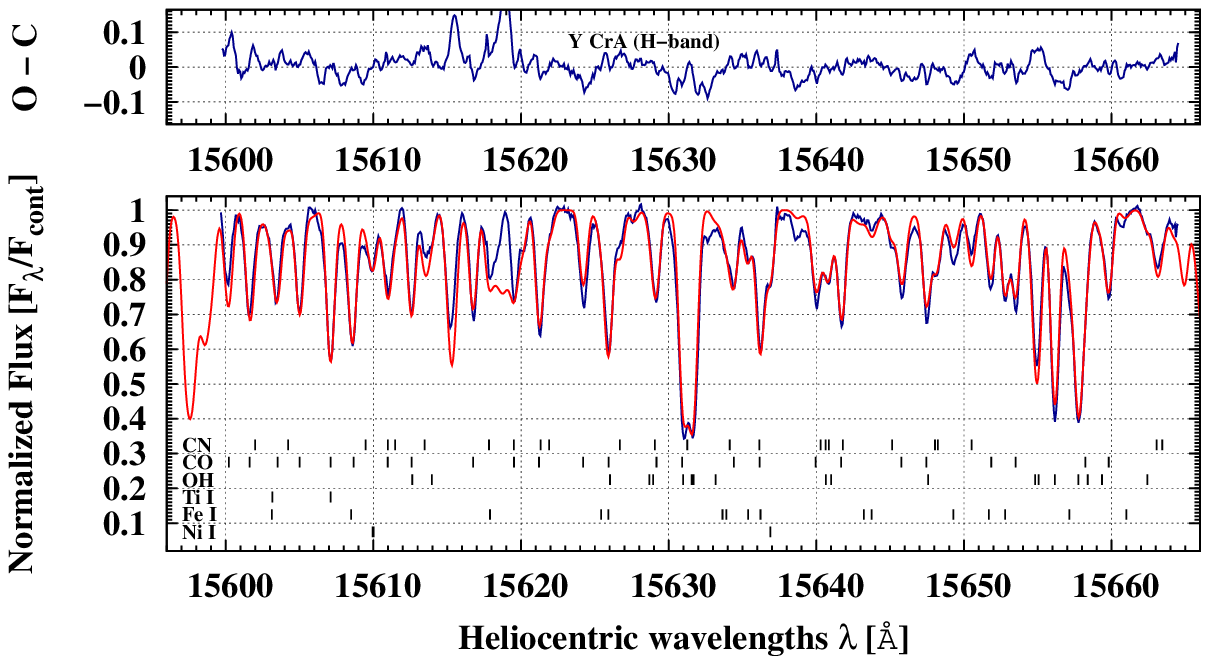}
  \caption{The $H$-band spectrum of Y\,CrA observed 2009 July (blue line)
and a synthetic spectrum (red line) calculated using the final abundances
(Table\,\ref{T4}).}
  \label{FB55}
\end{figure}
                                     
\begin{figure}
  \includegraphics[width=84mm]{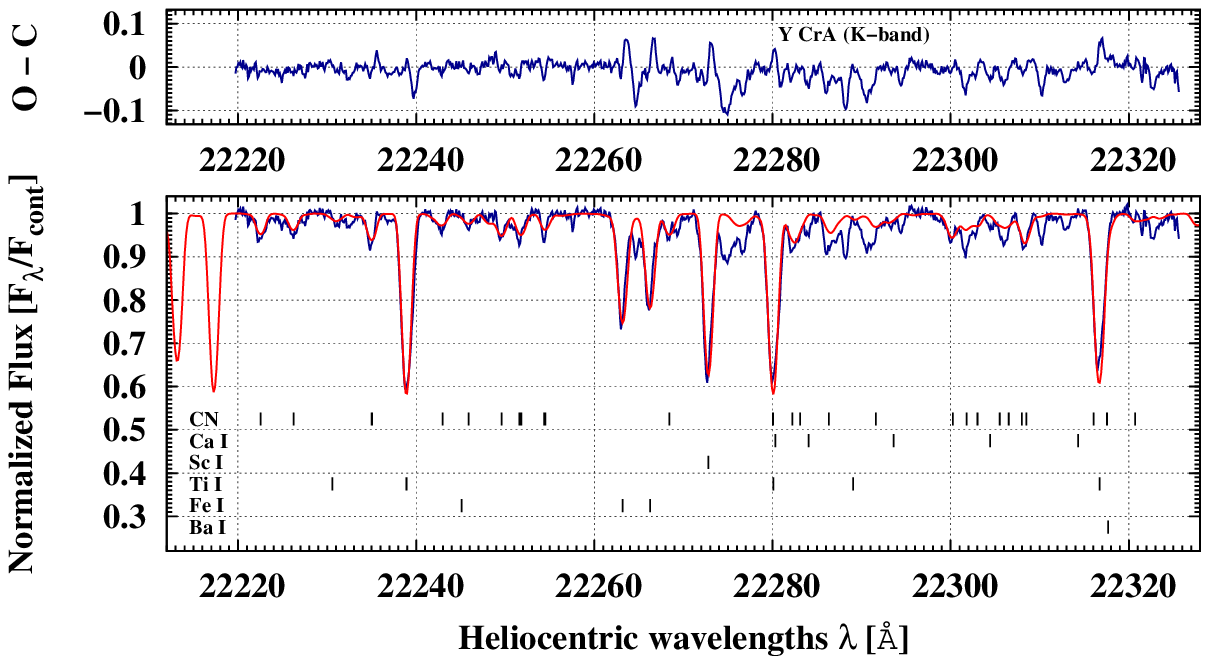} 
  \caption{The $K$-band spectrum of Y\,CrA observed 2003 April (blue line)
and a synthetic spectrum (red line) calculated using the final abundances
(Table\,\ref{T4}).}
  \label{FB56}
\end{figure}

\begin{figure}
  \includegraphics[width=84mm]{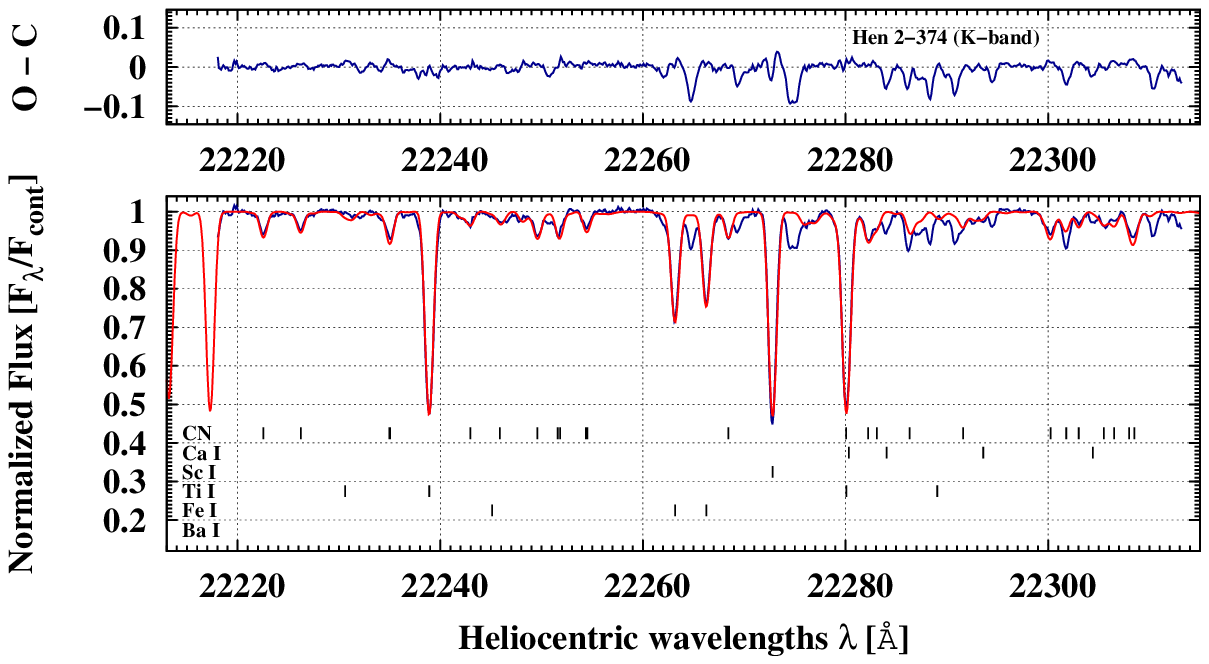} 
  \caption{The $K$-band spectrum of Hen\,2-374 observed 2004 April (blue
line) and a synthetic spectrum (red line) calculated using the final
abundances (Table\,\ref{T4}).}
  \label{FB57}
\end{figure}

\begin{figure}
  \includegraphics[width=84mm]{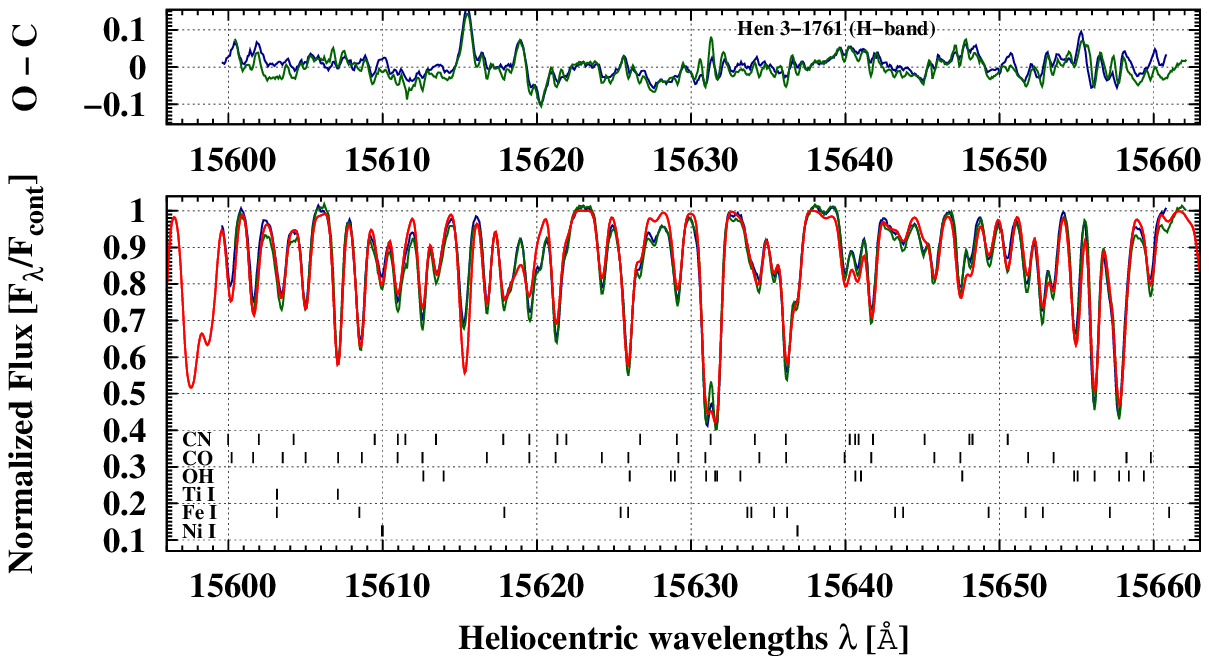}
  \caption{$H$-band spectra of Hen\,3-1761 observed 2009 June (blue line),
2010 June (green line), and a synthetic spectrum (red line) calculated using
the final abundances (Table\,\ref{T4}).}
  \label{FB58}
\end{figure}

\begin{figure}
  \includegraphics[width=84mm]{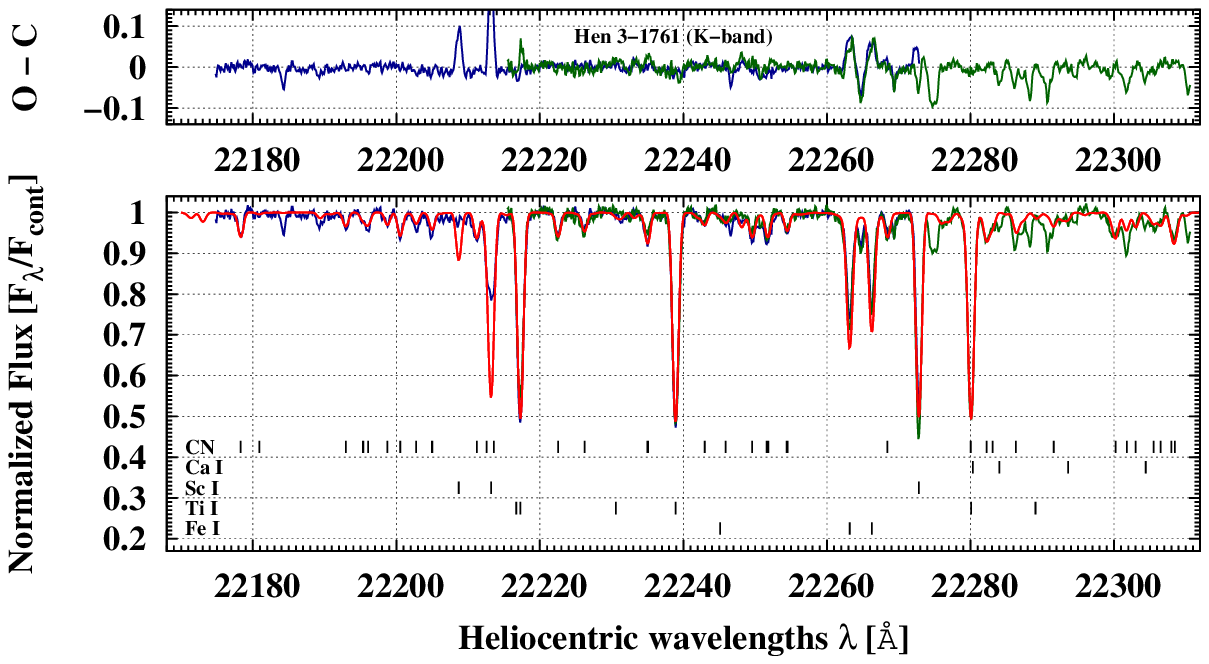}
  \caption{$K$-band spectra of Hen\,3-1761 observed 2003 August (blue line),
2004 April (green line), and a synthetic spectrum (red line) calculated
using the final abundances (Table\,\ref{T4}).}
  \label{FB59}
\end{figure}

\bsp

\label{lastpage}

\end{document}